\documentclass[fleqn,usenatbib]{mnras}

\usepackage{newtxtext,newtxmath}
\usepackage[T1]{fontenc}
\usepackage{ae,aecompl}

\usepackage{graphicx}	
\usepackage{amsmath}	
\usepackage{amssymb}	

\usepackage{multicol}
\usepackage{color}
\usepackage{multibib}

\newcites{app}{REFERENCES}
\defcitealias{2017ApJS..230...15M}{MS17}

\title[Uncertainties in Binary Pop Synth]{Evaluating the impact of binary parameter uncertainty on stellar population properties}

\author[E.R.Stanway et al.]{
E.~R.~Stanway,$^{1}$\thanks{E-mail: e.r.stanway@warwick.ac.uk}
A.~A.~Chrimes,$^{1}$
J.~J.~Eldridge,$^{2}$
and H.~F.~Stevance$^{2}$
\\
$^{1}$Department of Physics, University of Warwick, Gibbet Hill Road, Coventry, United Kingdom\\
$^{2}$Department of Physics, University of Auckland, Private Bag 92019, Auckland, New Zealand.
}

\date{Accepted 2020 April 22. Received 2020 April 22; in original form 2020 February 25}

\pubyear{2020}

\begin{document}
\label{firstpage}
\pagerange{\pageref{firstpage}--\pageref{lastpage}}
\maketitle

\begin{abstract}
Binary stars have been shown to have a substantial impact on the integrated light of stellar populations, particularly at low metallicity and early ages - conditions prevalent in the distant Universe. But the fraction of stars in stellar multiples as a function of mass, their likely initial periods and distribution of mass ratios are all known empirically from observations only in the local Universe. Each has associated uncertainties. We explore the impact of these uncertainties in binary parameters
on the properties of integrated stellar populations, considering which properties and timescales are most susceptible to uncertainty introduced by binary fractions and whether observations of the integrated light might be sufficient to determine binary parameters. We conclude that the effects of uncertainty in the empirical binary parameter distributions are likely smaller than those introduced by metallicity and stellar population age uncertainties for observational data. We identify emission in the He\,II\,1640\AA\ emission line and continuum colour in the ultraviolet-optical as potential indicators of a high mass binary presence, although poorly constrained metallicity, dust extinction and degeneracies in plausible star formation history are likely to swamp any measurable signal.
\end{abstract}




\section{Introduction}\label{sec:intro}

Stellar population and spectral synthesis (SPS) models provide the bridge between unresolved observations of galaxies or stellar clusters, and interpretation of their stellar content. Such models are routinely fit to both photometric and spectroscopic data, yielding stellar masses, population ages, star formation histories and sometimes metallicity or even redshift  \citep[see e.g.][]{1975MmSAI..46....3T,1976ApJ...203...52T,1995ApJS...99..173L,2003MNRAS.344.1000B,2009ApJ...699..486C,2010ApJ...712..833C,2013ARA&A..51..393C}. Each SPS model is constructed from models for the time evolution of individual stars, tracking changes in both their physical properties and their observable characteristics. A simple stellar population (SSP) accounts for the fragmentation of an initial star-forming region into stars according to an initial mass function (IMF) and traces the evolution of the resulting single-aged stellar population. A composite stellar population (CSP) combines SSPs of different ages and masses to construct a population with a known star formation history. 

The implicit assumption underlying this process is that the evolution and resultant atmospheres of individual stars are well understood, and that their distribution in initial mass can be described by a simple function. While the initial mass function is largely a valid construct for the ensemble of typical stellar populations in the local Universe, there is increasing evidence for variation between galaxies \citep[see][]{2018PASA...35...39H}, and low mass systems present a challenge since their initial mass function must be sampled stochastically rather than statistically \citep[see][]{2012MNRAS.422..794E}. Similarly, there are regimes in which theoretical understanding of stellar evolution is still undergoing active and rapid development. In particular, the evolution of stars at very low metallicity, at very high mass, undergoing rotation or influenced by binary interactions, remain subject to significant uncertainties \citep[e.g.][and references therein]{2012ARA&A..50..107L,2013ApJ...764..166D,2015ApJS..220...15P,2017ASPC..508..121V,2017PASA...34...58E,2020MNRAS.491.3479C,2020arXiv200104476S}. The  determination that the majority of massive stars interact with a binary companion during their evolutionary lifetime \citep{2012Sci...337..444S,2013A&A...550A.107S,2014ApJS..215...15S}, combined with advancements in extragalactic astronomy which push towards the young, metal-poor stellar populations at ever higher redshifts  \citep[e.g.][]{2018ApJ...869..123S,2018ApJ...868..117S,2018ApJ...869...92R,2019ApJ...871..128T}, now require these uncertainties to be addressed, or at least acknowledged, in SPS modelling and the fitting of galaxy data. In particular, interest has focused on the role of binary and multiple stars in shaping the evolution of a population. This has been shown to have a particularly significant effect on the ionizing photon production and its time evolution \citep[e.g.][]{2016MNRAS.456..485S,2019A&A...629A.134G}, and is evidently crucial in interpreting the rates of stellar compact mergers \citep[e.g.][]{2015ApJ...814...58D,2016MNRAS.462.3302E,2020MNRAS.493L...6T}

The Binary Population and Spectral Synthesis  \citep[BPASS,][]{2017PASA...34...58E,2018MNRAS.479...75S} project produces simple stellar population models from a custom grid of detailed stellar evolution tracks which incorporate not only the evolution of isolated single stars, but also binary interactions resulting from a range of binary initial mass ratio and initial separations. The use of detailed stellar models allows the response of individual stars to interactions, mass loss and certain rotational effects to be traced in detail, rather than approximated. This is important in  regimes in which the response of a star does not vary smoothly with mass, mass ratio, etc. and in determining the mass of core-collapse remnants created at the end of stellar evolution. However, as a consequence, the detailed approach lacks the speed of synthetic model populations based on algorithmic interpolation between a sparser grid of models \citep[e.g.][]{2002MNRAS.329..897H,2018MNRAS.473.2984I}. 

A typical detailed binary population and spectral synthesis model requires about 8 hours of processor time to run (assuming the underlying stellar models already exist and have been pre-processed to the correct format and resolution). The current BPASS data release model set (as of v2.2) incorporates 9 representative initial mass functions (IMFs), with or without including binary evolution tracks, each calculated at 13 metallicities - a total of 208 models which must be run independently, with the results output at 51 time steps. Our recent studies on gravitational wave transient rates have also involved trialling two different supernova kick velocity models (doubling the number of models again), albeit for a single IMF \citep{2020MNRAS.493L...6T}. In order to make this manageable with current computing resources, each model is calculated with only one set of assumptions for the properties of the binary population as a function of mass. However as computing power, and more importantly, the number of cores and RAM available per core, continues to increase, more ambitious model runs become possible.

In \citet{2019A&A...621A.105S} we explored the impact of varying our initial mass function, for a grid of variant models which were truncated at 1\,Gyr to reduce the run-time, but otherwise followed the BPASS v2.2 prescription described by \citet{2018MNRAS.479...75S}. In particular, we assumed that the binary population parameters as a function of mass were independent of metallicity, and remained fixed as the initial mass function varied. Here we reverse these assumptions: we fix the initial mass function and instead question the impacts of the binary population parameters and the uncertainties associated with them. Previous work in this field has been limited and has either considered simple uniform binary fractions with mass \citep{2018ApJ...867..125D} or has made use of rapid population synthesis methods \citep[e.g.][]{2015ApJ...814...58D} and has considered primarily resolved stellar populations or stellar transients.

In this paper we quantify uncertainties on key aspects of a binary stellar population and spectral synthesis model grid which arise directly as a result of uncertainties on the input binary population parameters. In section \ref{sec:params} we vary the initial binary parameters used as an input by resampling the empirical constraints given their known uncertainty distribution. In section \ref{sec:results} we evaluate the impact of such variation on the output properties of the integrated stellar light, considering ionizing photon production, photometric colour and the ultraviolet-optical spectrum. In section \ref{sec:base} we set aside the observational constraints on binary fraction and vary this distribution through a larger range in order to search for signatures which may be observable, particularly if binary fraction evolves towards low metallicity.  The stellar populations contributing most to the resulting variations and uncertainties are discussed in section \ref{sec:discussion}. Finally we present our conclusions in section \ref{sec:conc}.


\section{Binary population parameters and their uncertainties}
\label{sec:params}

\subsection{Binary Parameters}\label{sec:defs}

The models presented here are calculated as a full population and spectral synthesis as described in \citet{2017PASA...34...58E} and \citet{2018MNRAS.479...75S}, although each model is truncated at an age of 1\,Gyr. Where a star in a binary undergoes a supernova, we estimate the survival probability of the binary and weight models for the subsequent evolution of its stars assuming the Hobbs et al (2006) neutron star kick distribution, sampled 2000 times (sufficient to populate all but the rarest subsequent evolution pathways; a full BPASS data release model typically uses 100,000 kick iterations). All models presented here are calculated with the default initial mass function for single and primary stars ``imf135\_300" - a broken power law model with a slope of -1.3 for primary stars between 0.1 and 0.5\,M$_\odot$ and a slope of -2.35 between 0.5\,M$_\odot$ and an upper mass cut-off of 300\,M$_\odot$. Models are evaluated at three selected metallicities: Z=0.002, Z=0.010 and Z=0.020 (where Z is the metallicity mass fraction, and Z=0.020 is conventionally Solar metallicity). For the sake of brevity, these are labelled as ``z002", ``z010" and ``z020" respectively where appropriate. The same binary distribution parameters are taken to apply at all metallicities.

The binary populations implemented by v2.2 of the BPASS models are defined through the application of empirical constraints on five aspects of the population statistics, and these in turn are constrained by 65 individual quantities, each derived from a meta-analysis of the observed local stellar binary population and associated with an uncertainty by \citet[][hereafter MS17]{2017ApJS..230...15M}. These comprise the binary fraction as a function of primary mass (five empirical constraints), the companion probability as a function of period (three relevant period ranges, in five mass bins: 15 empirical constraints), and the dependence of companion mass ratio on mass and period, modelled as a broken power law (giving two power law slopes, each in three relevant period ranges, in five mass bins: 30 empirical constraints). A fourth period bin exists in the \citetalias{2017ApJS..230...15M} data, but stars in this regime would be considered single in BPASS, and this data is only utilised if a functional form is constructed as a function of period. We also utilise an additional 15 constraints on excess twin fraction (i.e. the number of systems with a mass ratio near unity, as a function of primary mass and period) although these are not considered further here, due to the limitations of BPASS stellar evolution models in precise handling of equal mass binaries. 

The current BPASS v2.2 implementation uses the best-fit values from table 13 of \citetalias{2017ApJS..230...15M}. Specifically:
\begin{enumerate}
    \item We implement a binary fraction for the purposes of BPASS as the complement of the single star fraction (i.e. $1-f_\mathrm{sin}$) at the median masses of the five stellar mass bins considered by \citetalias{2017ApJS..230...15M}. Masses are linearly interpolated between these fixed mass points. The value given at 16\,M$_\odot$ is loosely extrapolated to unity at 50\,M$_\odot$ and applied at all primary masses above this, while that at 1\,$M_\odot$ is extrapolated to lower masses (with the binary fraction constrained to lie between zero and unity in all cases).
    \item We implement a period distribution in each \citetalias{2017ApJS..230...15M} mass bin given by step-wise interpolation between the frequencies given at log (P/days) = 1, 3 and 5. A second linear interpolation is then performed between adjacent \citetalias{2017ApJS..230...15M} mass bins at fixed period, in order to populate a BPASS mass-period grid.
    \item We implement a mass ratio distribution by linearly interpolating the upper and lower mass-ratio power-law indices separately between fixed empirical data points. Again interpolation is performed initially in period at each \citetalias{2017ApJS..230...15M} mass bin and then between mass bin centres to populate the full BPASS grid in period and mass.
    \item Similarly the fraction of twin systems is interpolated first  as  a  function  of  period  and  then  of  mass,  and  an appropriate additional weighting  is added to our highest mass-ratio bin ($q=0.9$).
\end{enumerate}
Care is taken at each stage to ensure that the initial period and mass ratio distributions at each mass sum to unity, and that the weighting of each primary or single star is drawn from the underlying initial mass function, together with the binary fraction for its mass.

\subsection{Uncertainties on inputs}\label{sec:uncertainties}

The current population synthesis models do not take into account the uncertainties on these measurements, but simply apply the best fit values reported by \citetalias{2017ApJS..230...15M}. 
As Figs~\ref{fig:fbin} and \ref{fig:fperiod} demonstrate, the uncertainties on the \citetalias{2017ApJS..230...15M} best-fit binary population parameters can be considerable. For example, the companion frequency for O-type stars with log(P/days)=3 is 0.32$\pm$0.11, while the power law slopes on the mass ratio distributions have typical uncertainties of $\pm$0.4 or more. BPASS v2.2, as described above, is also restricted to linear interpolation between fixed mass and period points, rather than considering any functional relationship between parameters.

\citet{2017ApJS..230...15M} suggested that the fraction of multiples at each log(initial primary mass) could be fit with a linear function, while the period distribution at each mass is well approximated by a normal distribution in log(Period). Fitting such functions and implementing them directly as a function of mass and period mitigates the scatter associated with each individual measurement and would remove some of the unphysically abrupt changes in binary parameters which result from our interpolation, but rely on these being the correct functional forms.

Given the computational burden of each synthesis model, undertaking the tens of thousands of iterations likely to be required by a Bayesian MCMC approach to calculate the posterior probability distribution of all modelled values remains unfeasible. This is particularly true given how little is known about the prior probability distributions on each possible input parameter. Instead we apply a bootstrap resampling of the empirical binary distribution parameters in order to establish the likely impact of their uncertainties.

\subsection{Calculating Perturbed Models}\label{sec:method}

\begin{figure}
	\includegraphics[width=\columnwidth]{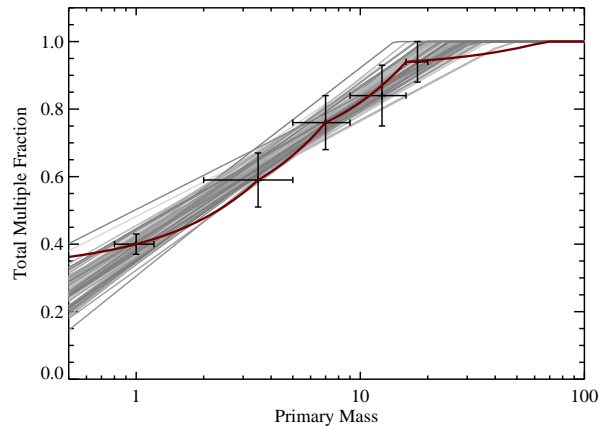}
    \caption{Interpolation of \citetalias{2017ApJS..230...15M} multiple fractions onto the BPASS mass grid. Points with error bars are the canonical values of \citetalias{2017ApJS..230...15M} in each mass bin. The thick line is the result of the BPASS v2.2 interpolation algorithm. Fainter lines show trials in which the fixed points have been perturbed through random sampling of their uncertainty distributions before fitting with a linear dependence of binary fraction on log(primary mass).}
    \label{fig:fbin}
\end{figure}

\begin{figure*}
	\includegraphics[width=0.65\columnwidth]{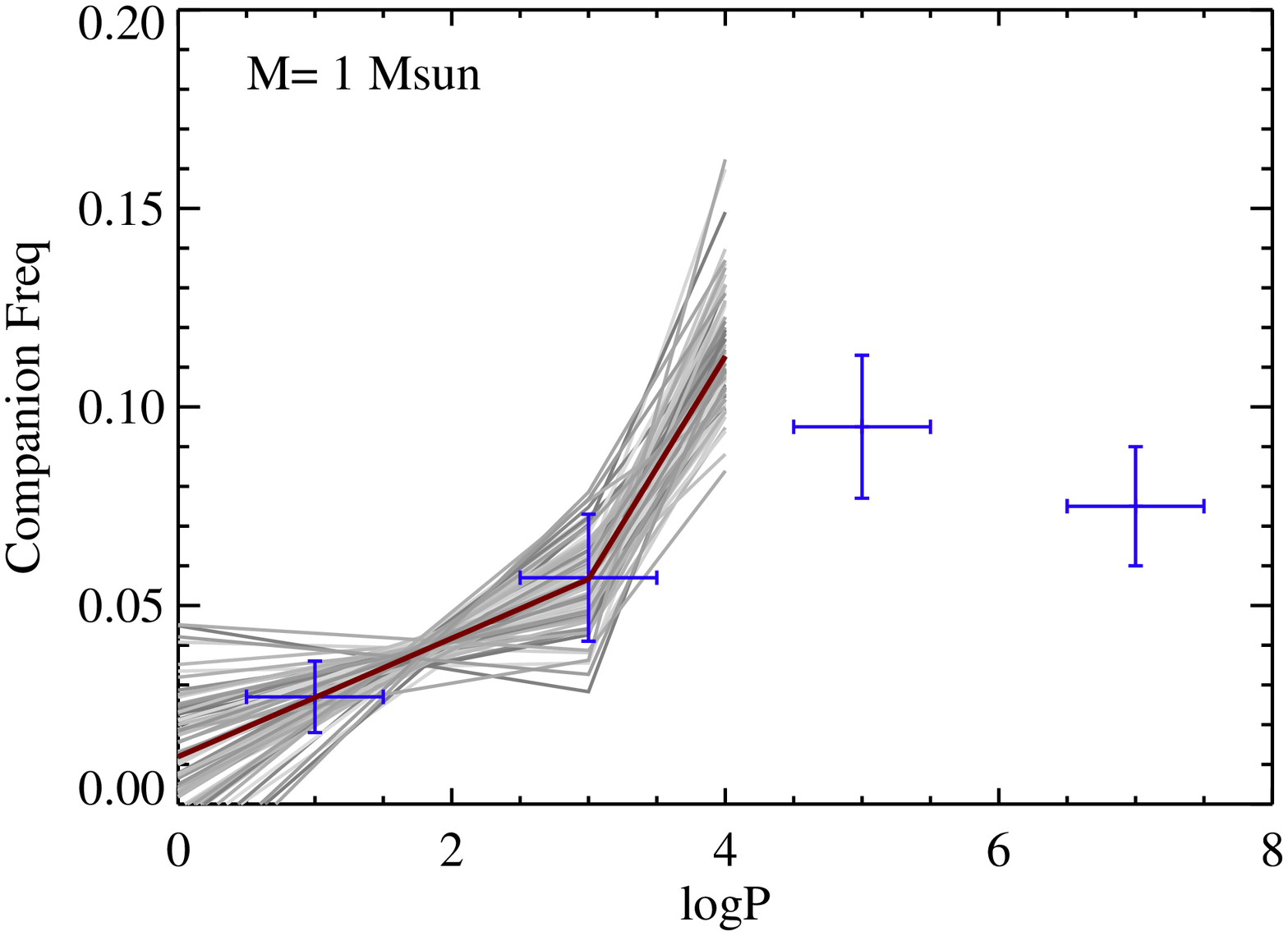}
	\includegraphics[width=0.65\columnwidth]{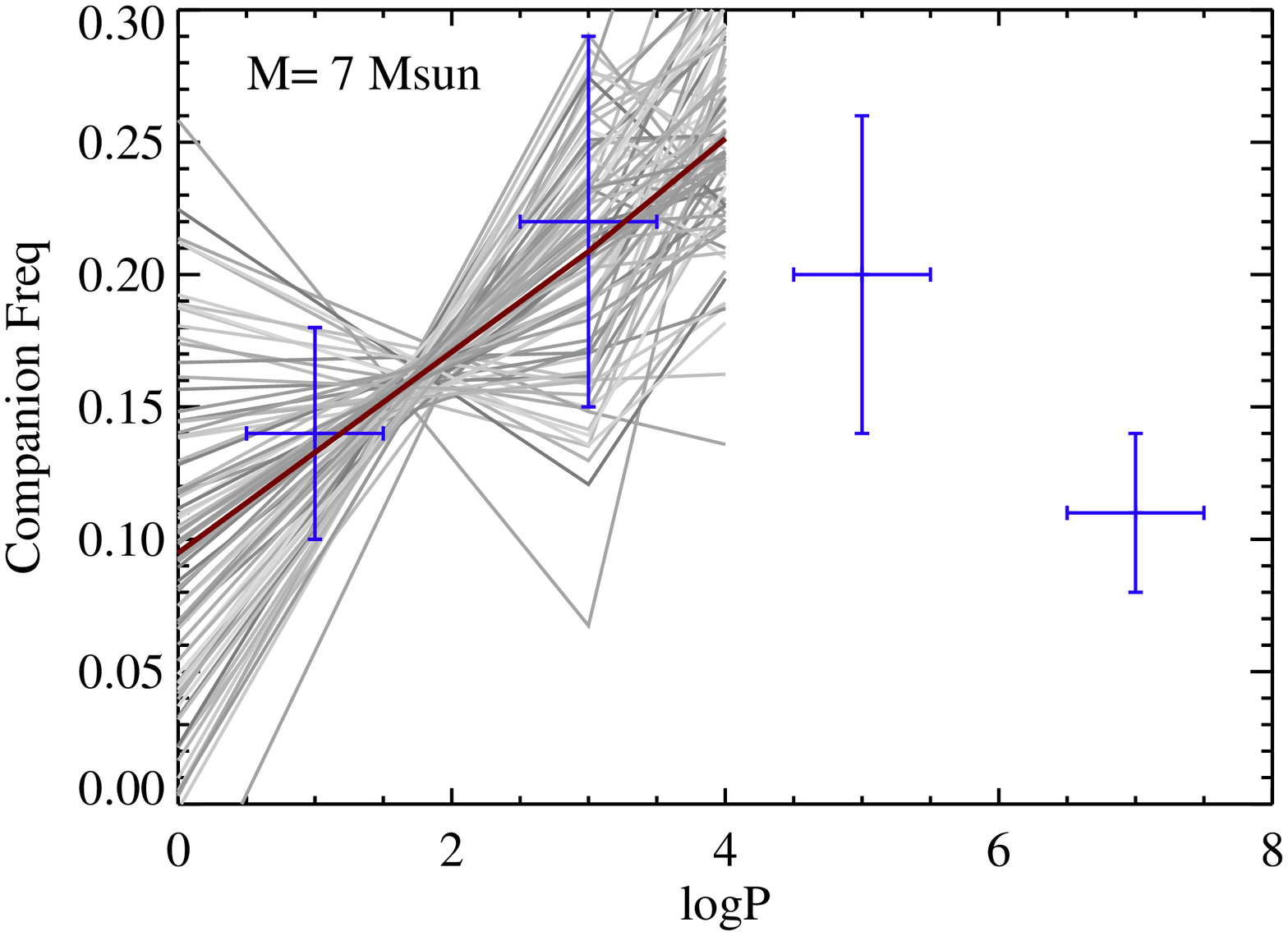}
	\includegraphics[width=0.65\columnwidth]{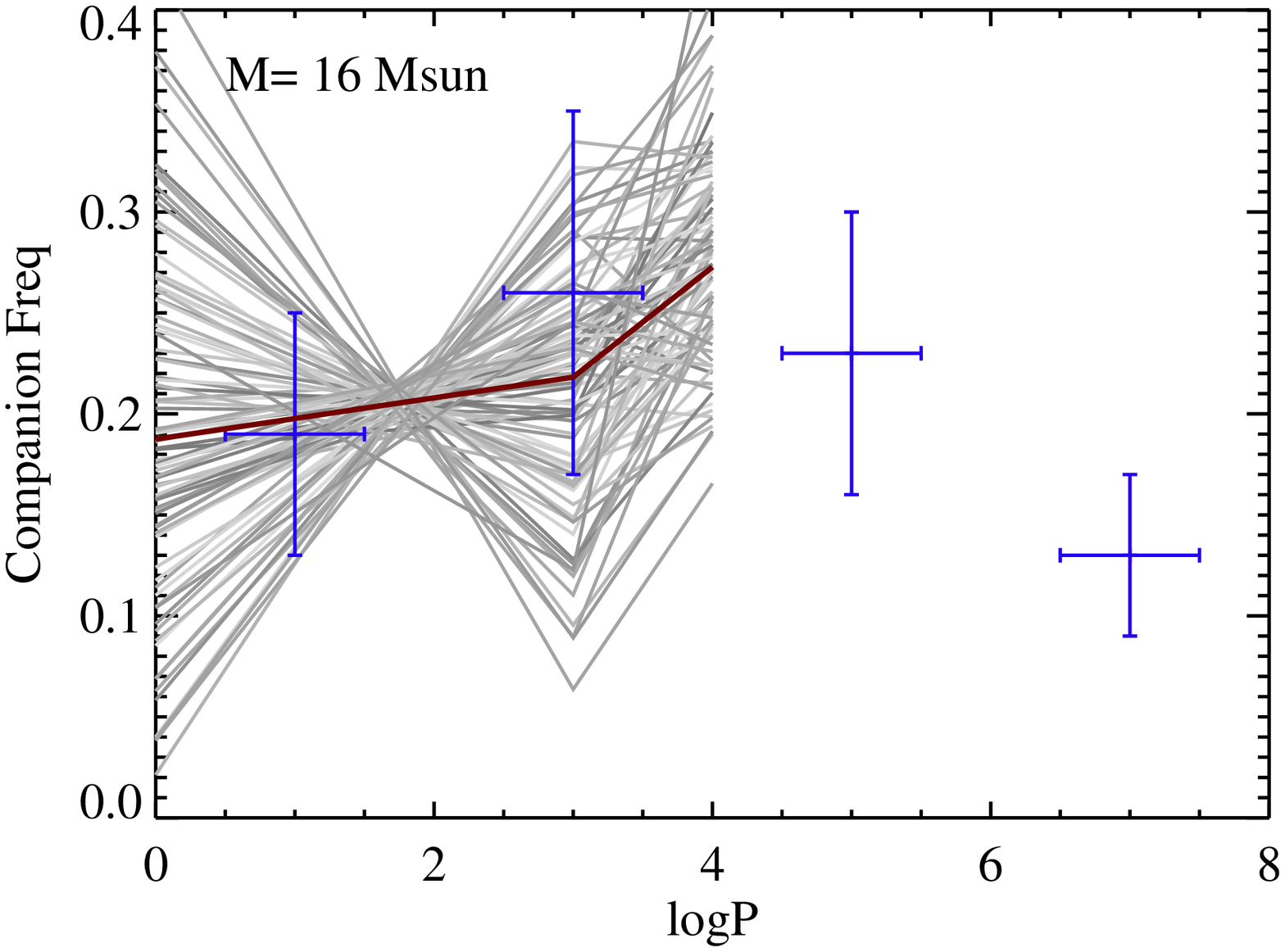}
	\includegraphics[width=0.65\columnwidth]{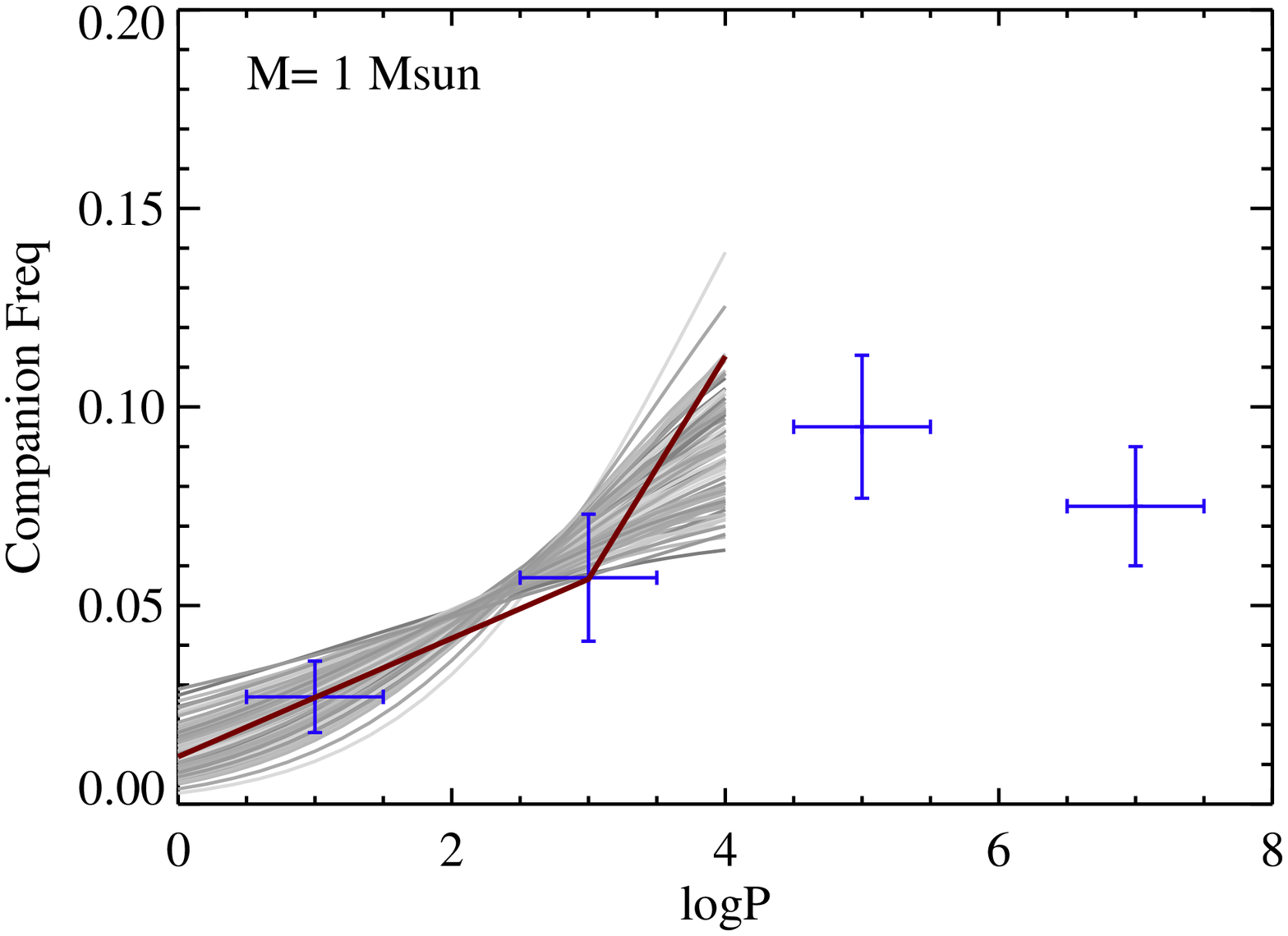}
	\includegraphics[width=0.65\columnwidth]{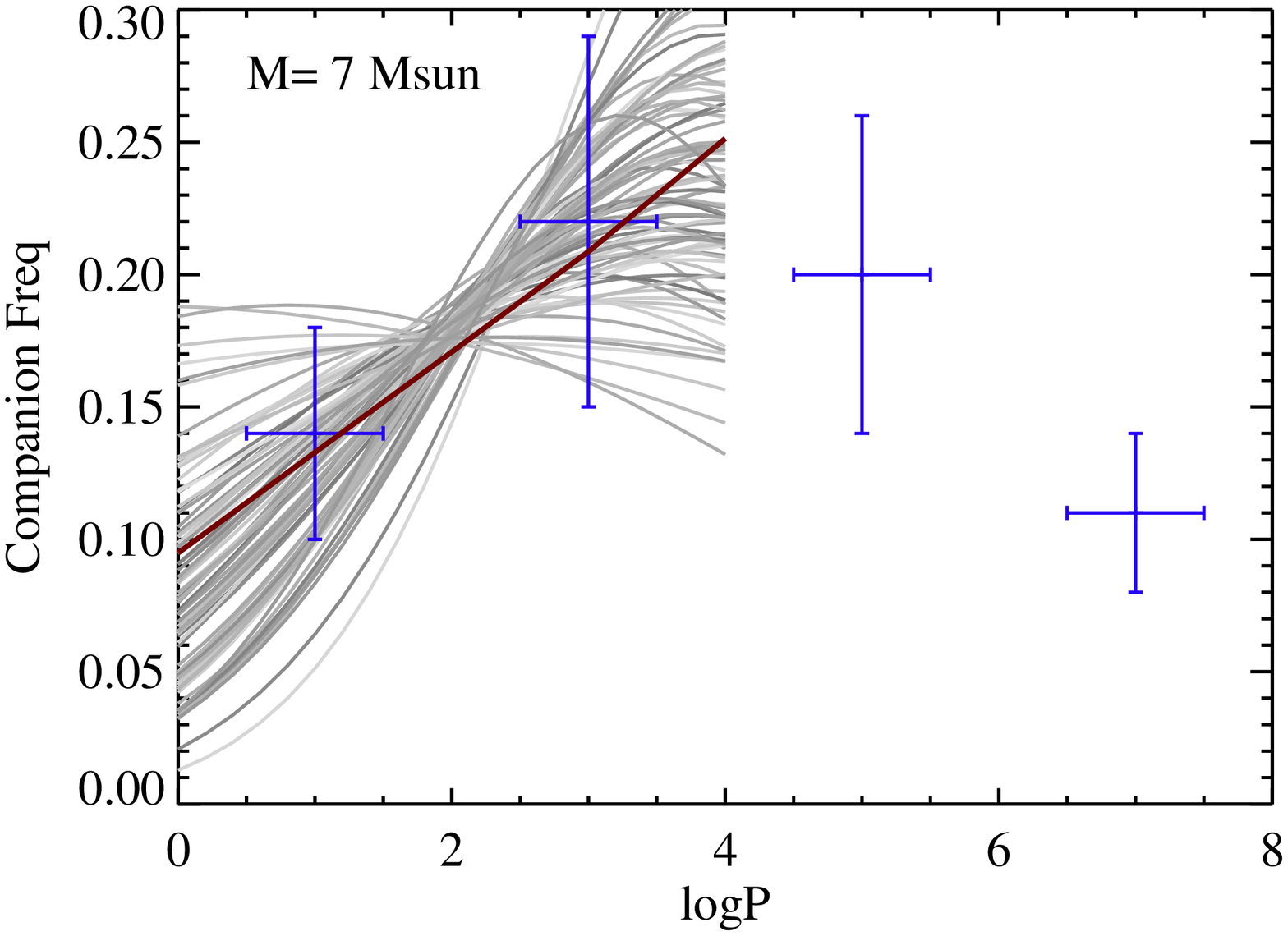}
	\includegraphics[width=0.65\columnwidth]{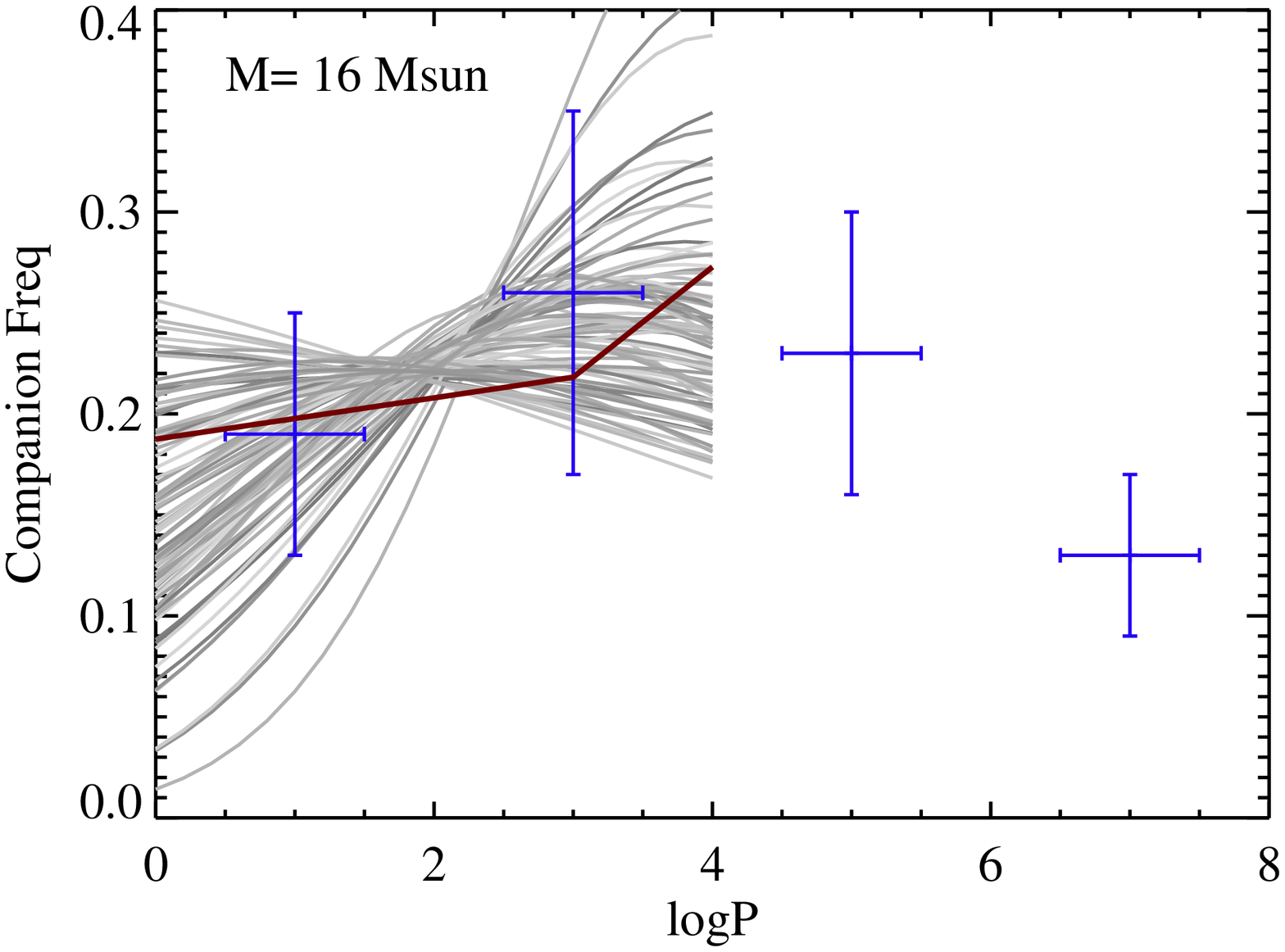}
    \caption{Two methods for interpolating the \citetalias{2017ApJS..230...15M} initial period frequencies as a function of mass into distributions on the BPASS grid. Faint lines show trials in which the fixed points have been perturbed through random sampling of their uncertainty distributions before fitting or interpolation. Top: linear interpolation between fixed points; bottom: a gaussian fit to log(P) at each mass. Points with error bars are the canonical values of \citetalias{2017ApJS..230...15M} in this mass bin. The thick line is the result of the BPASS v2.2 interpolation algorithm. BPASS treats binaries wider than log(P/days)=4 as effectively single.}
    \label{fig:fperiod}
\end{figure*}

We evaluate an extensive set of variant stellar population and spectral synthesis models, each with a slightly different set of input binary population parameters. To define these models, perturbations are randomly drawn from a Gaussian distribution with a width scaled to match the uncertainty reported on the parameter by \citetalias{2017ApJS..230...15M}. These values represent a combination of Poisson detection statistics, measurement uncertainties, fitting uncertainties and systematic uncertainties, reported by \citetalias{2017ApJS..230...15M} as a symmetric 1\,$\sigma$ error range on each parameter. The perturbations are applied to the best fit parameters and a new synthesis performed, either using the existing interpolation scheme as described above, or by fitting a functional form to the newly-perturbed parameters. We vary categories of parameters separately (leaving remaining parameters untouched at the BPASS v2.2 defaults unless otherwise stated), before performing a simultaneous perturbation to all the input parameters.

\begin{figure}
	\includegraphics[width=\columnwidth]{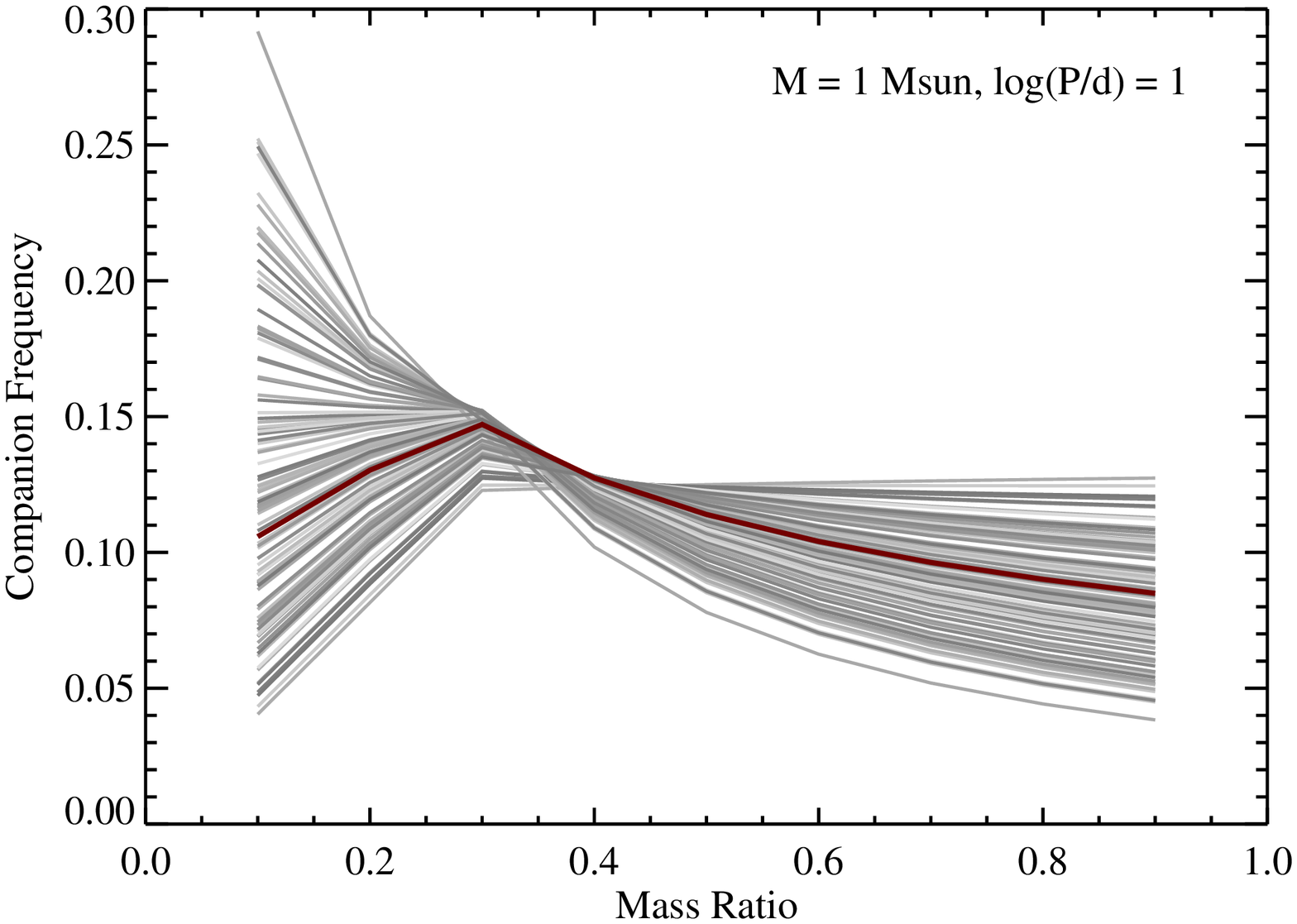}
	\includegraphics[width=\columnwidth]{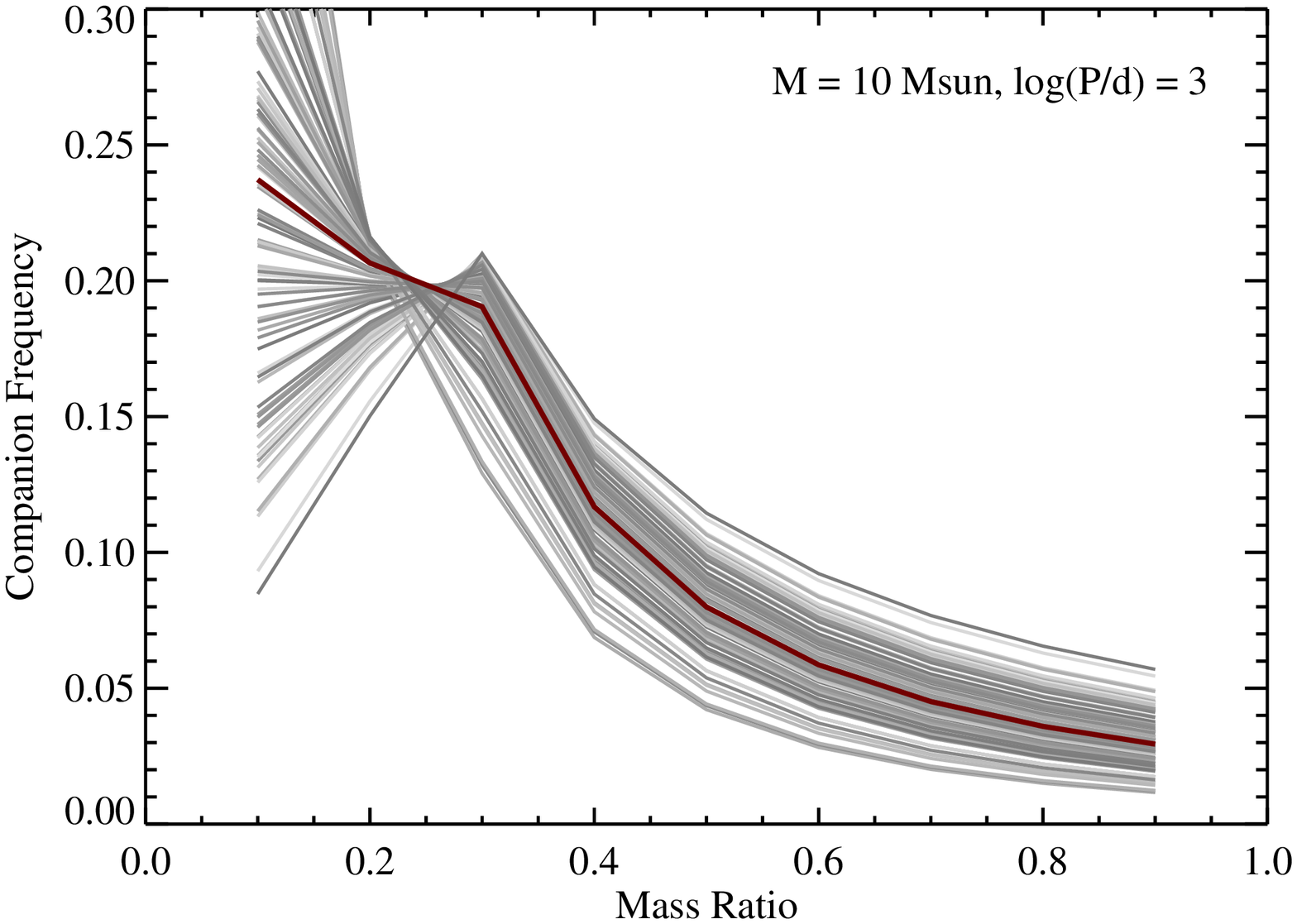}
    \caption{Examples of interpolation of \citetalias{2017ApJS..230...15M} initial mass ratio distributions onto the BPASS mass grid. \citetalias{2017ApJS..230...15M} provide two power law slopes of the form $p_q\propto q^\gamma$ at each of five mass bins and four initial periods. The thick line is the result of the BPASS v2.2 interpolation algorithm. Faint lines result from perturbing each of the 20 initial power law slopes (simultaneously) before interpolation. In each case, the sum of the mass ratio frequencies is normalised to unity.}
    \label{fig:fq}
\end{figure}

Variant models are calculated as follows:
\begin{enumerate}
    \item We perturb the binary fraction as a function of initial mass only. Mass ratio and period distributions remain fixed. A linear fit is used to binary fraction versus log(primary mass). 120 iterations are calculated per metallicity. The resultant binary fraction distributions are shown in Fig.~\ref{fig:fbin}.
    \item We perturb the period distribution simultaneously in all five mass bins. Binary fraction and mass ratio distributions are fixed. Step-wise linear interpolation is used between mass and period bins. 120 iterations are performed per metallicity. The resultant binary period distributions at three masses are shown in Fig.~\ref{fig:fperiod} (upper panels).
    \item We perturb the period distribution simultaneously in all five mass bins. Binary fraction and mass ratio distribution remain fixed. A Gaussian functional form is fit to perturbed values of log(P) at each mass, where these are linearly interpolated between mass bins as necessary. 120 iterations are calculated per metallicity. The resultant binary period distributions at three masses are shown in Fig.~\ref{fig:fperiod} (lower panels).
    \item We perturbing the two mass ratio power law indices simultaneously in all five mass bins, three period bins and two mass ratio bins. Binary fraction and period distribution remain fixed, with 120 iterations per metallicity. The resultant binary mass ratio distributions at two combinations of mass and period are shown in Fig.~\ref{fig:fq}.
    \item We simultaneously perturb all of the above. 300 random draws are taken from the uncertainty distributions on each empirical parameter, at each metallicity. Input distributions are calculated from the perturbed parameters using a linear fit to the binary fraction with log(primary initial mass), a Gaussian fit to the distribution in log(Period) at each mass, and the perturbed broken power law for the mass ratio distribution at each mass and period. Uncertainties are assumed to be uncorrelated.
\end{enumerate}

As the figures demonstrate, the observational uncertainties in binary fraction as a function of mass are relatively small, allowing for little spread in the models except at the high and low mass limits of the observational data. By contrast, the uncertainties on the period distribution as a function of mass permit substantial variation, particularly when a step-wise linear interpolation method is employed. This relatively crude approach permits fractions of close, high mass stars ranging from near zero to twice the canonical fraction, with potentially large impacts on the output population. If a Gaussian fit is used instead, the more extreme models become disfavoured, and the evolution in period far smoother and likely more physical. Uncertainties on the mass ratio distribution are also large, particularly for mass ratios approaching zero - the distributions for mass ratios $q>0.3$ are generally well constrained, with a slowly declining frequency as the mass ratio approaches unity. By contrast those at $q<0.3$ show substantial variation permitted by current observational uncertainties, ranging from a decline with increasing mass ratio to a rapid growth.

We note that the uncertainties on binary parameters are likely correlated in reality (i.e. a lower binary fraction at a given mass may be degenerate with a period distribution biased towards larger separations). Thus the estimates on output scatter from variant sample (v) represent a worst-case scenario given current observational constraints.

Outputs of each variant model include stellar type number counts, an integrated stellar spectrum, and various parameters derived from these, including the ionizing photon flux and colours. Models output SSPs at intervals in log(age/years)=0.1 between $10^6$ and $10^9$ years, these can be combined to produce complex CSPs given a star formation history. Older populations are not considered here.

\section{Results}\label{sec:results}

\subsection{Ionizing photon output and UV continuum}\label{sec:nion}

A key output from an SPS model is the ionizing photon production rate, and its efficiency compared to continuum emission longwards of the 912\AA\ Lyman break. For any stellar population which incorporates young stars, the ionizing spectrum is likely to be reprocessed by nebular gas in the interstellar medium, and heavily modify the resulting spectrum. A full calculation of the observed spectrum requires the stellar spectra derived from SPS codes to be passed through an independent radiative transfer code. We do not undertake this here, but instead consider the ionizing photon production directly as a function of metallicity for a case in which a stellar population is undergoing star formation at a constant rate of 1\,M$_\odot$\,yr$^{-1}$. Since it takes time for the rate of stellar birth to balance that of stellar death, and so for the output spectrum to stabilise, we evaluate the photon production at a fixed epoch of 100\,Myr after the onset of star formation.

\begin{figure*}
	\includegraphics[width=\columnwidth]{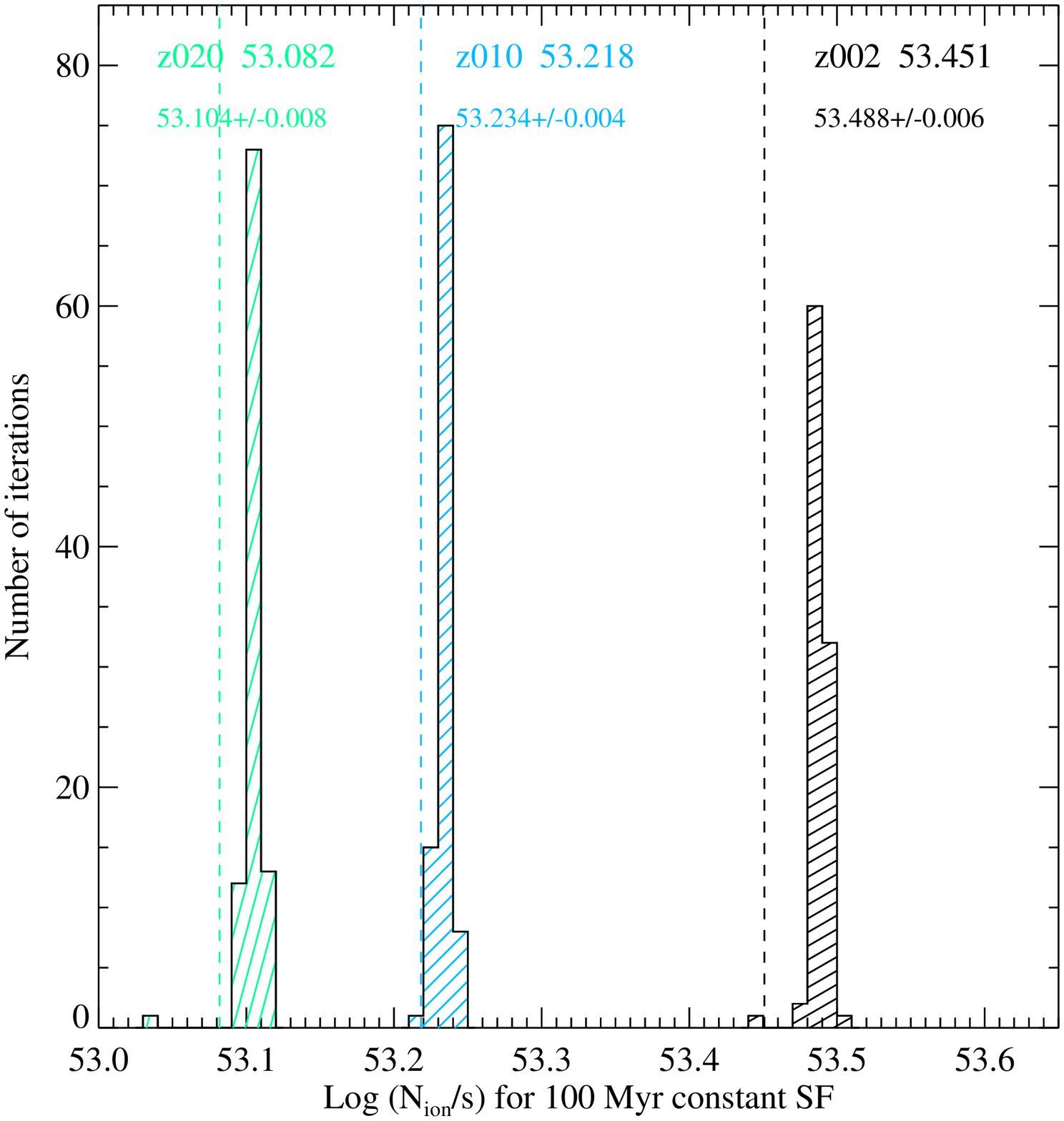}
	\includegraphics[width=\columnwidth]{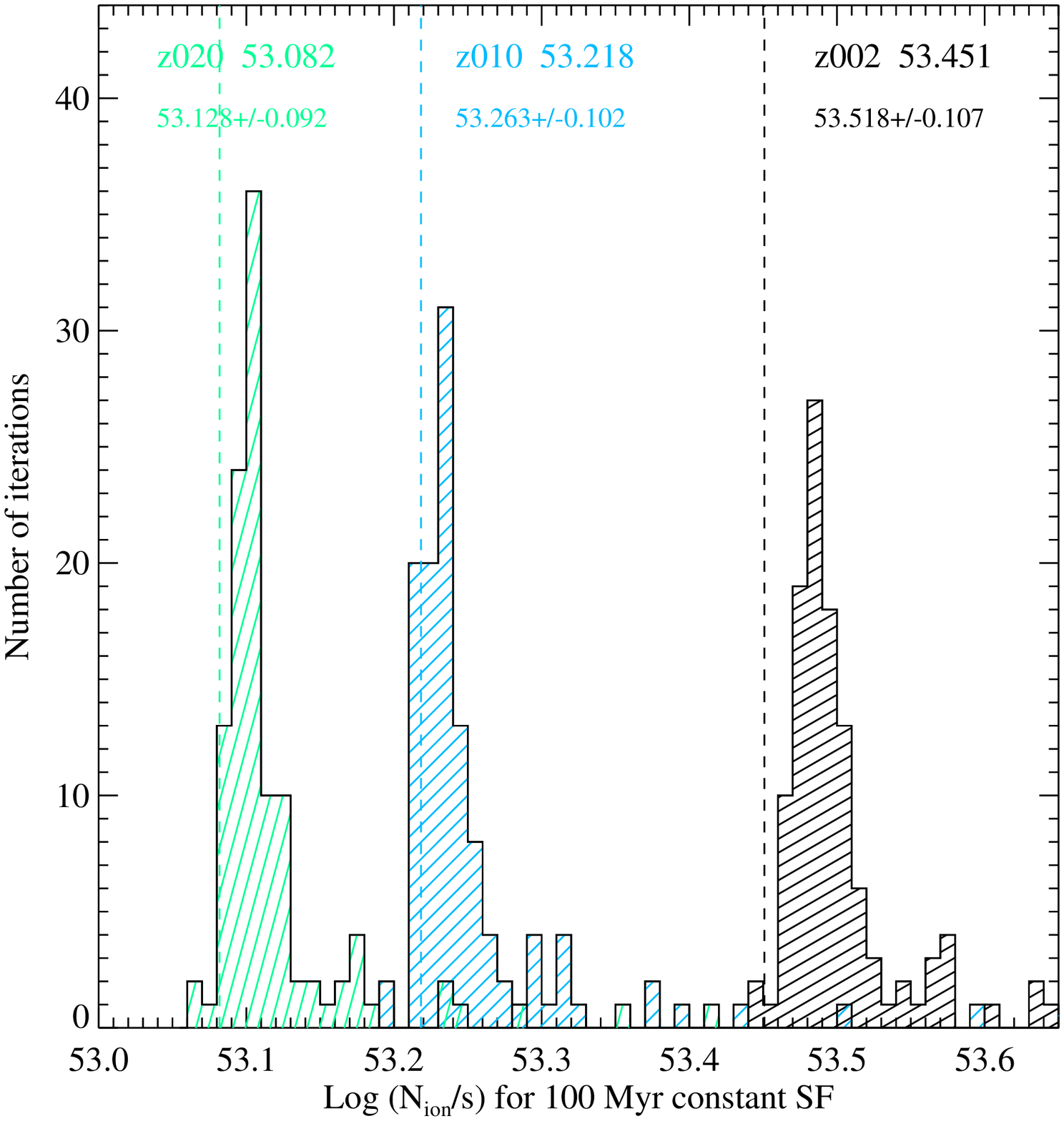}
	\includegraphics[width=\columnwidth]{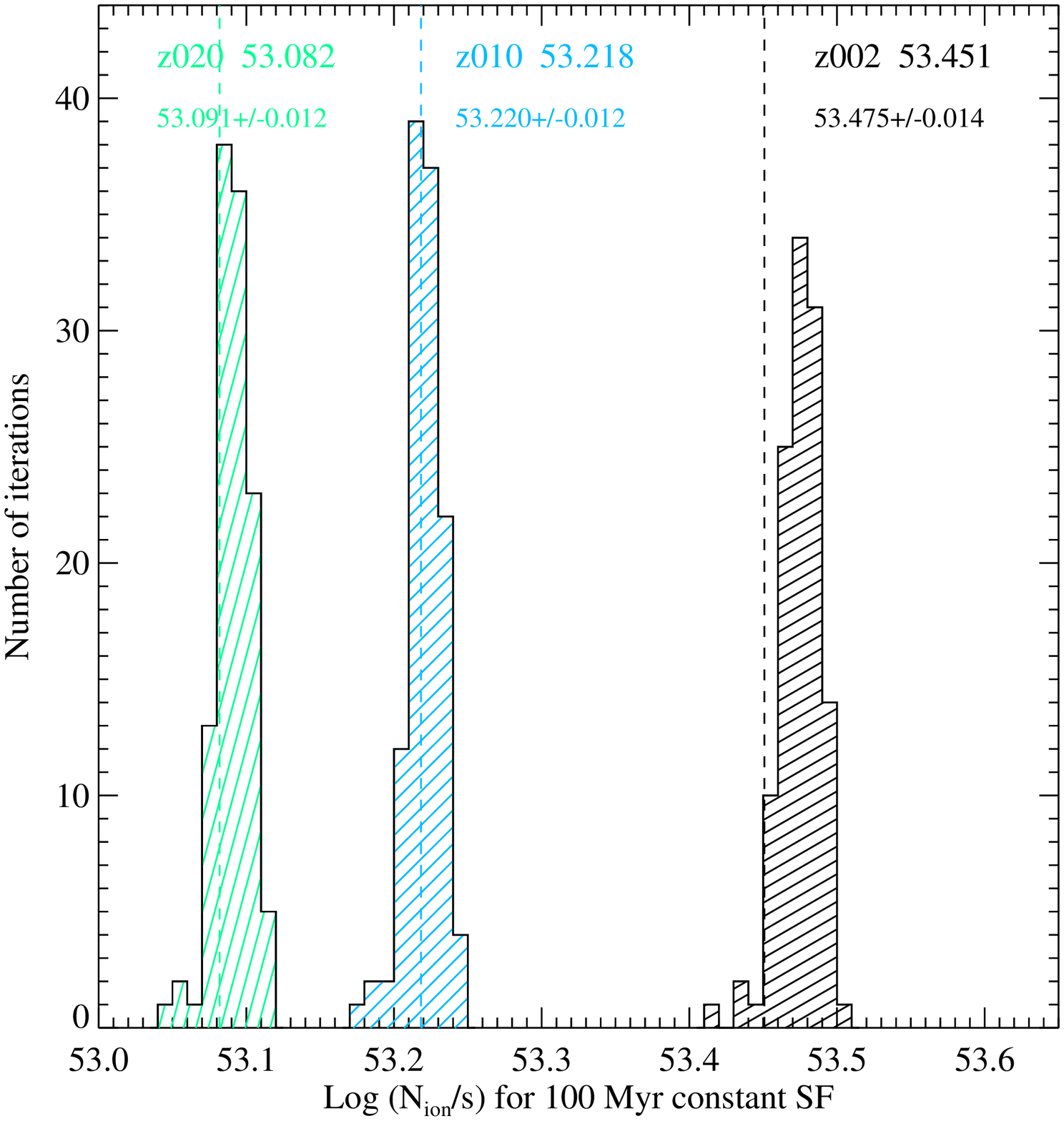}
	\includegraphics[width=\columnwidth]{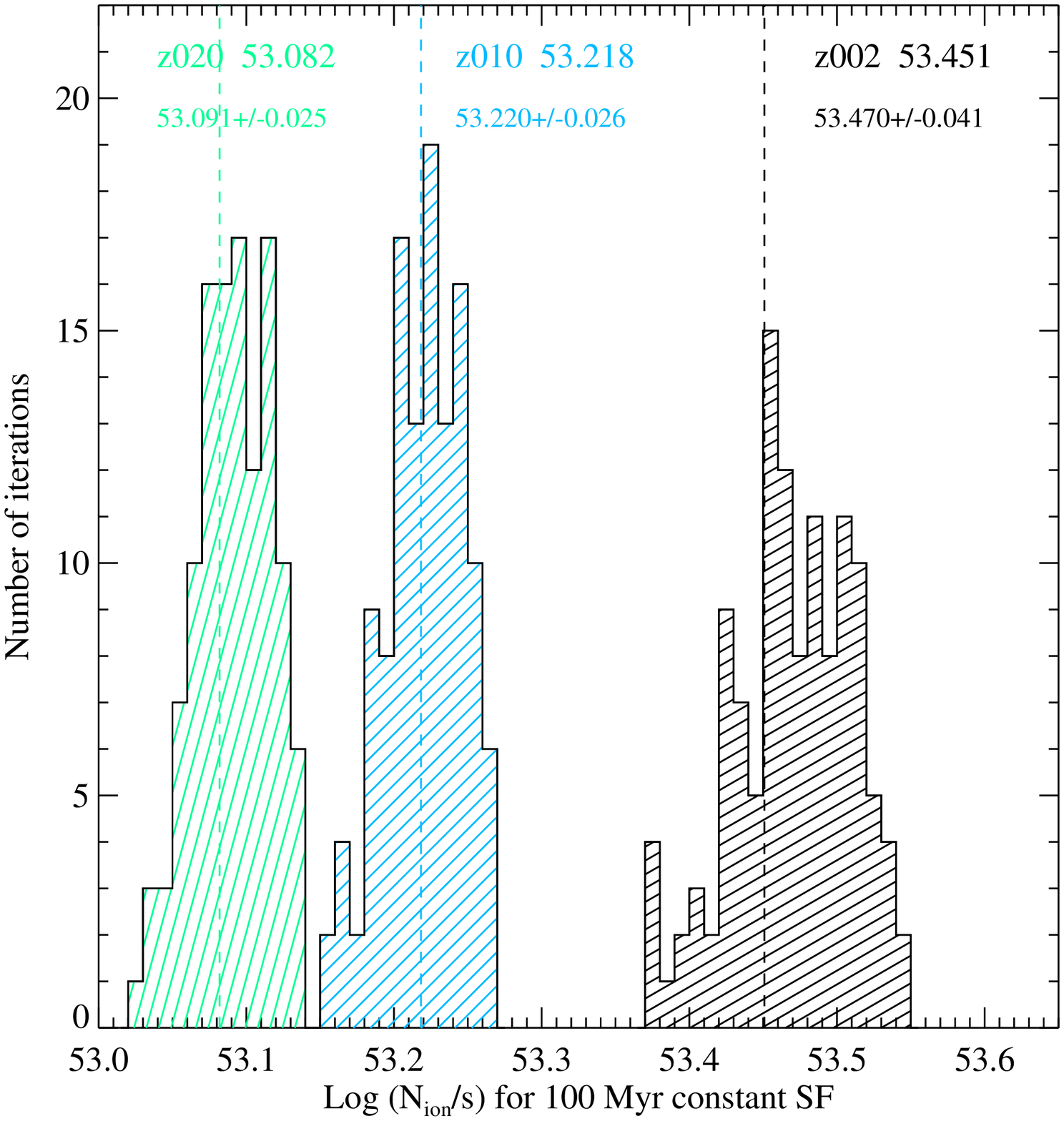}
    \caption{The variation seen in the ionizing photon production rate, for populations forming stars at a constant rate of 1\,M$_\odot$\,yr$^{-1}$, evaluated after 100\,Myr, at three different metallicities. Top left: varying binary fraction (linear fit), top right: varying period distribution (stepwise fit), bottom left: varying period distribution (log-gaussian fit), bottom right: varying mass ratio parameters. In each case other parameters are kept fixed using the extant BPASS v2.2 prescription. Dashed vertical lines indicate the fiducial BPASS v2.2 values which are given on the top line.}
    \label{fig:cnion}
\end{figure*}

The variation in the ionizing photon output in this continuous star formation case is illustrated in Fig.~\ref{fig:cnion} and tabulated in Table~\ref{tab:cnion} for variants (i) to (iv) above. In each case, the scatter introduced by variation in the binary inputs is well below the differences between the three metallicities considered here, and typically below $0.05$\,dex (except in the case of set (ii)). The effect of switching from step-wise interpolation of the binary fraction between mass bins, as implemented in BPASS v2.2, to a linear fit to the fraction as a function of mass is a small, metallicity-dependent systematic shift of [0.04, 0.02, 0.02] dex at Z=0.002, 0.010 and 0.020 respectively. This is larger than the random error introduced by perturbations on the binary fractions themselves, which are $<$0.01 dex at all metallicities.

The uncertainties introduced by varying the binary period distribution (while holding other parameters fixed) depend on the method of interpolation. The step-wise linear interpolation approach currently adopted in the BPASS v2.2 model grids leads to both a systematic offset in the ionizing photon rate at each metallicity and an increased standard deviation (although this is driven primarily by large outliers which have either very large or very small numbers of interacting binaries at the highest masses), both systematic and random uncertainties are comparable in scale at $\sim$0.1\,dex. By contrast, a log-gaussian interpolation approach leads to a much smaller scatter in ionizing photon rates with perturbation of the input parameters, resulting in only $<$0.015\,dex variation, while also producing a more physically plausible input period distribution. 

By contrast, perturbing the power law indices of the mass ratio distribution results in a broader distribution of ionizing photon rates than any of the other parameters, with no clear systematic offset but standard deviations of [0.041, 0.026, 0.025] dex at Z=0.002, 0.010 and 0.020 respectively. While not obviously related, the large scatter may result from the interaction of this parameter with the IMF sampling in a binary population synthesis (see Appendix A in the supplementary information for a more detailed discussion of this point). During the population synthesis, primary stars are weighted by mass according to the selected IMF. The weighting for each primary mass bin is then subdivided between single and binary models, and the binary weighting is further divided between models with a range of periods and mass ratios. The mass of secondary companions are therefore not included in the initial IMF weighting, but are included in the final normalisation of the weighting scheme to a total of $10^6$\,M$_\odot$. Increasing the mean companion mass for low mass stars, or decreasing that for high mass stars, would have the effect of decreasing the total mass fraction occupied by stars above any given mass threshold, and thus decreasing the ionizing flux for the integrated population.

\begin{table}
	\centering
	
	\caption{Variation in ionizing photon production rate in the case of constant star formation at 1\,M$_\odot$\,yr$^{-1}$. The mean and standard deviation are given for each set of parameter variations described in the text.}
	\label{tab:cnion}
	\hspace*{-0.75cm}
	\begin{tabular}{rlccc} 
	\hline
		&Variant & \multicolumn{3}{c}{log (N$_\mathrm{ion}$/s)) at}\\
		&        & Z=0.002 & 0.010 & 0.020\\
		\hline
		      & BPASS v2.2         & 53.451 & 53.218 & 53.082\\
		  (i) & Binary fraction    & 53.488 0.006 & 53.234 0.004 & 53.104 0.008\\
		 (ii) & Period (stepwise)  & 53.518 0.107 & 53.263 0.102 & 53.128 0.092\\
		(iii) & Period (gaussian)  & 53.475 0.014 & 53.220 0.012 & 53.091 0.012\\
		 (iv) & Mass ratio         & 53.470 0.041 & 53.220 0.026 & 53.091 0.025\\
		  (v) & Combined           & 53.477 0.044 & 53.222 0.029 & 53.096 0.028\\
		\hline
	\end{tabular}
\end{table}

\begin{figure*}
	\includegraphics[width=\columnwidth]{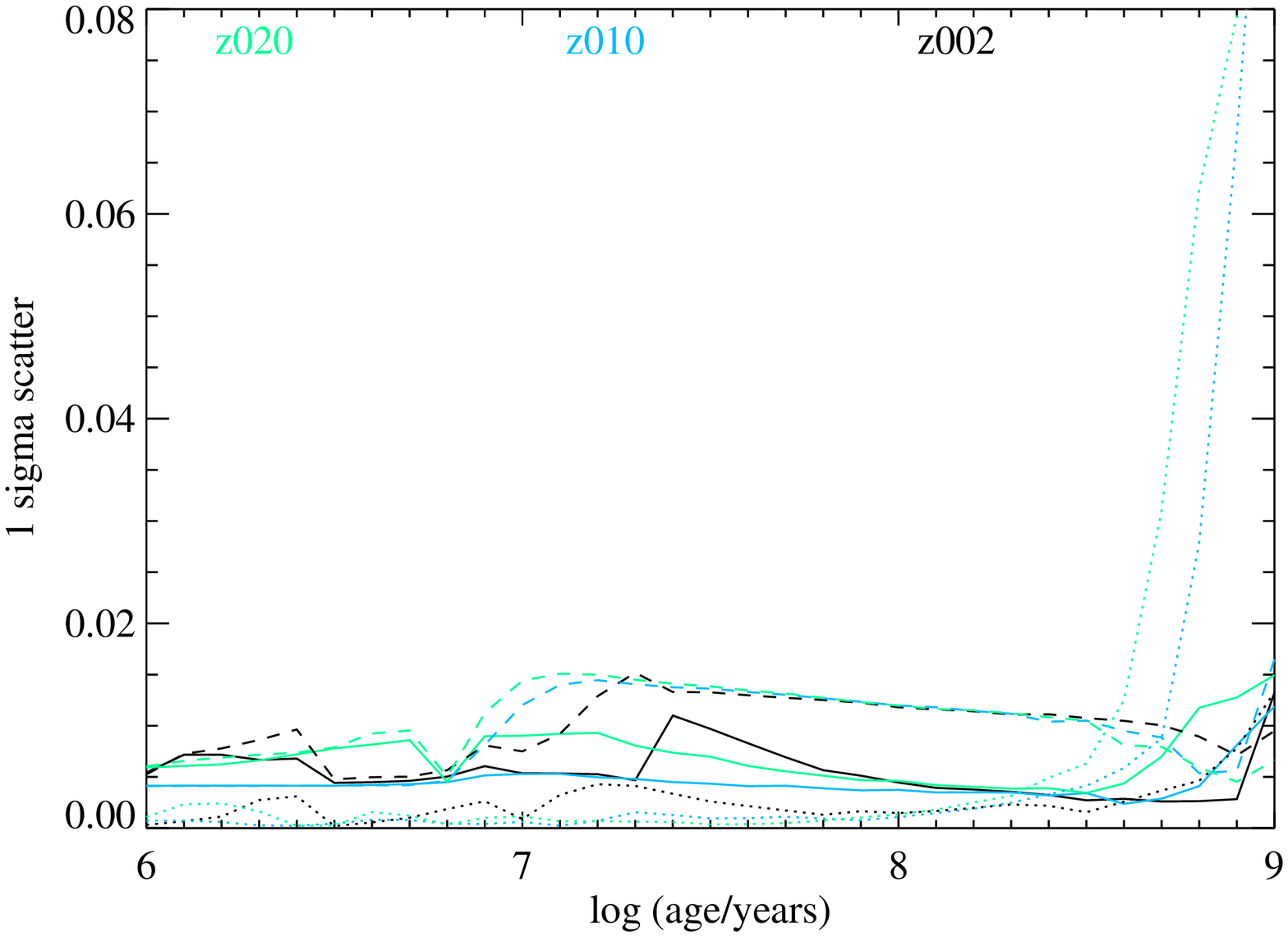}
	\includegraphics[width=\columnwidth]{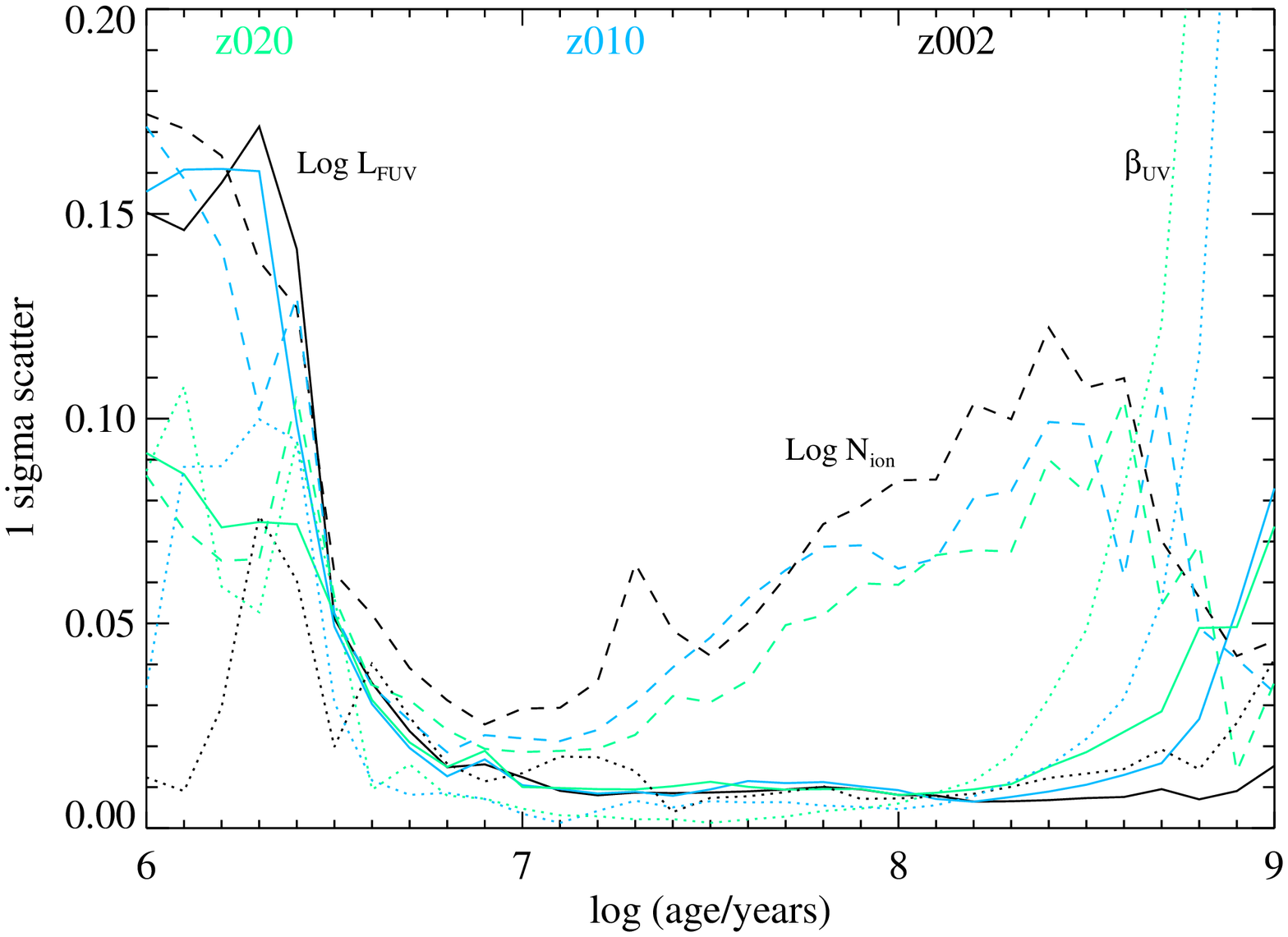}
	\includegraphics[width=\columnwidth]{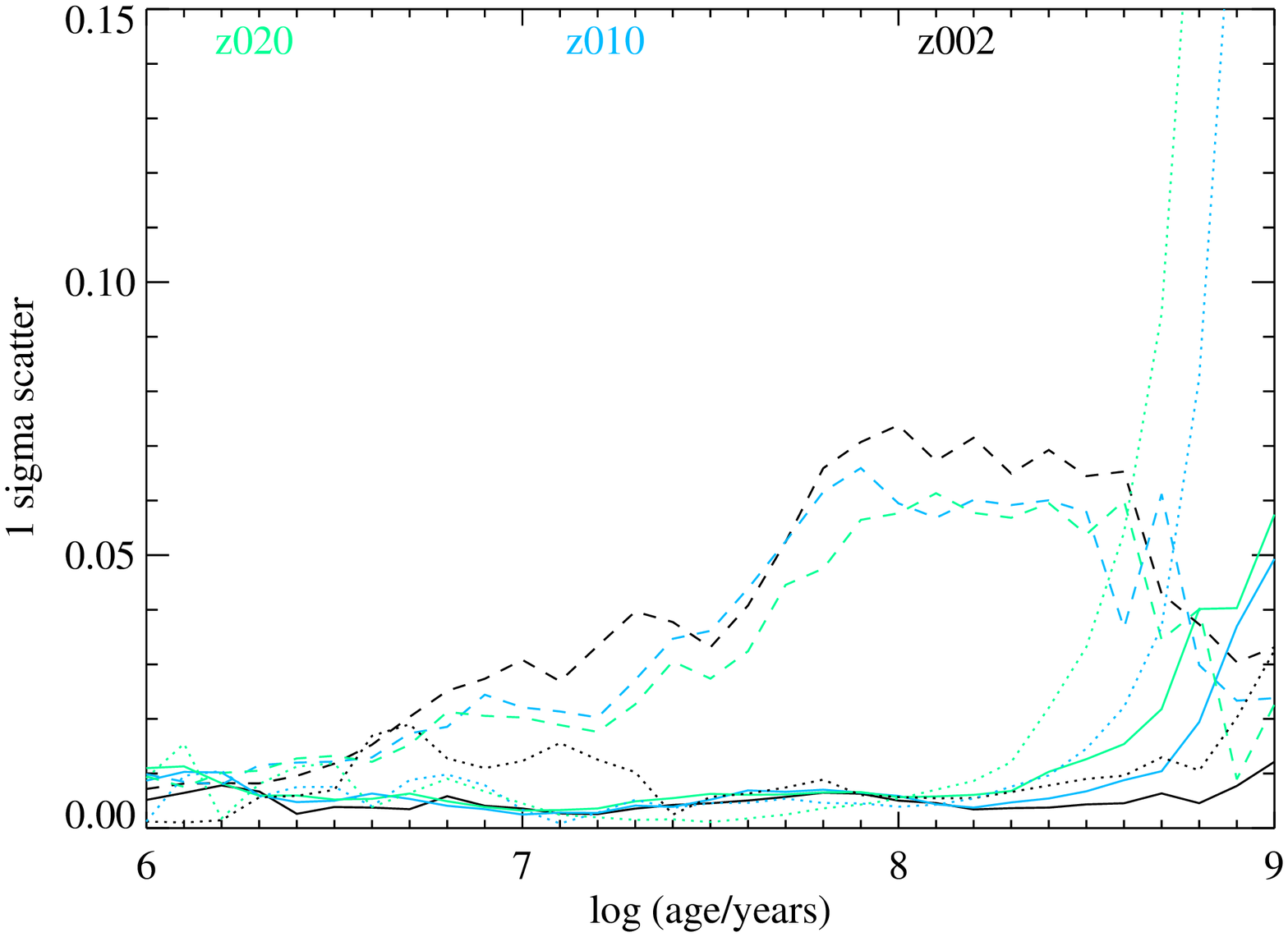}
	\includegraphics[width=\columnwidth]{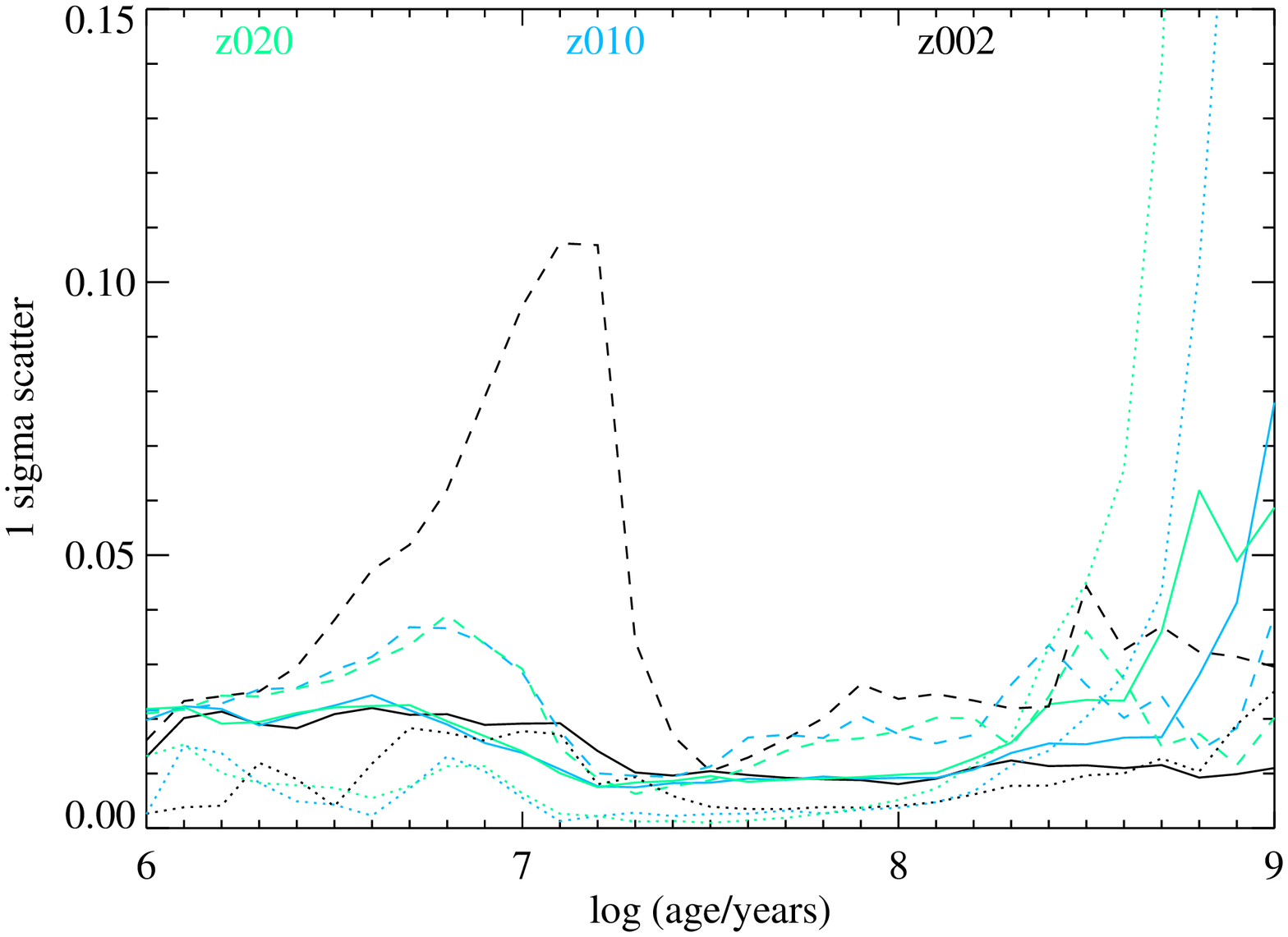}
    \caption{The standard deviation in ionizing photon production rate (in dex, dashed line), FUV-NUV spectral slope $\beta_\mathrm{UV}$ (dotted line) and the FUV luminosity (in dex, solid line) as a function of stellar population age for a simple stellar population, given variation of the binary input parameters as in Fig.~\ref{fig:cnion}. Top left: varying binary fraction (linear fit), top right: varying period distribution (step-wise fit), bottom left: varying period distribution (log-gaussian fit), bottom right: varying mass ratio parameters.}
    \label{fig:stdevs}
\end{figure*}

\begin{figure*}
	\includegraphics[width=\columnwidth]{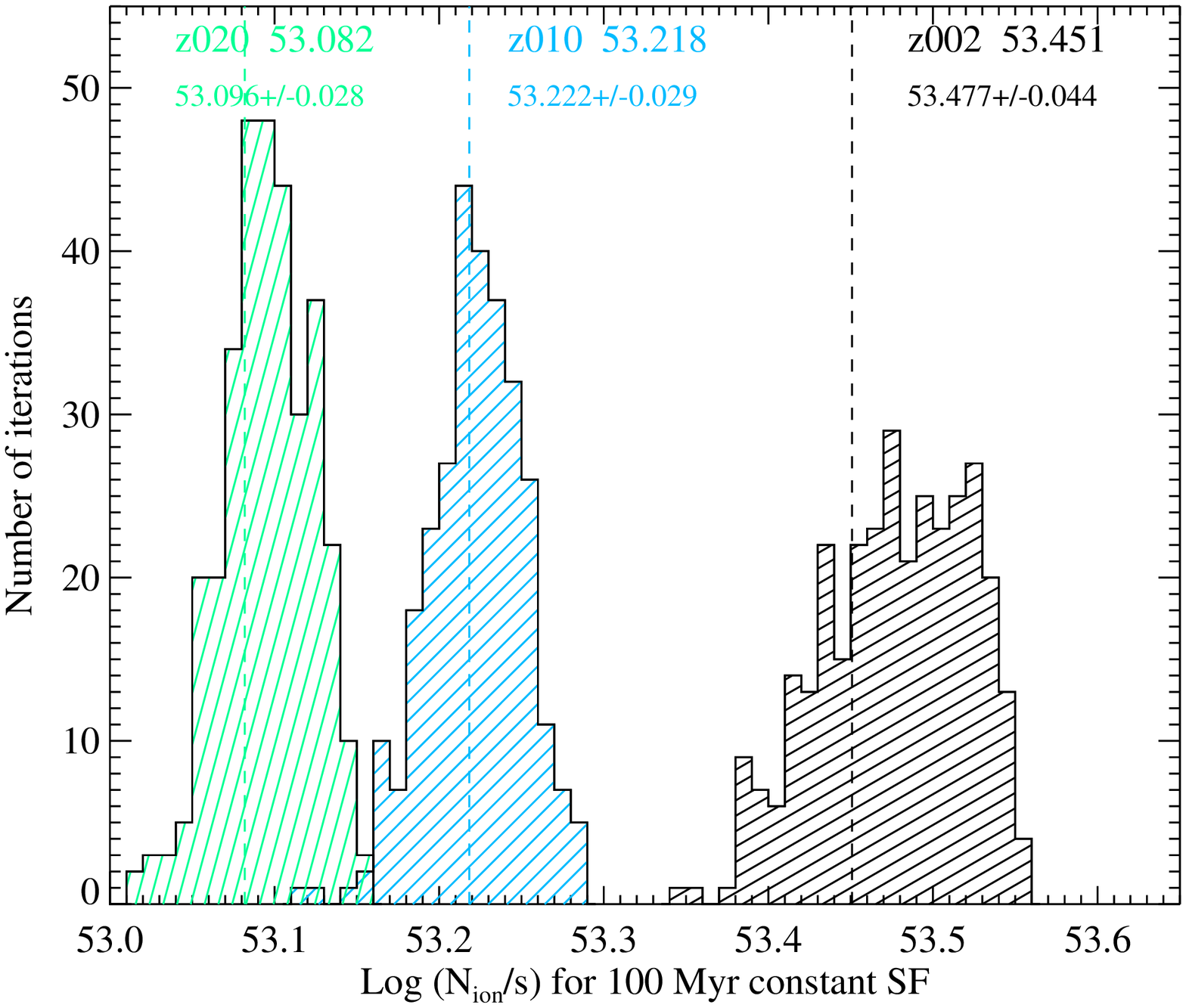}
	\includegraphics[width=\columnwidth]{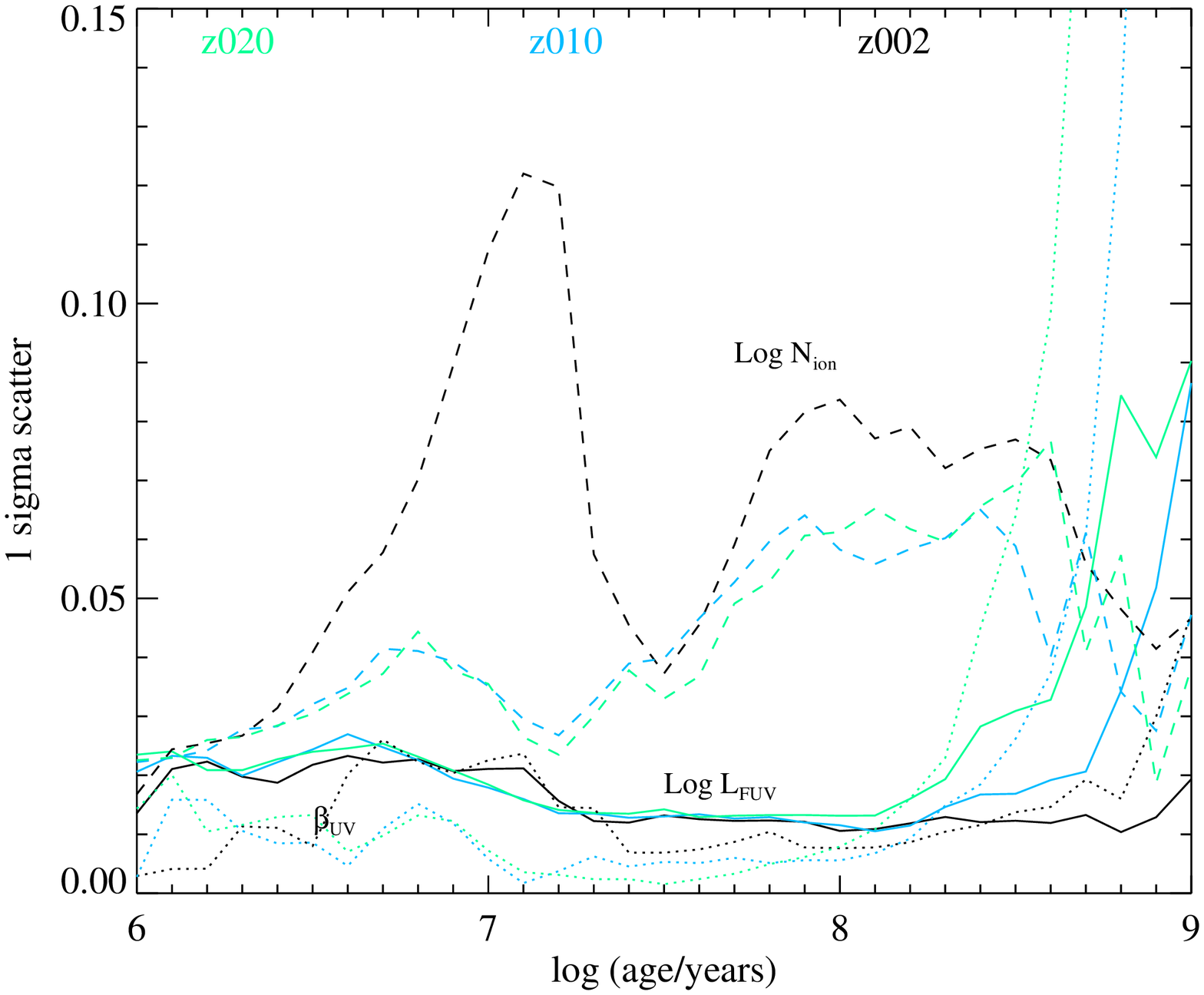}
    \caption{The uncertainty in ultraviolet properties when all binary parameters are varied simultaneously (i.e. variant (v)). The panels show the variation in ionizing photon production rate (left) and the standard deviation in photon production, UV spectral slope and UV luminosity (right) as in Figs.~\ref{fig:cnion} and \ref{fig:stdevs}.}
    \label{fig:mix}
\end{figure*}

The time dependence of the uncertainty on the ionizing photon production rate, the FUV continuum luminosity and the FUV-NUV spectral slope  for each of variants are shown in Fig.~\ref{fig:stdevs}, for the case on a single-aged, evolving starburst population (i.e. an SSP). Since the constant star formation case is effectively an integral of the emission from populations less than 100\,Myr in age, this allows us to examine at which ages the populations are most sensitive to binary parameters, and hence later consider the populations involved. 

The uncertainty on both the UV continuum spectral slope power law index $\beta_\mathrm{UV}<0.02$ and that on the rest-UV luminosity $L_\mathrm{UV}<0.02$\,dex for a simple stellar population at all three metallicities and ages less than about 500\,Myr (after which the UV flux is intrinsically very faint and the uncertainty on these parameters grows significantly). Before this point, neither $L_\mathrm{UV}$ nor $\beta_\mathrm{UV}$ show significant time or metallicity evolution in their uncertainties, except in the case of a step-wise interpolation of the period distribution. Additional uncertainty occurs at early times in this case, because the prescription allows for both zero close binary fractions and for very high numbers of close binaries - a range which is more restricted in the other variants - and hence there is also large scatter in the number of very massive,  luminous, blue stars which interact or merge early in their evolution.

Ionizing photon production rate shows more time dependence in its scatter between model variants. In all cases the lowest metallicity we consider, Z=0.002 (0.1\,Z$_\odot$), demonstrates the largest scatter in ionizing photon rate. This output is sensitive both to the fraction of massive stars at early times (log(age/yrs)$<$6.5), and the fraction of interacting binaries or their products at later times. 

As was the case for $L_\mathrm{UV}$ and $\beta_\mathrm{UV}$, the ionizing photon rate of variant (ii) shows large early time scatter due to large uncertainty in the close binary population of massive stars, which may not be physical. However both period interpolation variants also show a relatively large variance at late log(age/yrs)$\sim$8 at all three metallicities. This occurs because moving the stars from closer initial orbits to wider orbits, and vice versa, changes the fraction of relatively numerous intermediate mass stars which are likely to interact during their main sequence lifetimes or soon thereafter. A subset of these will either be stripped (primaries), or spun-up and rejuvenated (secondaries), and will become hotter and bluer, producing more ionizing photons. Interestingly, the uncertainty on the ultraviolet continuum does not track the ionizing photon production in this case, suggesting that most of the variation is occurring blueward of the Lyman limit, and thus an important role for stripped-envelope stars.

For variant (iv), the mass-ratio distribution, there is a short-lived rise in the uncertainty on ionizing photon production rate at the lowest metallicity and ages of log(age/yrs)$\sim$7. 

The same results when all parameter distributions are sampled simultaneously and independently (i.e. variant (v), the maximum possible scatter permitted by current observational constraints in the local Universe) are shown in Fig.~\ref{fig:mix}. Unsurprisingly the uncertainties on the observable parameters are slightly larger than in the variants where all but one distribution is held fixed. The time evolution of those uncertainties also show contributions from the combination of the same effects shown above, including a short-lived rise in uncertainty on the ionizing flux around log(age/years)$\sim7$ and a longer duration epoch of uncertainty at log(age/years)$\sim8$.  The uncertainty on the ionizing photon flux  varies from 0.02\,dex to almost 0.12\,dex over the first Gyr for a simple stellar population, depending on age and metallicity, but this variation is reduced in a continuously star forming population, where uncertainties on ionizing photon production are 0.03-0.04\,dex. The uncertainties on both $L_\mathrm{UV}$ nor $\beta_\mathrm{UV}$ remain relatively time and metallicity insensitive implying a typical uncertainty on these parameters of 0.02-0.03 (dex for $L_\mathrm{UV}$ and linearly in $\beta_\mathrm{UV}$) for binary SSPs.

\subsection{UV and Optical Absorption Lines}\label{sec:abs}

\begin{figure*}
	\includegraphics[width=\columnwidth]{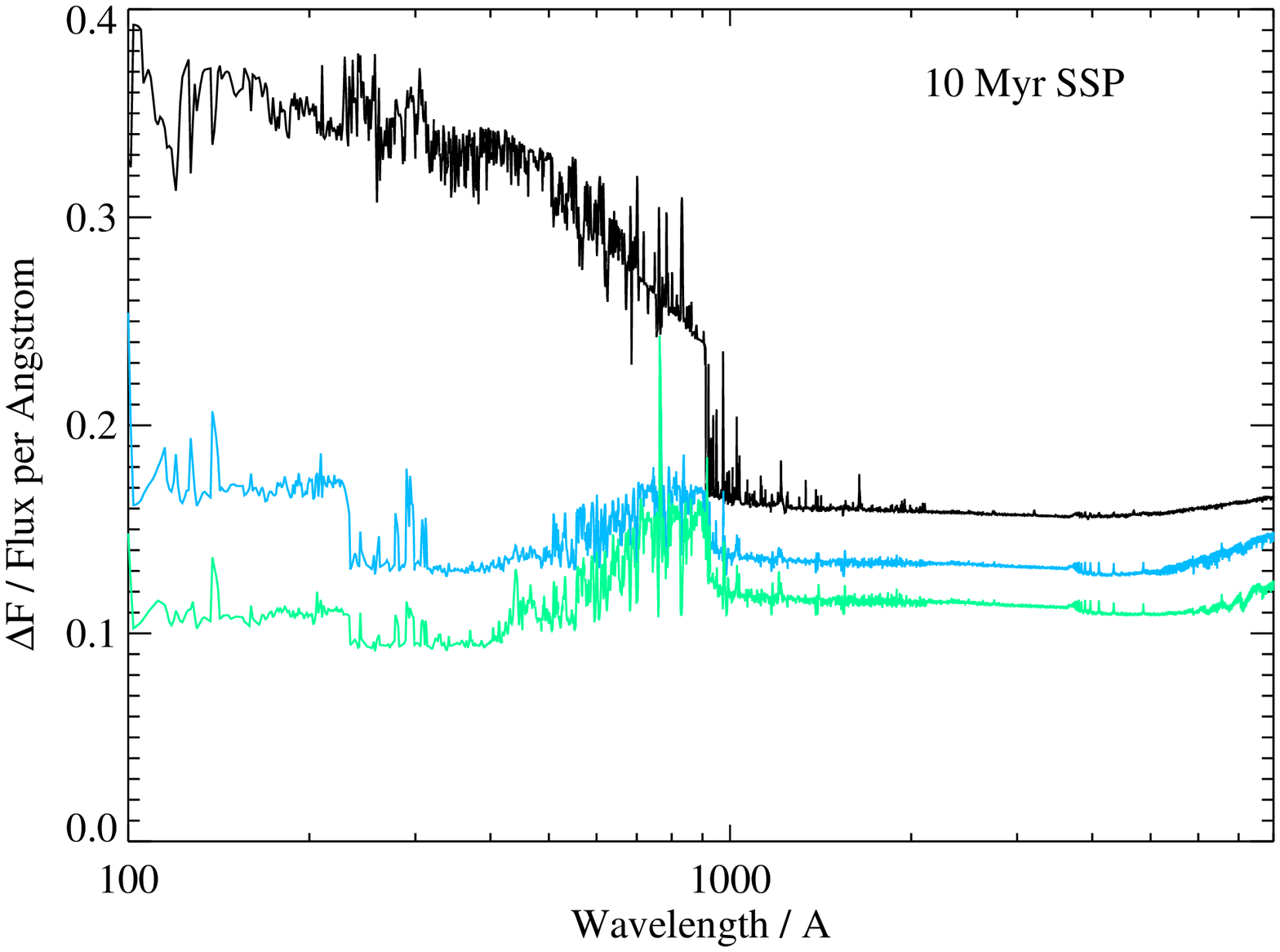}
	\includegraphics[width=\columnwidth]{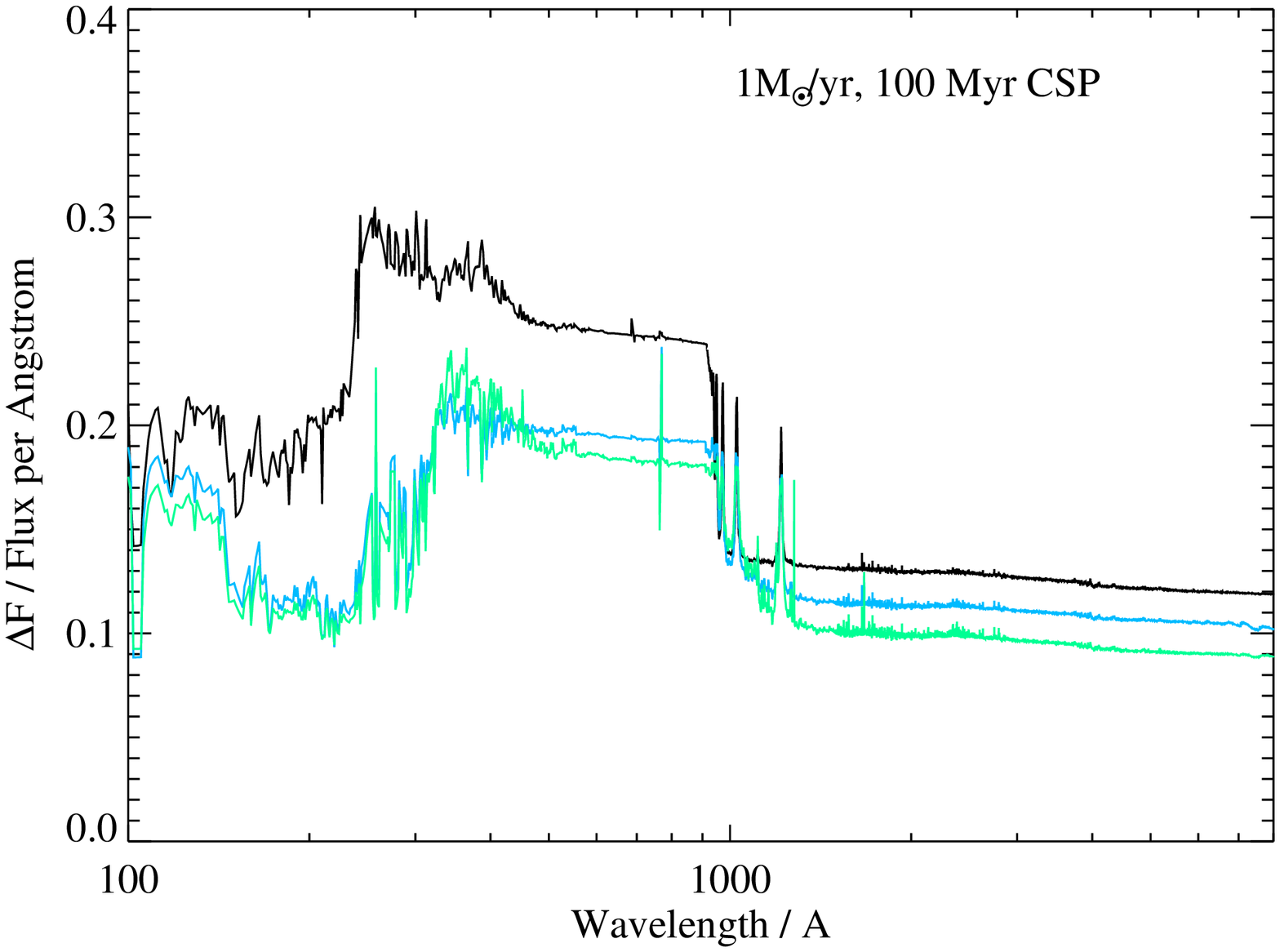}
    \caption{The fractional uncertainties on the stellar continuum flux, due to variation in the binary parameter distributions. Left: for a simple stellar population at age 10\,Myr; Right: for constant star formation at 1\,M$_\odot$\,yr$^{-1}$, measured after 100 Myr. The increase in uncertainty below the Lyman limit is visible in all three metallicities, while the variation longwards of the Lyman break is typically almost constant with wavelength.}
    \label{fig:specopt}
\end{figure*}

\begin{figure*}
	\includegraphics[height=0.71\columnwidth]{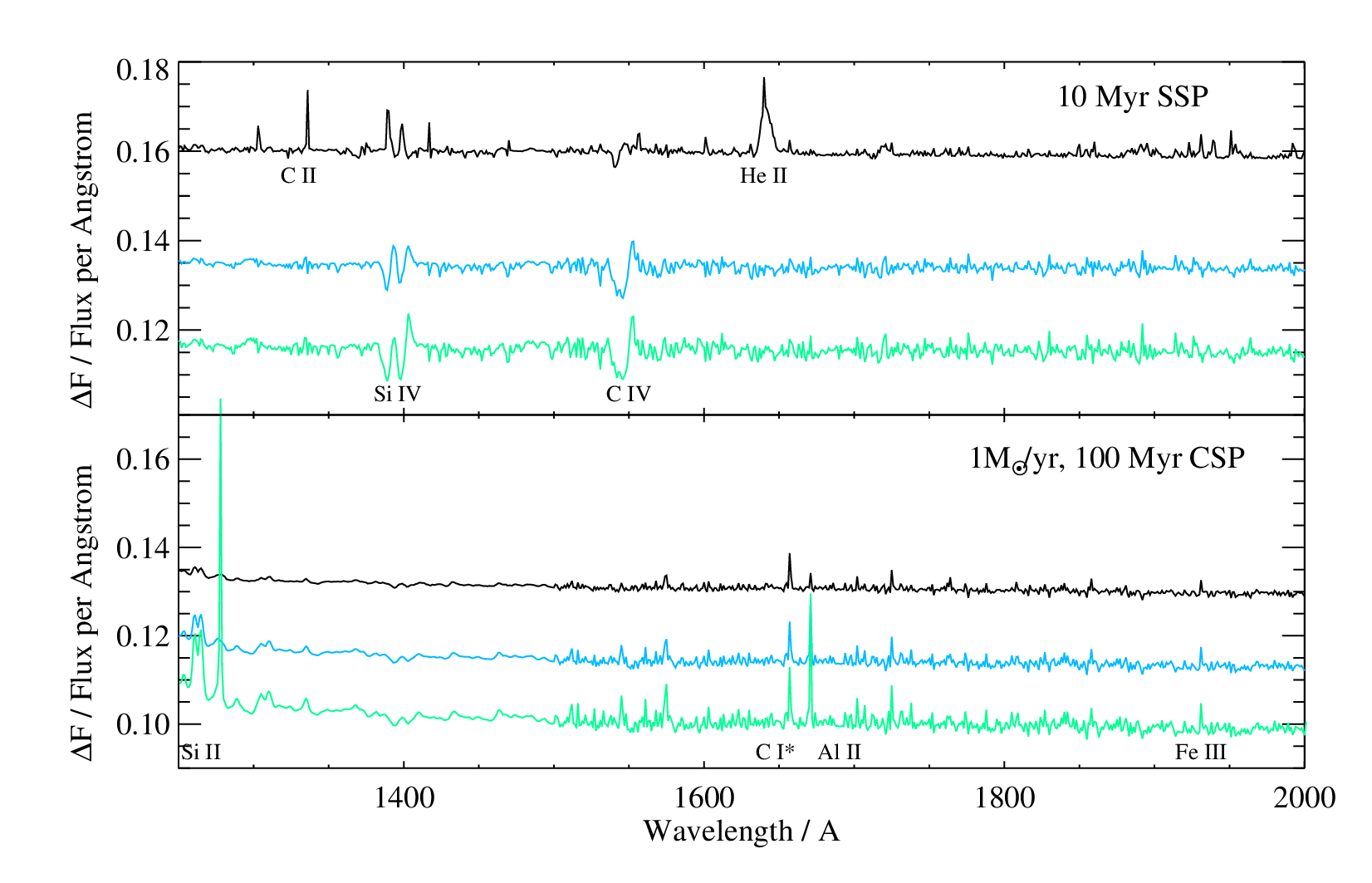}
	\includegraphics[height=0.71\columnwidth]{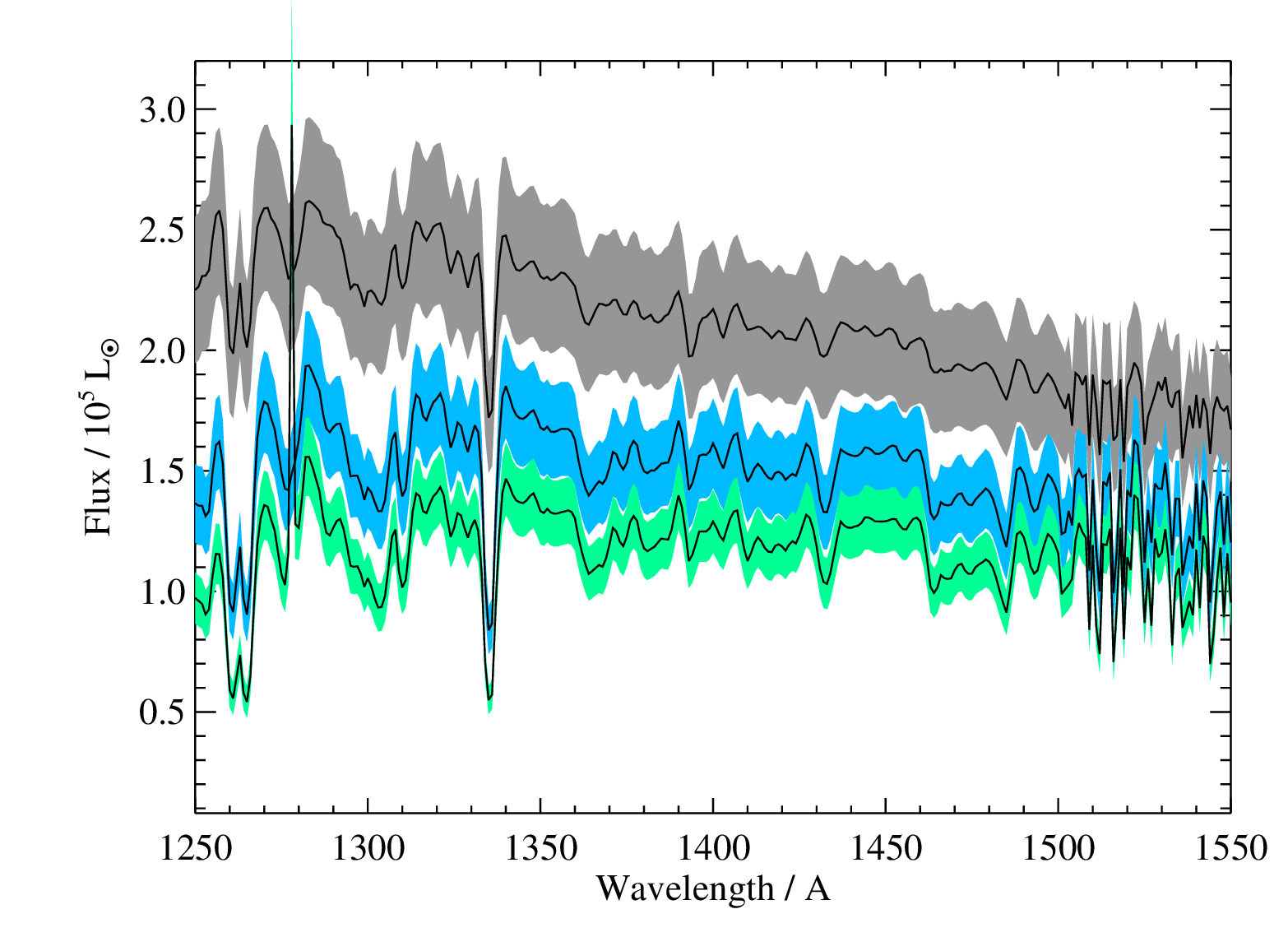}
    \caption{Left: The uncertainties on the ultraviolet continuum for a 10\,Myr old simple stellar population and for constant star formation at 1\,M$_\odot$\,yr$^{-1}$ measured after 100 Myr, due to variation in the binary parameter distributions. The one sigma uncertainty arising from binary parameter variations is shown as a fraction of the flux at each wavelength. Right: A zoom in to part of the UV spectrum, with the 1\,$\sigma$ uncertainty at each wavelength now shown as a shaded band.}
    \label{fig:speczoomuv}
\end{figure*}

The sensitivity of the synthetic population spectrum to variation in all binary parameters (i.e. variant (v)) is illustrated in Fig. \ref{fig:specopt}, where the one sigma flux uncertainty at each metallicity is shown as a function of wavelength for both a simple stellar population at 10\,Myr and for a population which has been constantly forming stars at 1\,M$_\odot$\,yr$^{-1}$ for 100 Myr. The largest uncertainties at all metallicities arise in the rest-frame far-ultraviolet, shortwards of the Lyman limit at 912\AA, and again are largest at low metallicity, reaching almost 30\,per\,cent at Z=0.002 and typically 20\,per\,cent at higher metallicities for a population with constant star formation. At wavelengths longwards of 1200\AA, perturbations in the binary parameters tend to introduce a systematic offset in the normalisation of the spectrum at any wavelength, with the uncertainty in flux per angstrom varying only slowly and of order 10-20\,per\,cent at all three metallicities. 

As Fig.~\ref{fig:speczoomuv} demonstrates, absorption lines and continuum strength have comparable uncertainties. An exception to this is seen in the strength of certain absorption lines in the ultraviolet, including Si\,II\,1261\AA, C\,I$^\ast$\,1657\AA\ and Al\,II\,1671\AA\ at the higher metallicities in a composite stellar population, and He\,II\,1640\AA\ and C\,II\,1335\AA\ at the lower metallicity for a young starburst. Each of these shows a slight excess uncertainty relative to the continuum, although this is unlikely to be useful as a diagnostic of binary parameters: the variation is still small (e.g. 12 rather than 10\,per\,cent scatter relative to the fiducial population), these lines are intrinsically very weak, and they are subject themselves to additional uncertainty due to factors such as $\alpha$-element enhancement and other abundance ratio or formation timescale effects. They are also likely to vary between different atmosphere models in use in the population synthesis (e.g. a different choice of input atmosphere models could plausibly change minor lines such as these) and so are unlikely to prove diagnostic given observational uncertainties. Small differences in the age or metal abundance of the stellar population will overwhelm any variation due to binary fraction. A simple stellar population (shown at 10\,Myr in Fig.~\ref{fig:specopt} and \ref{fig:speczoomuv}) exhibits larger typical uncertainties than a composite population, in which an effective time-average smooths over extremes in any given binary parameter set.

There is no single spectral feature longwards of 912\AA\ which is clearly and strongly diagnostic of variations in the binary parameter distributions, given the current level of uncertainty on binary parameters in the local Universe.

\subsection{Colours}\label{sec:colours}

\begin{figure*}
	\includegraphics[width=\columnwidth]{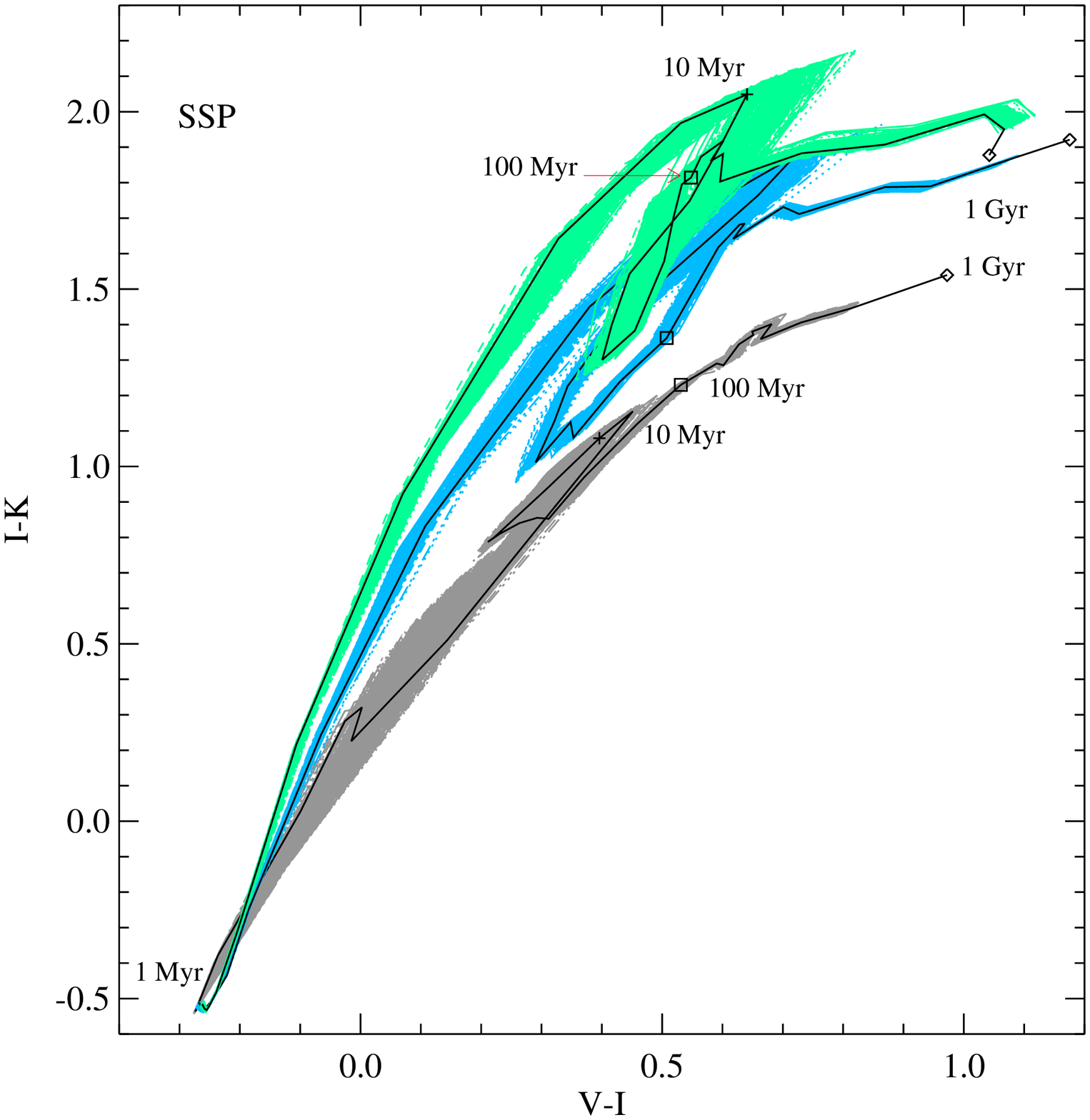}
	\includegraphics[width=\columnwidth]{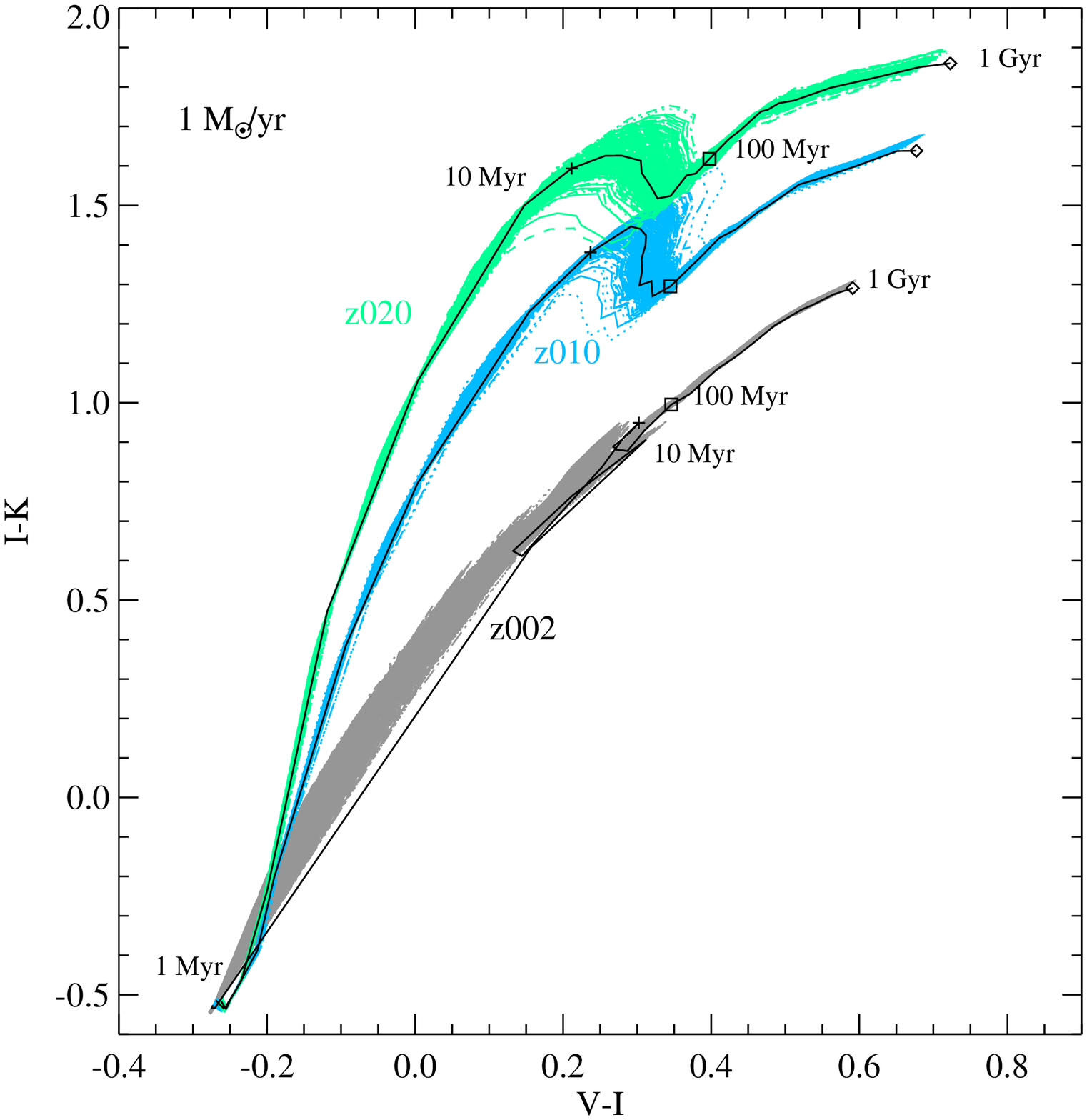}
	\caption{The variation in the V-I and I-K colour of the integrated light, as a result of uncertainty in the binary parameter distributions, for a simple stellar population as a function of age (left), and for a population forming stars at a constant rate of 1\,M$_\odot$\,yr$^{-1}$ as a function of time elapsed since the onset of star formation (right). The SSP model has been smoothed over three time bins for clarity and to account for the effects of discrete mass binning in the underlying stellar models. The fiducial BPASS v2.2 stellar population evolution is shown as a solid black line at each metallicity.}
    \label{fig:vicol}
\end{figure*}

Unsurprisingly, the small scatter in the continuum as shown in section \ref{sec:abs} results in a similarly small scatter in the predicted rest-frame colours of the integrated stellar light. This is demonstrated in Fig.~\ref{fig:vicol} which gives the distribution in $V-I$ vs $I-K$ colours for a sample of model variants, when all binary parameters are varied (i.e. variant (v)). The time evolution is given both for a simple stellar population as it ages from 1 Myr to 1 Gyr, and for a population with constant star formation as a function of time since onset of star formation. Note that the reddening effects of nebular continuum emission are not considered in this direct comparison. At most population ages, the uncertainty on rest-frame optical and near-infrared colour is  $<$0.05\,mag, significantly below the uncertainty that would be introduced by assumptions regarding stellar population age, star formation history and metallicity. However at stellar population ages between 10 and 100 Myr, binary products have a significant effect, causing the population to turn back towards bluer colours. The effect of varying the binary distribution parameters is to cause this blue loop to begin either slightly earlier (corresponding to a large interacting binary fraction at higher mass) or later (if interactions are less common at the highest masses). This leads to degeneracy between model age and the input binary parameter distributions, such that at Solar metallicity, this colour combination cannot distinguish the two. At significantly sub-Solar metallicities, the scatter in population colour is instead more pronounced at early ages, where the most massive stars dominate. 

In the constant star formation case, binary parameter uncertainties result in a $\sim$0.015\,mag variation in V-I colour at most ages, with the blue loop undertaken by the simple stellar populations leading to a larger colour range ($\sim$0.2\,mag of scatter) for high metallicity populations at ages between 10 and 100 Myr.

\section{Signatures of the Binary Fraction}\label{sec:base}

In the above tests, we have assumed that the empirical constraints on the binary population parameters derived from the local Universe apply at all metallicities. There are reasons to question that assumption. Metallicity likely affects the cooling and fragmentation of star forming molecular clouds. While direct evidence for the resultant binary fractions is difficult to obtain, there is now growing evidence that the close binary fraction in Solar-mass stars is higher at low metallicities \citep{2019ApJ...875...61M,2019MNRAS.482L.139E,2020arXiv200200014P}, which in turn would suggest a flatter relation between binary fraction and primary mass (or alternatively a period distribution more skewed towards close binaries at low mass). 

Given that the empirical uncertainty on the extant binary population parameters is unlikely to lead to a clear observational signature, it may be instructive to investigate whether such a deviation from the local parameters (for example in very low metallicity environments, or in the distant Universe) might be detectable. This question has been explored in past work by \citet{2018ApJ...867..125D} who used a grid of models in which the binary fraction varied between 0 and 1, but in which - crucially - the binary fraction was independent of mass (i.e. a weighted combination of single star models and models in which every star was in a binary). Their analysis focused on observables derived from resolved stellar populations, identifying the Red Supergiant (RSG) star population and its abundance relative to stripped stars as diagnostic of massive star multiplicity.  Here we extend a similar analysis to consider a range of mass-dependent binary fractions which differ significantly from the empirical constraints considered up to this point, and focus instead on plausible signatures in the integrated light of unresolved stellar populations.

\subsection{Extended Model Grid}

\begin{figure}
	\includegraphics[width=\columnwidth]{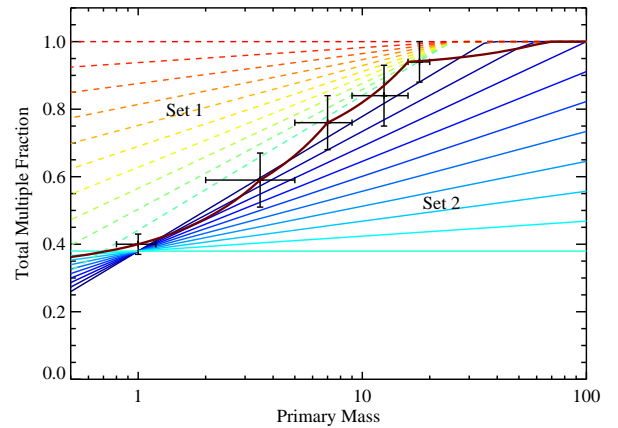}
    \caption{Multiple fractions tested in an experimental grid to examine possible observable signatures for binary populations. Each line indicates a model binary fraction distribution which either raises the binary fraction at low stellar mass (set 1, dashed lines) or lowers it at high mass (set 2, solid lines). Data points are drawn from \citetalias{2017ApJS..230...15M} (as in Fig.~\ref{fig:fbin}) and the thick line indicates the fiducial model applied in BPASS v2.2.}
    \label{fig:fbase}
\end{figure}

The parameter distribution used in BPASS v2.2 and shown in Fig.~\ref{fig:fbin} is characterised by a binary fraction that increases with the logarithm of stellar mass, from approximately 40\,per\,cent at 1\,M$_\odot$ to 100\,per\,cent at approximately 100\,M$_\odot$. To evaluate the effects of a substantial change to this distribution (i.e. one far exceeding the current empirical constraints) we evaluate 20 additional models at each metallicity with artificially modified binary fractions, as shown in Fig. \ref{fig:fbase}. 

These models each assume a linear relation between multiple (in this case binary) fraction and log(primary mass) and fall into two categories: ten models in which the low stellar mass binary fraction rises (set 1) and ten models in which the high mass binary fraction falls (set 2). 

We make no claims for the physical relevance of these models. While observational evidence may favour the set 1 scenario at low metallicities, as discussed above, the constraints are weak and it is difficult to be certain which regime may dominate in the distant Universe. Other binary parameters (i.e. binary mass ratio and initial period distributions) are held fixed at the fiducial values.

\begin{figure*}
	\includegraphics[width=\columnwidth]{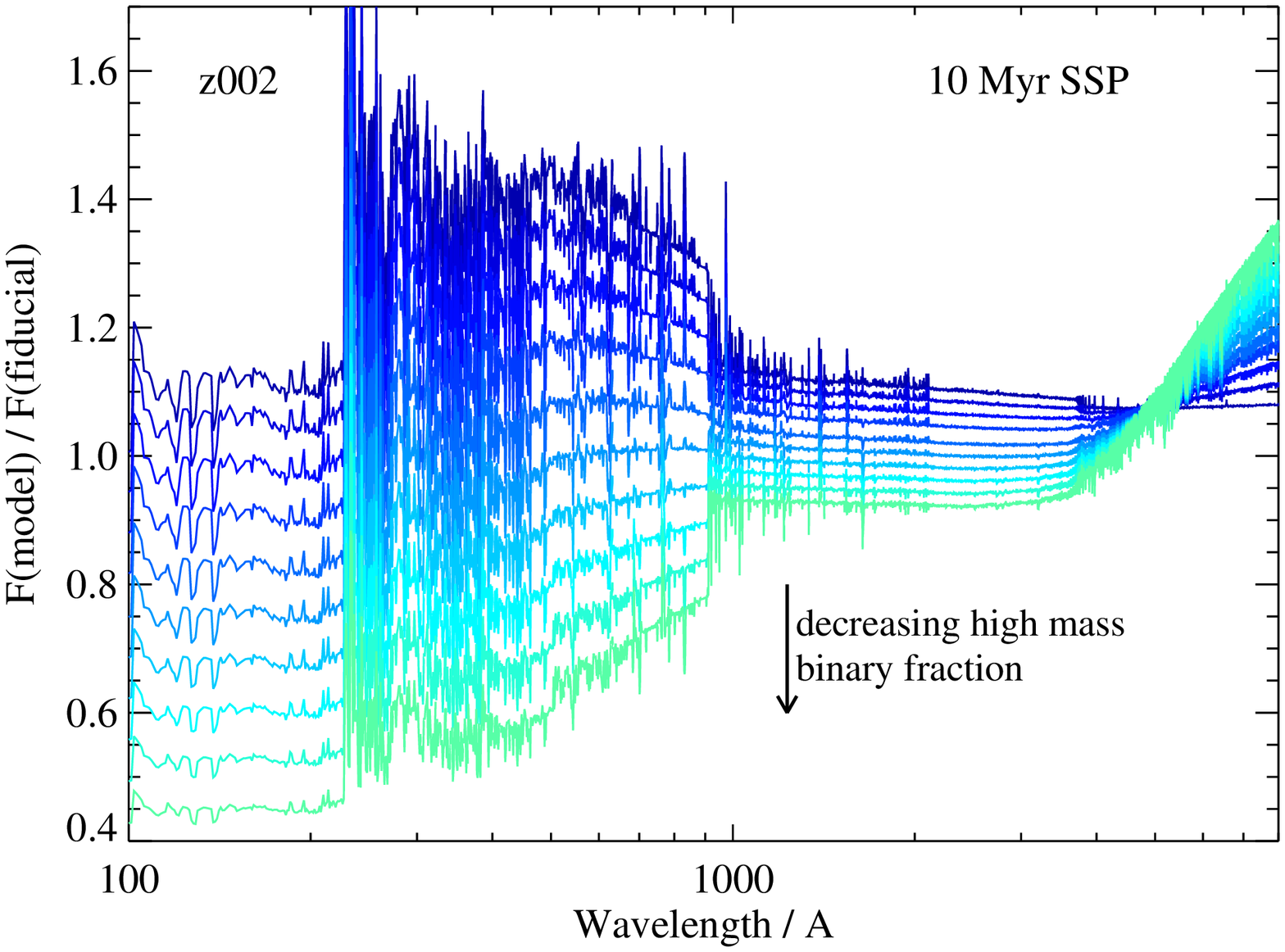}
	\includegraphics[width=\columnwidth]{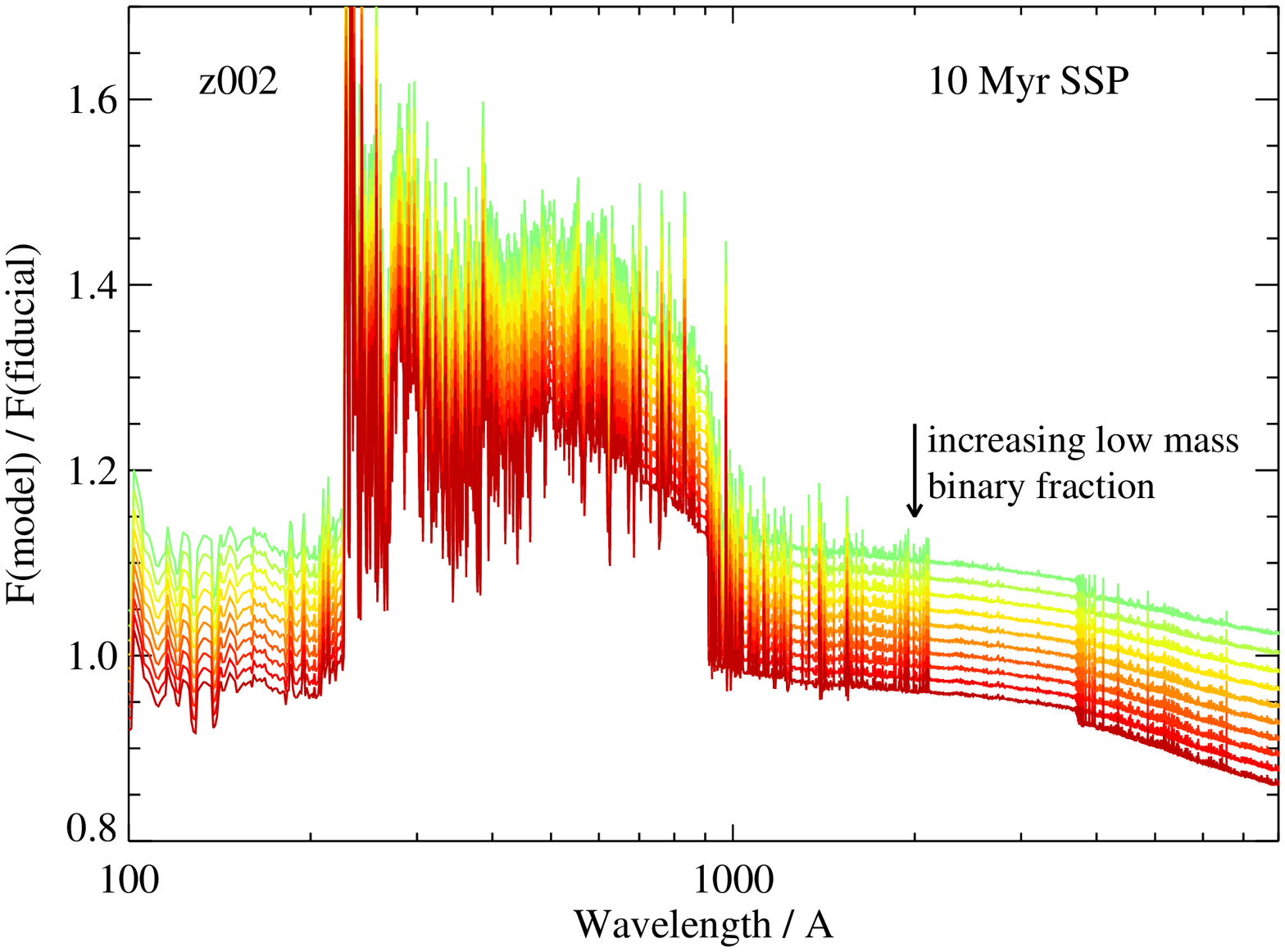}
	\includegraphics[width=\columnwidth]{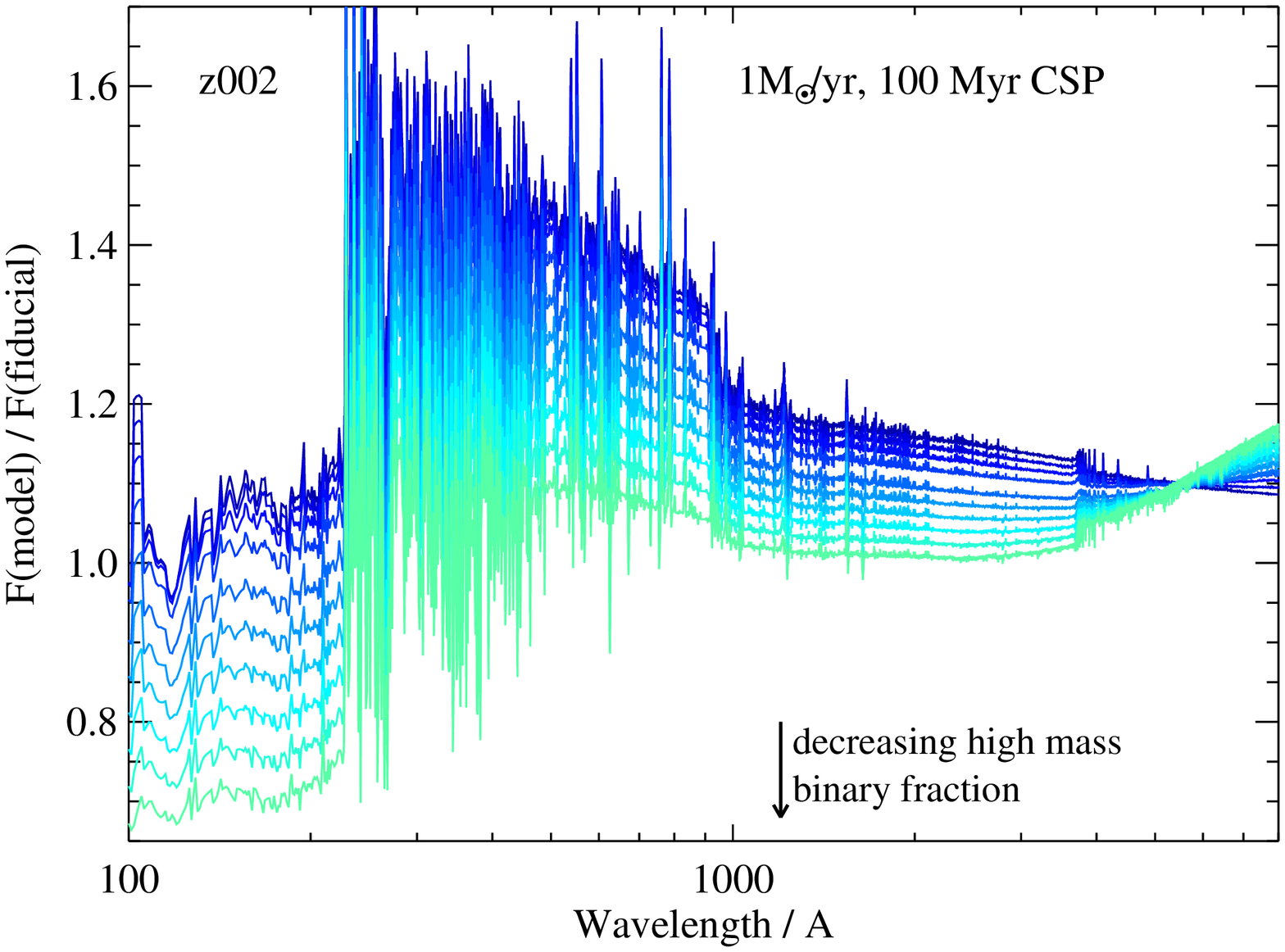}
	\includegraphics[width=\columnwidth]{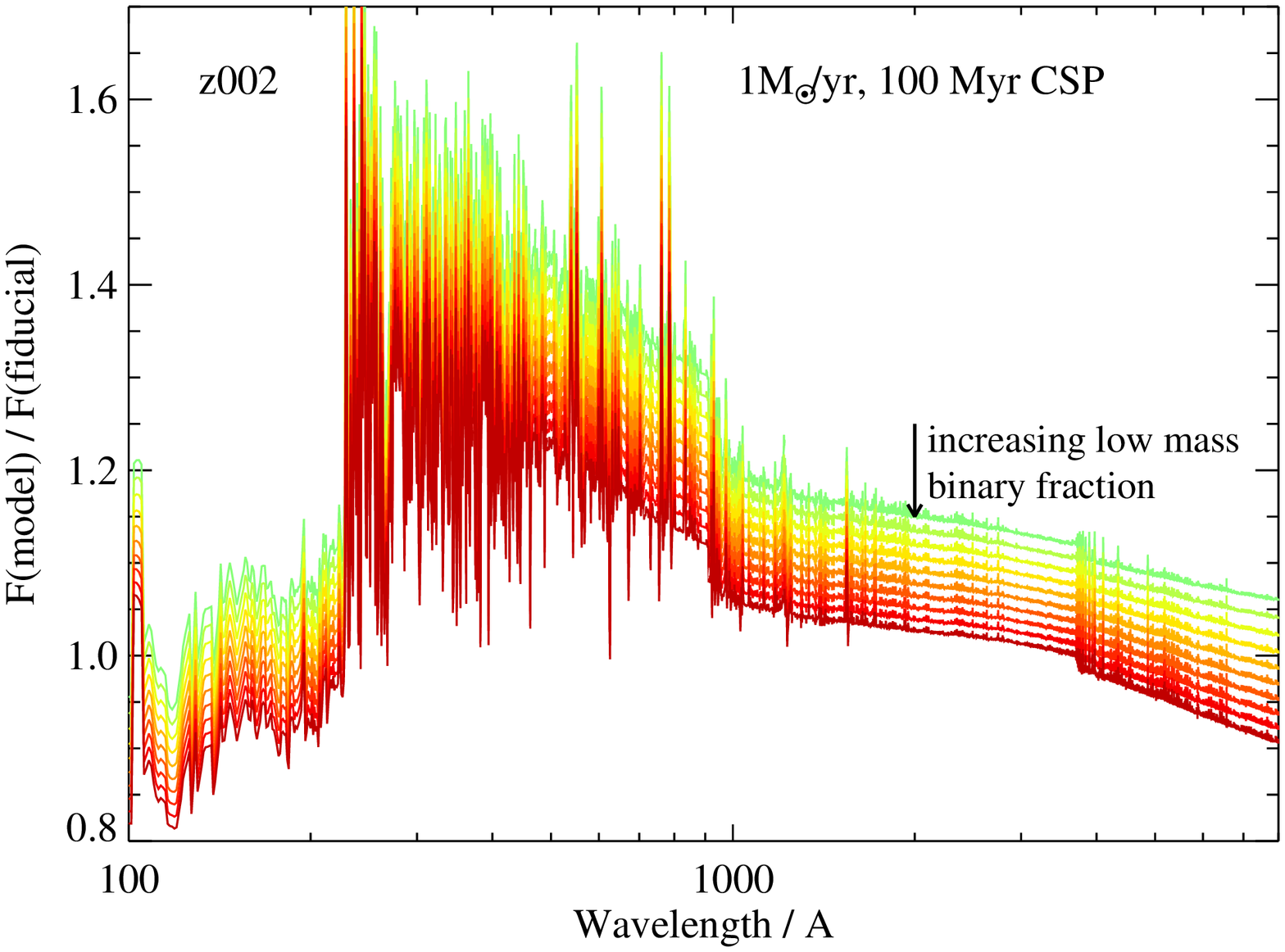}
    \caption{The effect of changing the binary fraction distribution on the ultraviolet-optical spectrum at Z=0.002. Left hand panels show set 2 (decreasing high mass binary fraction) and right hand panels set 1 (increasing low mass binary fraction) as defined in Fig.~\ref{fig:fbase}. The top row shows results for a 10\,Myr old SSP, while the bottom row shows the case for constant star formation. All models are normalised by the fiducial BPASS v2.2 model. Models are colour coded as in Fig.\ \ref{fig:fbase}.    }
    \label{fig:specbase}
\end{figure*}

\subsection{Stellar Continuum}

The ultraviolet and optical spectra resulting from these artificial variant models are shown in Fig. \ref{fig:specbase}, where results are shown at the lowest metallicity (Z=0.002) and each model is normalised by the flux as a function of wavelength in the fiducial BPASS v2.2.1 model set. The shift from a piece-wise interpolation to a linear function for binary fraction as a function of mass is such that none of these models precisely reproduces the BPASS v2.2 models. Nonetheless, clear trends are visible with binary fraction. Models with small high mass binary fractions unsurprisingly produce lower ionizing fluxes than those with large high mass binary fractions, and also show a lower continuum normalisation in the near-ultraviolet. 

Interestingly, increasing the binary fraction of low mass stars has a similar (although not identical) effect. As before, this likely arises from the application of the initial mass function in BPASS.  The broken power-law IMF used uniformly in all these models assigns a weighting to binary systems first by their primary mass, and then subdivides this weighting between variants with different mass ratios and period distributions. If the number of low mass binaries are increased, the weighting for each primary is unchanged, but the overall model normalisation must allocate mass to the (low mass) companions of low mass stars, which would otherwise have been allocated to high mass primaries and their (moderate-to-high mass) companions. Since the number of hot, luminous stars falls as a result, the overall model luminosity also drops. 

In the optical, the behaviour diverges between the two sets of binary fraction variants.  Models in which the low mass binary fraction is varied (set 1) shows the same trend in the optical as the ultraviolet. By contrast models in which the high mass binary fraction is decreased (set 2) show much redder spectra in the optical than those with a large high mass binary fraction. This results from a larger population of massive stars reaching the cool giant branch without being subject to stripping by a binary companion (i.e. a binary dependence in the Wolf-Rayet to Red Supergiant ratio), leading to a much redder integrated light spectrum.
In each case, this behaviour occurs in both young simple stellar populations, and in composite stellar populations forming stars at a constant rate, although in the latter case the variations between models are typically smaller. 

\begin{figure}
	\includegraphics[width=\columnwidth]{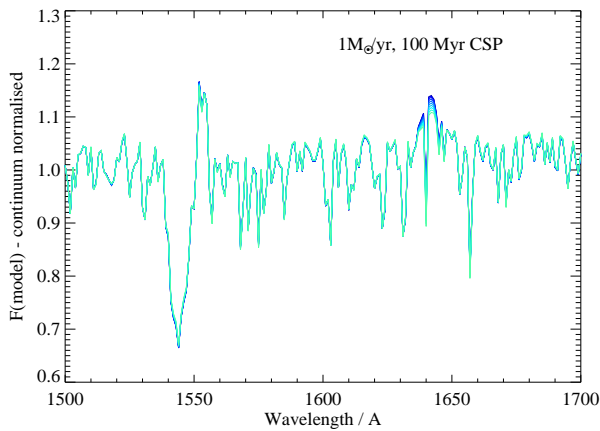}
    \caption{The effect of decreasing the high mass binary fraction on a representative region of the ultraviolet spectrum of a continuously star forming population at Z=0.002, highlighting the impact on the C\,IV\,1550\AA\ absorption feaure and the broad He\,II\,1640\AA\ emission feature. Models have been normalised relative to a smoothly varying continuum, and are colour coded as in Fig.\ \ref{fig:fbase}.}
    \label{fig:heii1}
\end{figure}

In addition to considering the continuum luminosity and shape, we can also measure the strengths of ultraviolet emission and absorption lines after normalisation by a pseudo-continuum (obtained by fitting a third-order polynomial to the wavelength range $\lambda=1000-2000$\AA). After this normalisation, the strengths of emission and absorption features relative to the continuum prove surprisingly insensitive to changes in the binary fraction, with modifications to line equivalent widths typically $<0.01$\AA\ (as Fig.~\ref{fig:heii1} demonstrates). This is particularly true in the continuously star-forming composite stellar spectra in which any variation in line strength is averaged out over time. A notable exception occurs only at the lowest metallicity under consideration (Z=0.002) and in the case of the broad, stellar-wind-driven He\,II\,1640\AA\ feature which arises predominantly from massive stars with stripped atmospheres. This shows a clear trend in strength with high mass binary fraction (see Fig.~\ref{fig:heii1}) but no clear dependence on low mass binary fraction variation.

\begin{figure*}
	\includegraphics[width=\columnwidth]{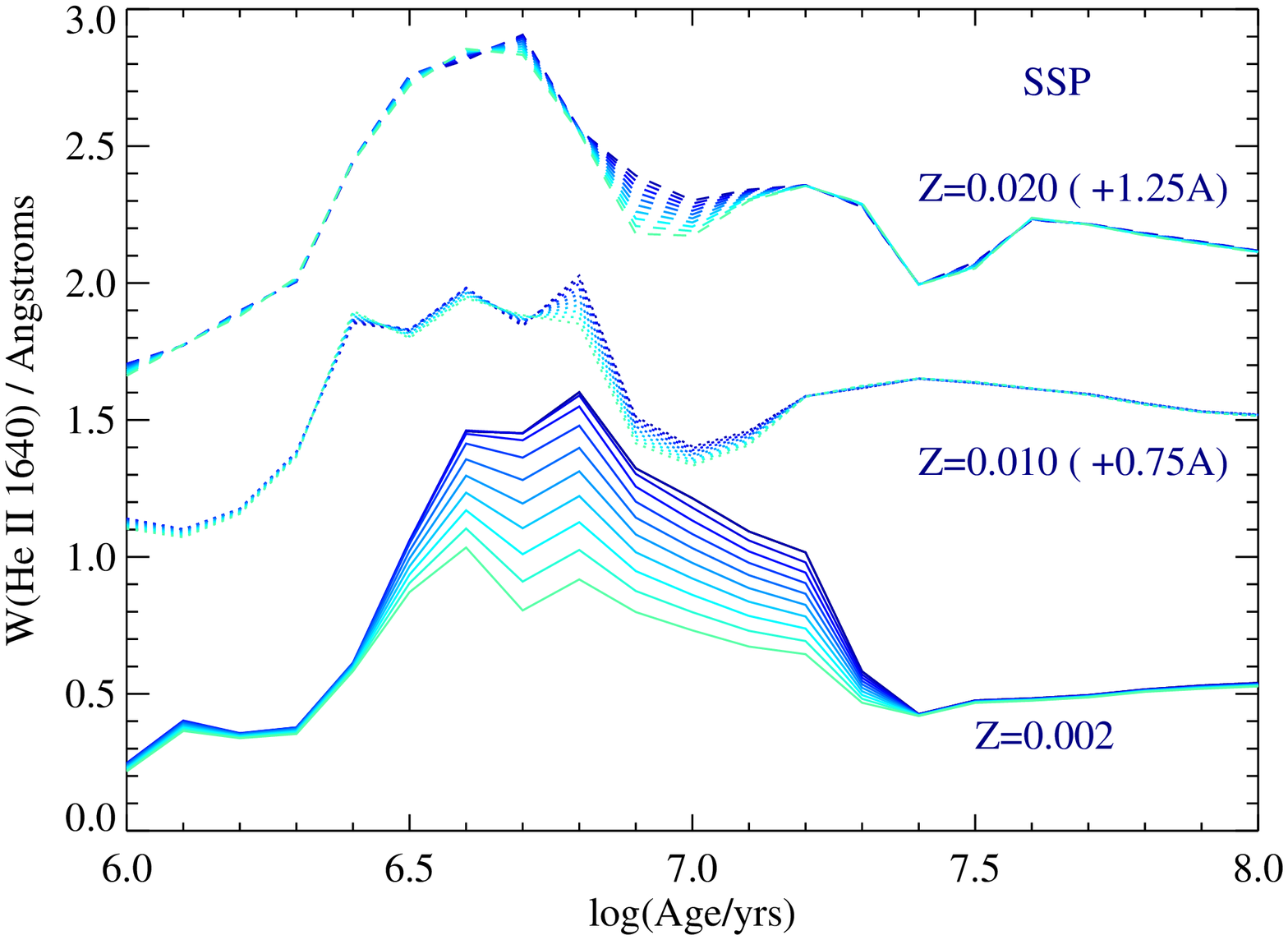}
	\includegraphics[width=\columnwidth]{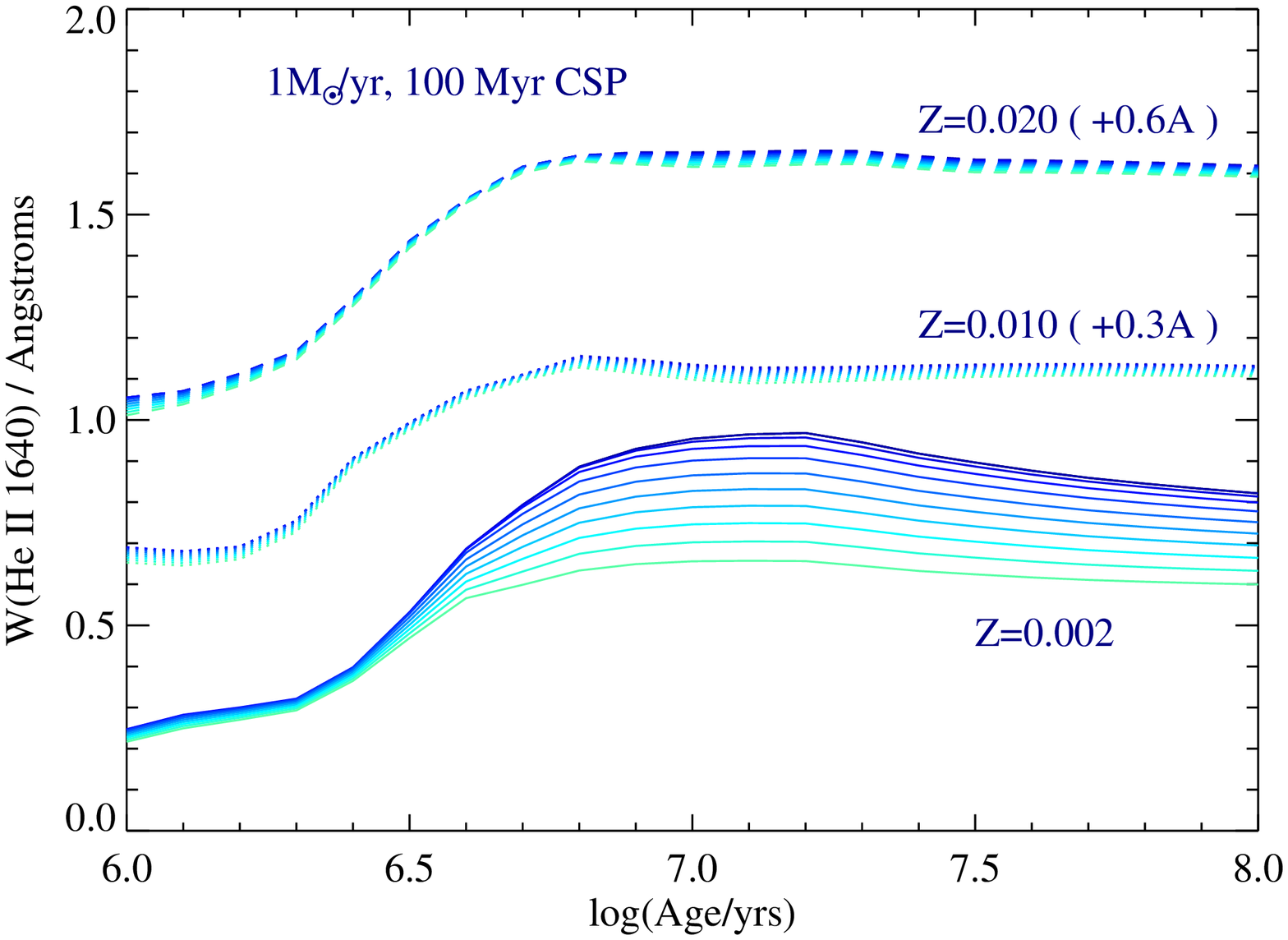}
    \caption{The effect of decreasing the high mass binary fraction on the broad He\,II\,1640\AA\ stellar emission feature equivalent width. The evolution of a constantly star-forming CSP is shown in the right hand panel, while that of an SSP is shown on the left. Models are colour coded as in Fig.\ \ref{fig:fbase}. Set 1 (increasing low mass binary fraction) models show little scatter and would overlie the highest set 2 (reducing high mass binary fraction) models.}
    \label{fig:heii2}
\end{figure*}

The origin of this dependence is suggested by Fig.~\ref{fig:heii2} which presents the time evolution of the line equivalent width, evaluated as a summation of the excess flux within $\pm$1000\,km\,s$^{-1}$ of the line centre. It demonstrates that the dependence in the composite population arises primarily from simple stellar populations at ages between 3 and 20\,Myr. Unsurprisingly this is the epoch at which the ultraviolet spectrum is dominated by massive stars (at around 10-30\,M$_\odot$) which are approaching the end of their main sequence lifetimes. In the high metallicity cases, these stars lose their hydrogen envelopes primarily as the result of strong radiation-driven winds. By contrast, at low metallicities, the primary mechanism for stripping is through binary mass transfer onto a companion. This both delays the emission of He\,II line flux and causes it to be  dependent on the binary fraction at low metallicity. 

It should be noted however that the difference between the largest high mass binary fractions considered here ($\sim$1) and the lowest ($\sim$0.4) changes the line equivalent width from 0.6\AA\ to only 0.8\AA\ for a constantly star forming population, and from 0.9 to 1.6\AA\ for a simple stellar population at peak emission, although the latter is only clearly diagnostic if the age of the population is constrained to around 0.1\,dex or better. Degeneracies in interpretation of this line as a binary indicator are likely to be substantial due to star formation history and metallicity uncertainties. We also note that the atmosphere models in this very low metallicity, high mass, stripped star regime remain an active area of study \citep[e.g.][]{2019A&A...621A..85H,2018A&A...615A..78G,2019A&A...629A.134G}. As a result, while the stellar-wind-driven broad He\,II is sensitive to massive binaries, it is unlikely to be a clear and unambiguous indicator of their quantitative fractions in realistic observations.

\subsection{Ionizing fluxes}

\begin{figure*}
	\includegraphics[width=\columnwidth]{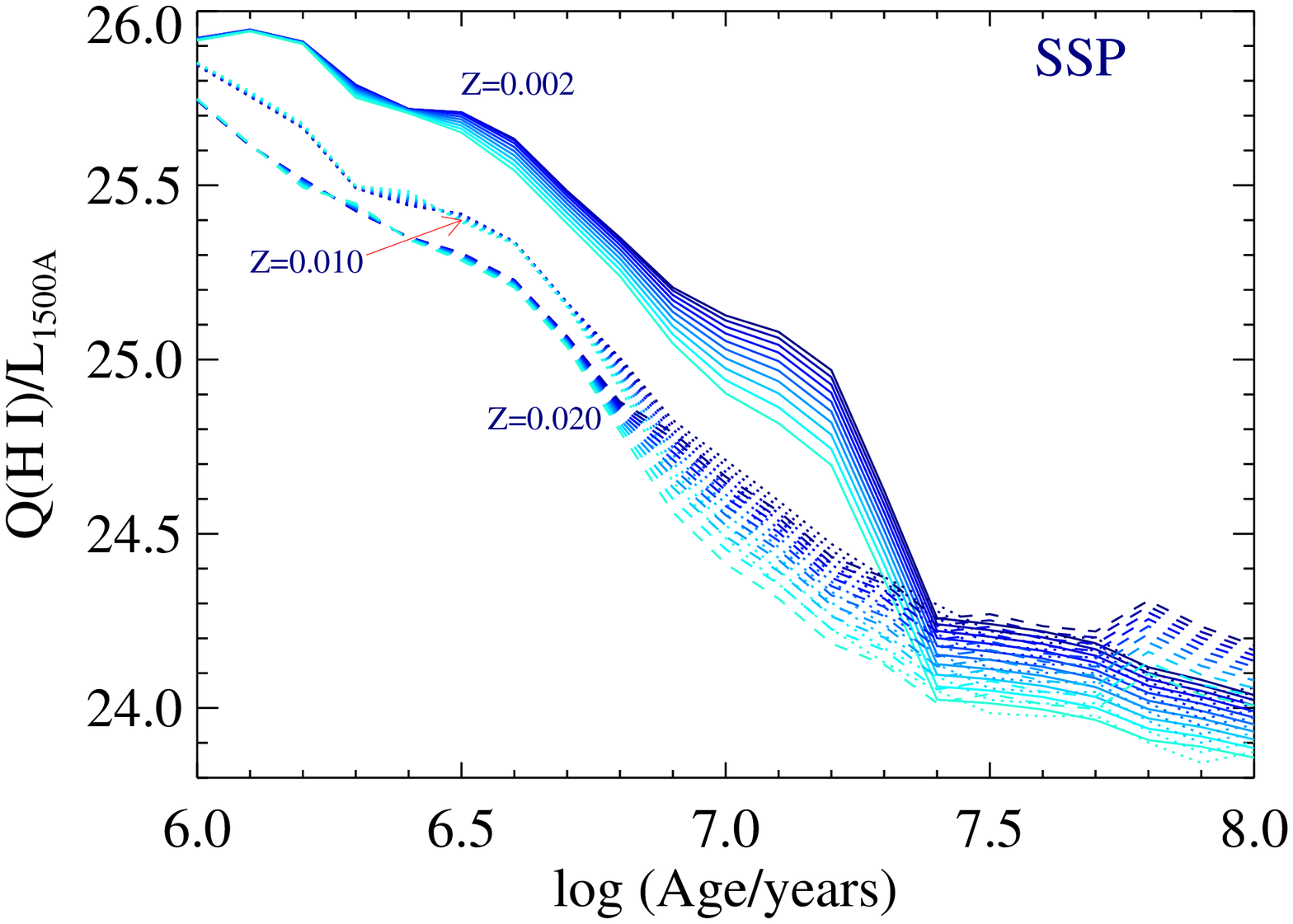}
	\includegraphics[width=\columnwidth]{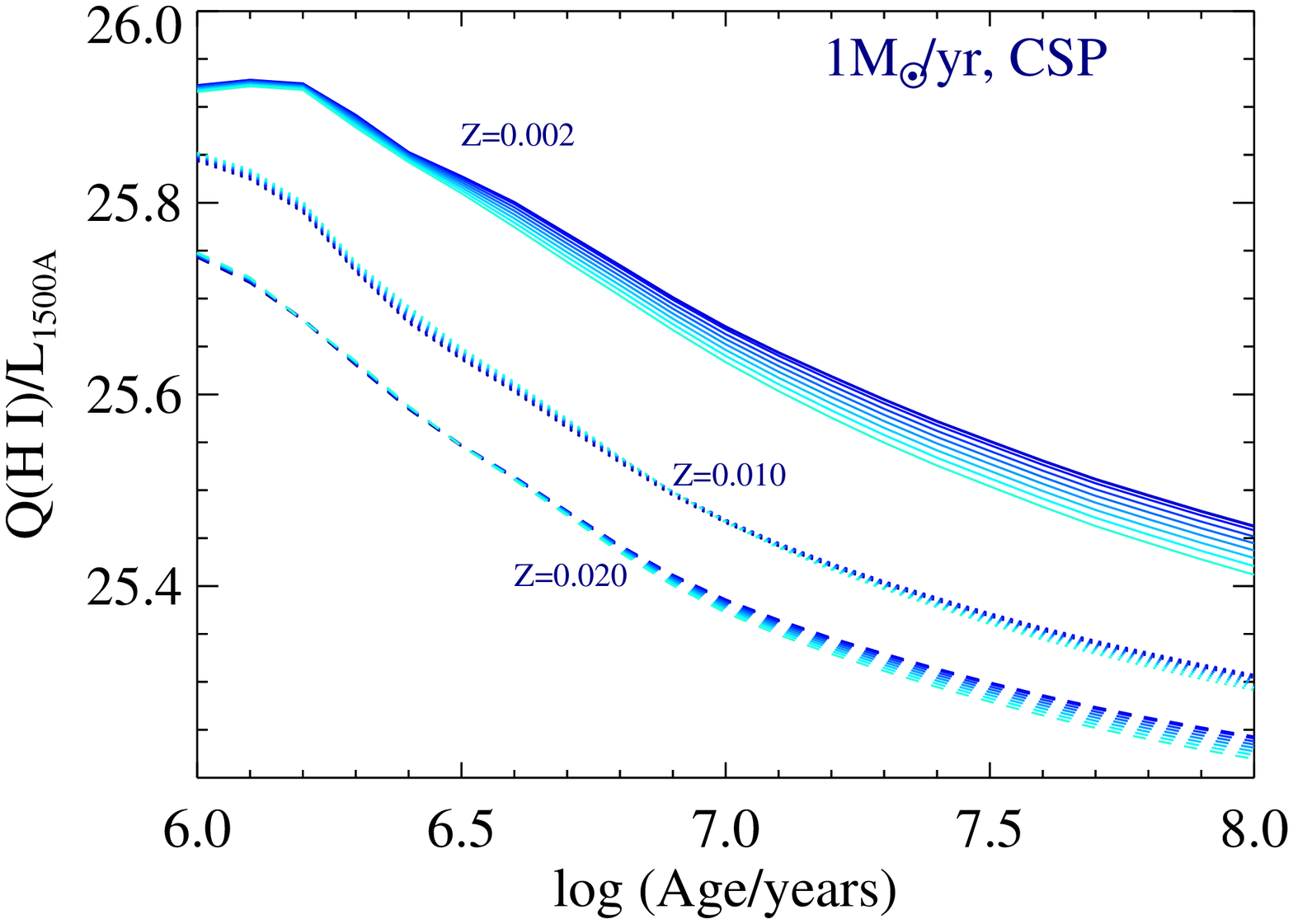}
    \caption{The strength of ionizing photon flux shortwards of 912\AA. Flux is shown relative to the 1500\AA\ UV continuum luminosity, i.e. the widely-used $\xi_\mathrm{ion}$ parameter. Models are colour coded as in Fig.\ \ref{fig:fbase}. Set 1 (increasing low mass binary fraction) models are essentially indistinguishable from the highest set 2 (reducing high mass binary fraction) models.}
    \label{fig:xi}
\end{figure*}

If the stellar spectrum shows little dependence on binary fraction (over the range considered here) then can we instead identify strong changes in the ionizing spectrum (i.e. shortwards of the Lyman limit at 912\AA)? Fig.~\ref{fig:specbase} suggests that the ionizing spectrum varies significantly between models in both set 1 and set 2. However this is primarily in terms of normalisation, just as is the case with the ultraviolet-optical stellar continuum. This means that the overall number of ionizing photons emitted by a population of known age and star formation history will be higher when the high mass stellar binary fraction increases, but the observable UV continuum is also stronger. The ionizing photon production efficiency parameter $\xi_\mathrm{ion}=\dot{N}_\mathrm{ion}/$L$_UV$ can be measured as a ratio between the observable 1500\AA\ continuum and the strength of hydrogen recombination emission lines (as a proxy for ionizing flux). This parameter shows a maximum difference between models in set 2 of  $<0.05$\,dex for a continuously star forming population at Z=0.002 and $<0.03$\,dex at Z=0.020, as shown in Fig.~\ref{fig:xi}. While slightly larger variation can be seen in simple stellar populations, this is highly time dependent. 

At any given age, populations with lower binary fractions amongst high mass stars show lower ionizing flux. The models in set 2 diverge significantly after log(age/years)$\sim$6.8 in a simple stellar population model and thereafter show a $\sim0.2$\,dex spread in $\xi_\mathrm{ion}$. Interestingly, Fig.~\ref{fig:xi} demonstrates that binary fraction dominates over metallicity in determining the ionizing photon production efficiency at ages above log(age/years)$=$7.4 (25 Myr). If simple stellar populations can be identified in this age regime, then $\xi_\mathrm{ion}$ may prove a useful probe of binary fraction. However, as Fig.~\ref{fig:xi} also demonstrates, any more complex star formation history drowns out this signal (as would any significant variation in binary mass ratio and period distributions) leading to ambiguity between population age, metallicity and binary fraction.

\begin{figure*}
	\includegraphics[width=\columnwidth]{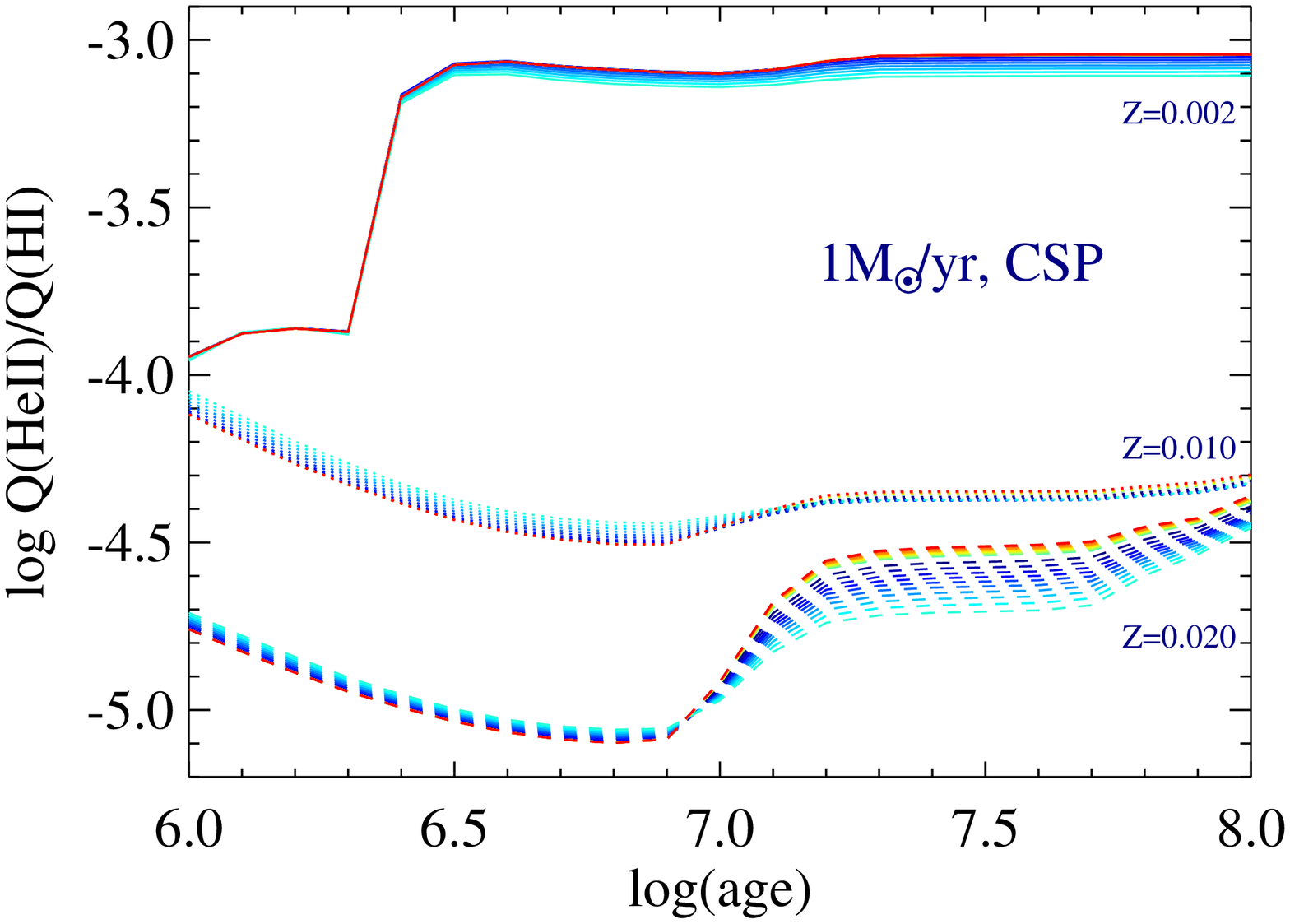}
	\includegraphics[width=\columnwidth]{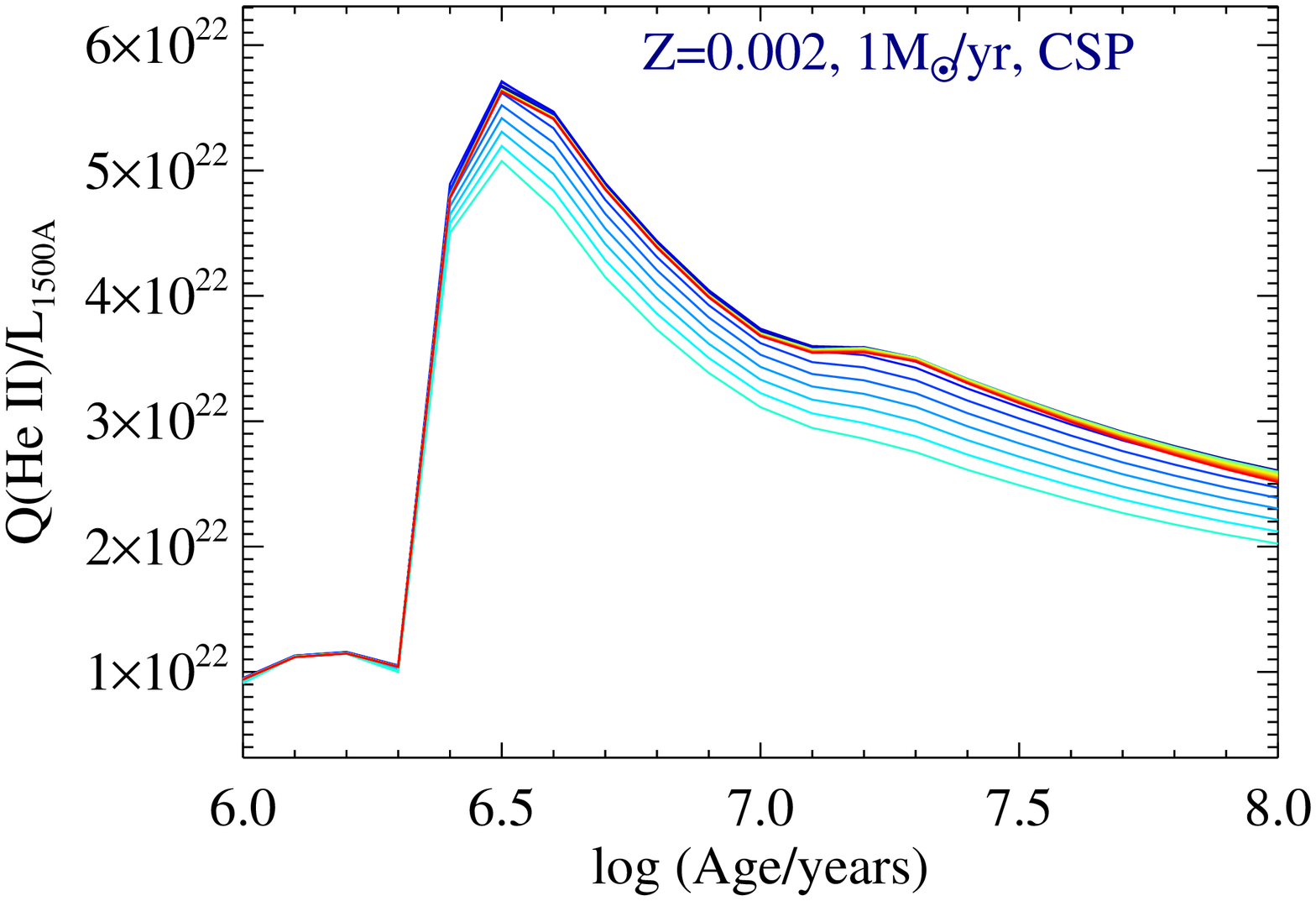}
    \caption{The strength of ionizing photon flux capable of powering the narrow He\,II\,1640\AA\ nebular emission line (i.e. emitted below 227\AA). Flux is shown relative to the H\,I ionizing flux (below 912\AA) and relative to the 1500\AA\ UV continuum luminosity, i.e. the He\,II equivalent of the widely-used $\xi_\mathrm{ion}$ parameter. Models are colour coded as in Fig.\ \ref{fig:fbase}. Set 1 (increasing low mass binary fraction) models are essentially indistinguishable from the highest set 2 (reducing high mass binary fraction) models.}
    \label{fig:heii3}
\end{figure*}

The helium ionizing photon flux (i.e. integrated light emitted below 227\AA) is also dependent on massive binary fraction. The time evolution of this flux, which will power a narrow nebular emission line component overlaying the stellar He\,II\,1640\AA\ line discussed above, is shown in Fig.~\ref{fig:heii3}. For a low metallicity composite stellar population with constant star formation, by the age of 100\,Myr the ratio of helium ionizing flux to UV continuum, $\xi_\mathrm{He\,II}$, varies by 25\,per cent, although the variation is substantially smaller at higher metallicities.

\subsection{Prospects for Observation}\label{sec:obs}

While the He\,II\,1640\AA\ line (and also its counterpart in the optical at 4686\AA) emerges as a potential binary indicator both through its broad stellar wind-driven component and the narrow recombination component emitted by ionized nebular gas, the prospects for using this line as a binary fraction indicator remain poor, particularly for sources in the distant Universe, where the metallicities and population ages are expected to favour binary population effects. 

The He\,II emission feature has been identified in distant galaxies ever since the first stacked spectra of galaxies at $z\sim3$ were constructed with sufficient signal to noise \citep{2003ApJ...588...65S}. As this and studies since have demonstrated, typical ultraviolet-selected star-forming galaxies in the early Universe show evidence for weak, broad He\,II emission {\it en masse} \citep[e.g.][]{2003ApJ...588...65S,2016ApJ...826..159S,2019MNRAS.487.2038C}. A handful of objects in the distant Universe have also been identified as individual emitters of strong but usually narrow (nebular) He\,II emission  \citep[e.g.][]{2010ApJ...719.1168E,2019MNRAS.482.2422S,2014ApJ...790..144B}. Similar strong emission features are also seen in both the ultraviolet and optical He\,II lines in extreme, low metallicity starbursts \citep[e.g.][]{2019ApJ...878L...3B,2018MNRAS.480.1081K}.

Such strong He\,II emission is by no means universal even in young starbursts: \citet{2019arXiv191109999S}, working with extremely deep spectra from the VANDELS survey suggest that only about 20 galaxies show strong (W(He\,II)$>0.5$\AA) narrow emission, and 6 broad emission, amongst a much larger sample (899, $<z>\sim3.5$) of galaxies with similar observed properties, masses and inferred star formation rates. Typical uncertainties on the line flux are at the 25 to 40 per cent level, even for these detections. Similarly, \citet{2019A&A...624A..89N} identify a small sample of 13 individually detected sources amongst almost eight hundred galaxies observed in deep (>9\,hr) VLT/MUSE spectroscopic fields. 

This suggests that such emission is highly stochastic, occurring in short-lived bursts, more consistent with the $\sim$10\,Myr timescale for emission from simple stellar populations shown in figure 13 (left), rather than a slow but continuous star formation episode. Such stochasticity in the star formation history will make the interpretation of any given line detection as a binary indicator challenging. The time variation in the SSP line strength on 10 Myr timescales exceeds the difference in line strength introduced by changing the binary fraction, even if the line is measured at sufficient signal-to-noise to distinguish between binary fraction models (i.e. uncertainties $<<$10 per cent, beyond most current observations in the distant Universe, and challenging given the faintness of the lines even at low redshift). 

A further complication is hinted at by the strengths of the observed emission lines. In extreme cases, these reach equivalent widths of W(He\,II)$>10$\AA\ in the rest-frame. As we discussed in \citet{2019A&A...621A.105S}, such lines are beyond the capacity of any current stellar population synthesis model, including BPASS, to reproduce and imply the presence of a source of ionizing photons other than stellar photospheres. This, together with the strong metallicity dependence observed in X-ray binary models \citep[e.g.][]{2019ApJ...885...65F}, has motivated ongoing work by us and others to incorporate XRB populations in stellar population synthesis (Eldridge, Stanway et al, in prep). Since emission spectra of such sources are not implemented in the current version of BPASS (or any other code) we caution against over-interpretation of the He\,II feature. 

A full analysis of nebular radiative transfer from these ionizing populations lies well beyond the scope of this paper. However, we note that the comparison of flux below 912\AA\ (hydrogen ionizing) and that below 227\AA\ (helium ionizing) may fail to detect changes in the ionizing spectrum \textit{shape} between these limits. Other nebular emission lines may be sensitive to this behaviour, for example the ratios between the semi-forbidden C\,III] doublet, the He\,II line, Lyman-$\alpha$ and [O\,III] are diagnostic of the relative flux below the respective critical wavelengths for these features. All of these features save Lyman-$\alpha$, itself a resonantly scattered line and difficult to interpret, are weak in the spectrum of typical high redshift stellar populations. Nonetheless  spectroscopic surveys such as VANDELS \citep{2018MNRAS.479...25M} are now achieving signal to noise ratios $\sim20$ per resolution element. The prospect of NIRSPEC on the {\em James Webb Space Telescope} extending similar high precision measurements to larger samples and into the rest-frame optical, suggests that detailed reconstruction of the ionizing spectrum (and the role of binaries in shaping it) may be possible in the near future. 

\begin{figure*}
	\includegraphics[width=\columnwidth]{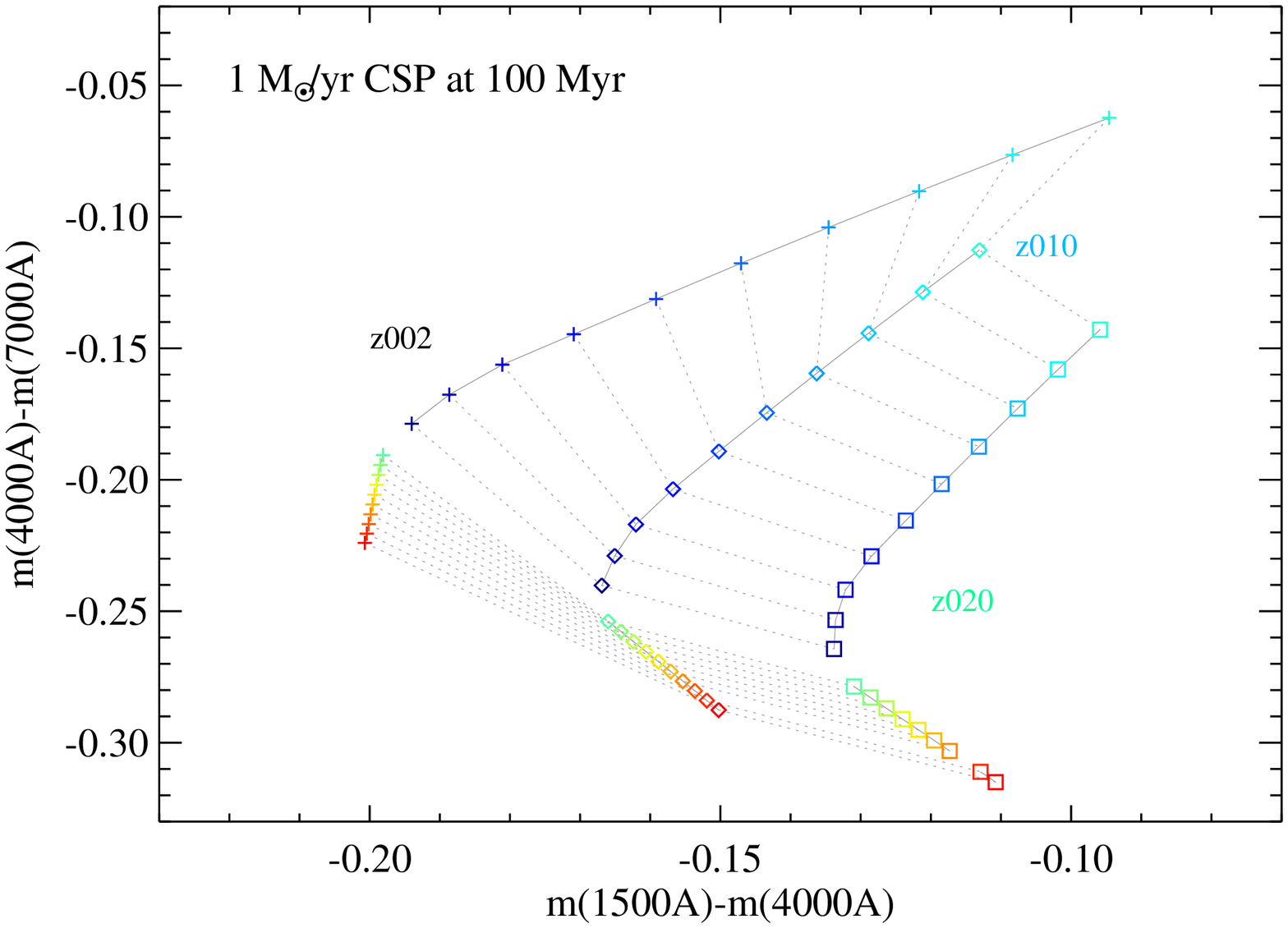}
	\includegraphics[width=\columnwidth]{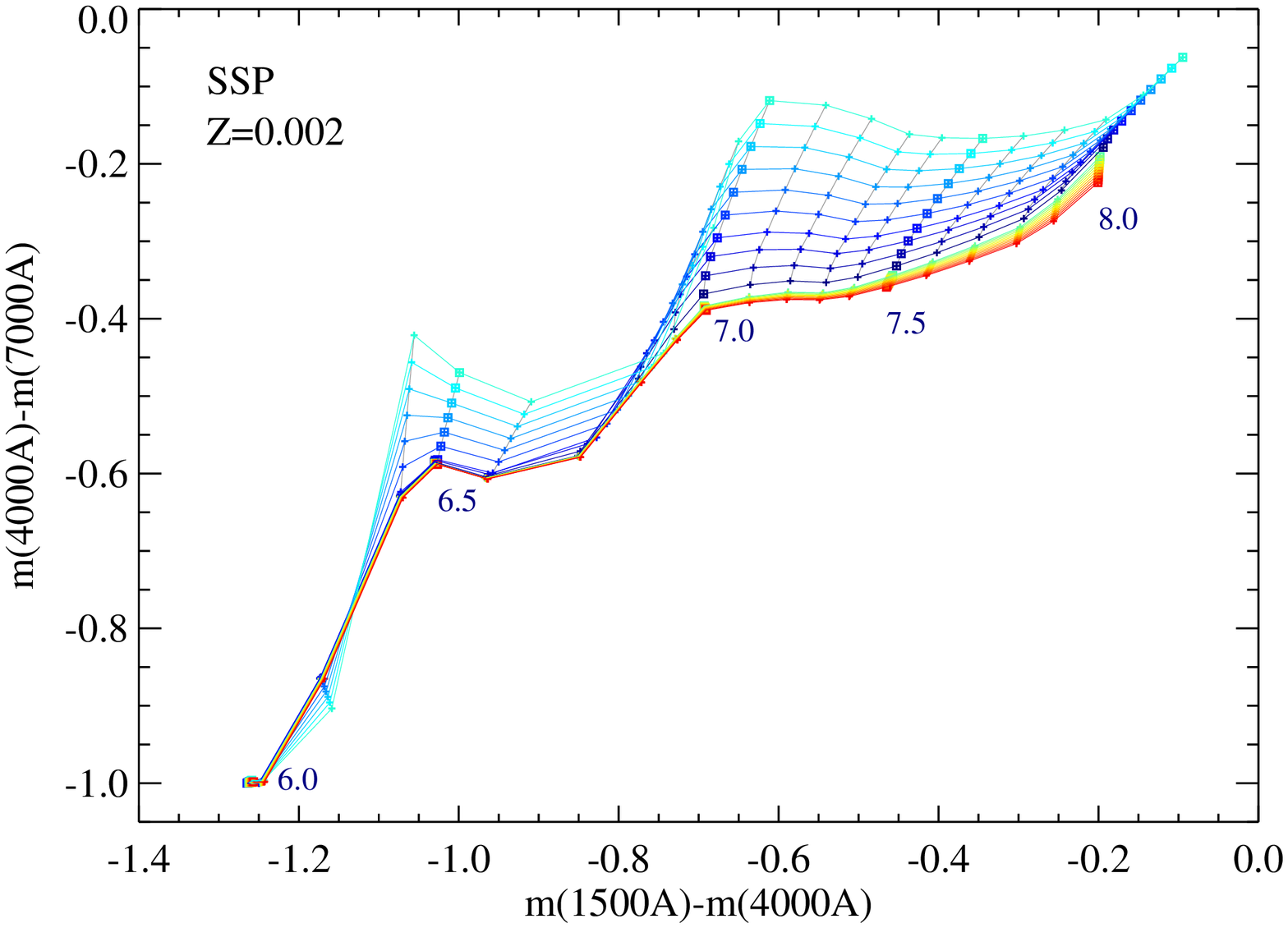}
	\includegraphics[width=\columnwidth]{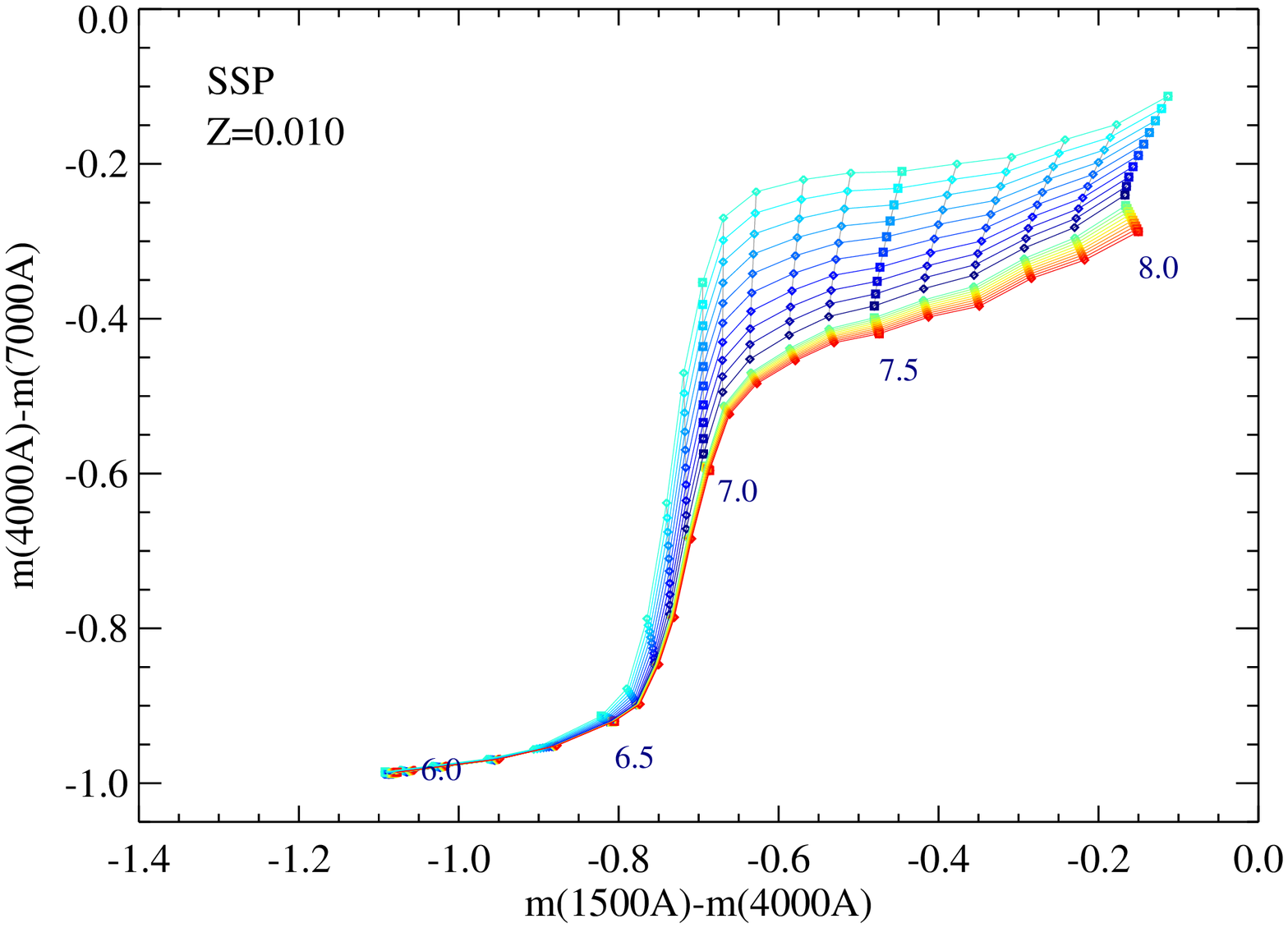}
	\includegraphics[width=\columnwidth]{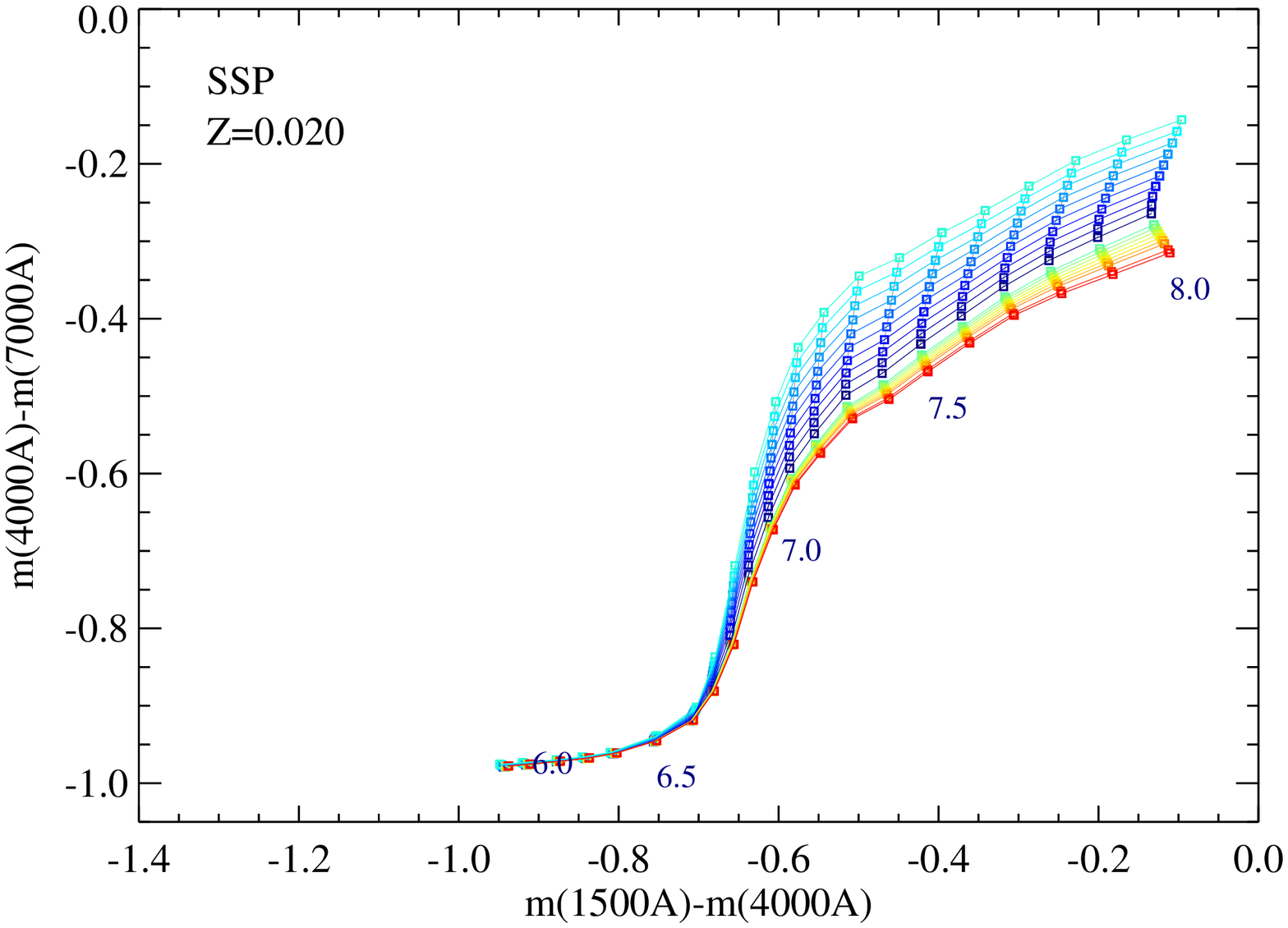}
    \caption{The synthetic photometric colours of populations for stellar populations with varying binary fraction, colour coded as in Fig.\ \ref{fig:fbase}. Top-hat pseudo-filters of 100\AA\ width are centred at 1500, 4000 and 7000\AA. Colours are given in AB magnitudes, and shown for all three metallicities given a continuously star forming population at 100\,Myr (top left) and for simple stellar populations as a function of age for the three metallicities separately: Z=0.002 (top right), Z=0.010 (bottom left), Z=0.020 (bottom right). Selected age steps are annotated with log(age/years). In all cases, the 4000-7000\AA\ colour for set 2 is higher than that for set 1. No nebular gas reprocessing is considered}
    \label{fig:newcol}
\end{figure*}

Potentially more promising as a high mass binary fraction diagnostic is the rest-frame ultraviolet through optical colour, identified in Fig.~\ref{fig:specbase}. In Fig.~\ref{fig:newcol} we estimate photometric colours for the variant models of section \ref{sec:base}. Given that the details of filter profile will depend on instrument and target redshift, we model the observations as simple top-hat filters with a width of 100\AA, centred at rest-frame wavelengths of 1500, 4000 and 7000\AA. At $z=5$, such a combination might be approximated by combining NIRCAM filters F090W (or HST/WFC3 F606W), F162M and F300M, while at $z=3$ the equivalent combination would be F090W, F250M and F430M (note: broader filters will typically reduce photometric spread between models).

The top left panel of Fig.~\ref{fig:newcol} suggests that this colour combination (or a similar filter selection) is indeed a potential diagnostic of both metallicity (at the level of precision explored here) and high mass binary fraction for continuously star forming populations, if $\sim$0.01\,mag photometric precision can be achieved. Distinguishing between low mass binary fractions requires significantly higher (millimag) precision. 
Unfortunately, as the remaining panels of Fig.~\ref{fig:newcol} make plain, such a simple interpretation will be compromised by any young stellar populations or complex star formation history. The predicted colours of simple stellar populations at different metallicities overlap over much of their parameter space, introducing a stellar age-metallicity-binary fraction degeneracy. 

\begin{figure}
	\includegraphics[width=\columnwidth]{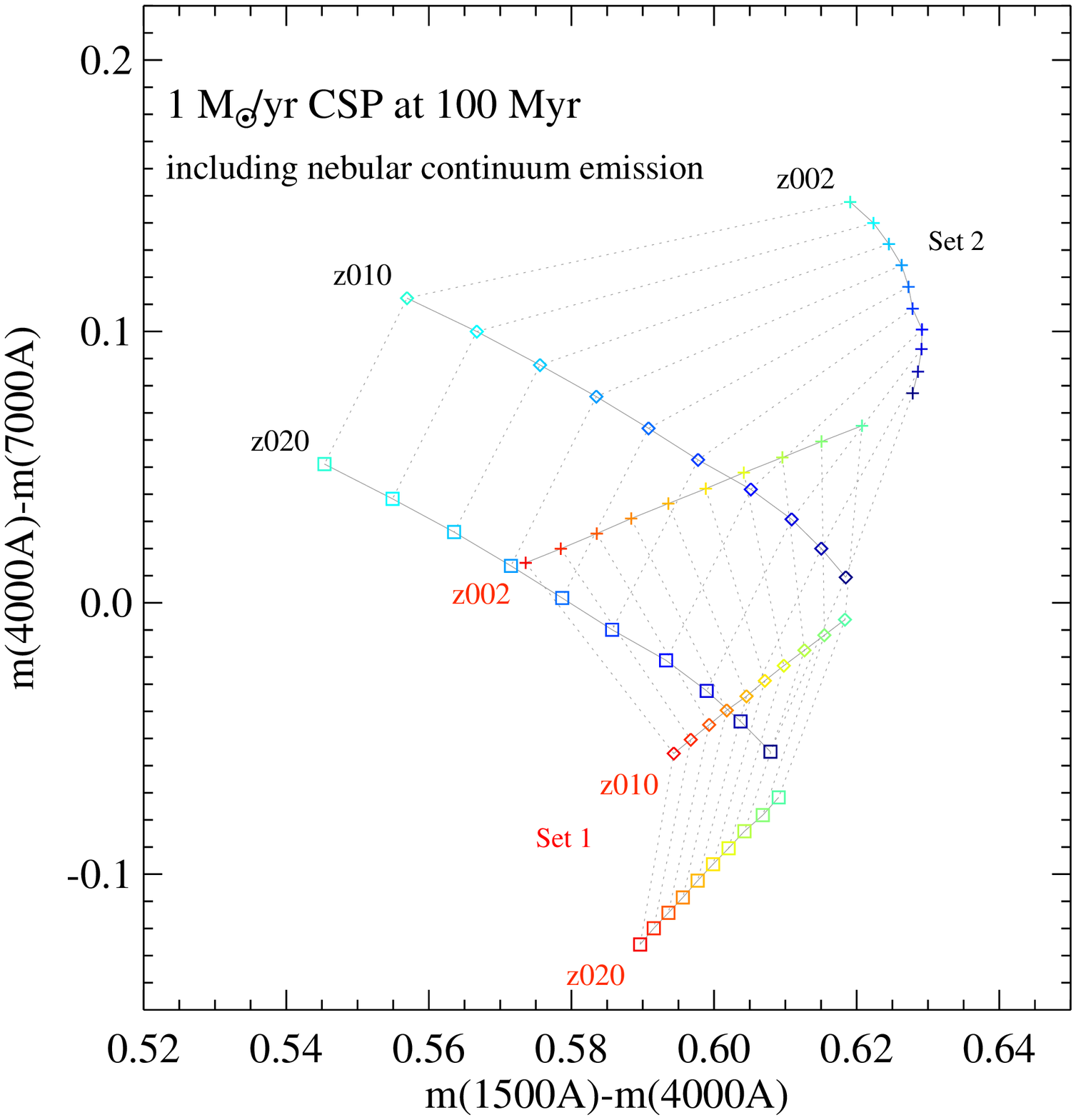}
    \caption{The synthetic photometric colours of populations for stellar populations with varying binary fraction, colour coded as in Fig.\ \ref{fig:fbase}. Colours are given in AB magnitudes, and shown for all three metallicities given a continuously star forming population at 100\,Myr. The reprocessing of the stellar spectrum by nebular gas is included, as described in section \ref{sec:obs}.}
    \label{fig:newcol_neb}
\end{figure}

We also note that the colours presented in Fig.~\ref{fig:newcol} are for the stellar continuum only and do not include the reddening effects of the nebular gas continuum. The effects of nebular gas reprocessing are demonstrated (for continuous star formation and a single selected set of gas parameters - $n_e=10^2$\,cm$^{-3}$, spherical, complete covering, fixed geometry, dust grain depletion\footnote{Modelled with \sc{CLOUDY} c17.01 \citep{2017RMxAA..53..385F}}) in Fig.~\ref{fig:newcol_neb}. This acts as a reddening term to reduce the spread in predicted photometric colour. The models still form a grid which is potentially diagnostic of metallicity and binary fraction given $\sim$0.01\,mag photometric precision. However this is challenging to achieve in UV to optical colours due to systematic calibration uncertainties (more typically a few percent), corrections for dust extinction (with uncertainties often $>$0.1\,mag in the ultraviolet), and the degeneracies introduced by star formation history,  with additional possible uncertainties due to the gas parameters also introduced.

Upcoming observations, such as the planned NIRCam-NIRSpec Galaxy Assembly Survey and The Cosmic Evolution Early Release Science (CEERS) Survey, will plausibly be able to eliminate some of the more extreme potential variants in binary fraction at moderate redshift, through a combination of extremely high signal to noise and uniformly calibrated imaging, the use of multiple lines to determine dust extinction, and high precision spectroscopic data. Nonetheless the difficulty of distinguishing systems with high binary fractions from those with lower metallicities or slightly younger stellar populations will not be easily solved. It remains to be seen whether combining full spectral-fitting (including nebular gas radiative transfer) with photometric spectral energy distributions will be sufficient to break these degeneracies. The analysis presented here suggests that doing so may be challenging, but further analysis may point to more hopeful strategies in future.

\section{The effects of uncertainties on stellar populations}\label{sec:discussion}

The time evolution of the variant models explored above has already provided hints regarding the components of the stellar populations most vulnerable to uncertainties in the input distributions. However, it is also informative to consider the variation in these populations directly.

\begin{figure*}
	\includegraphics[width=\columnwidth]{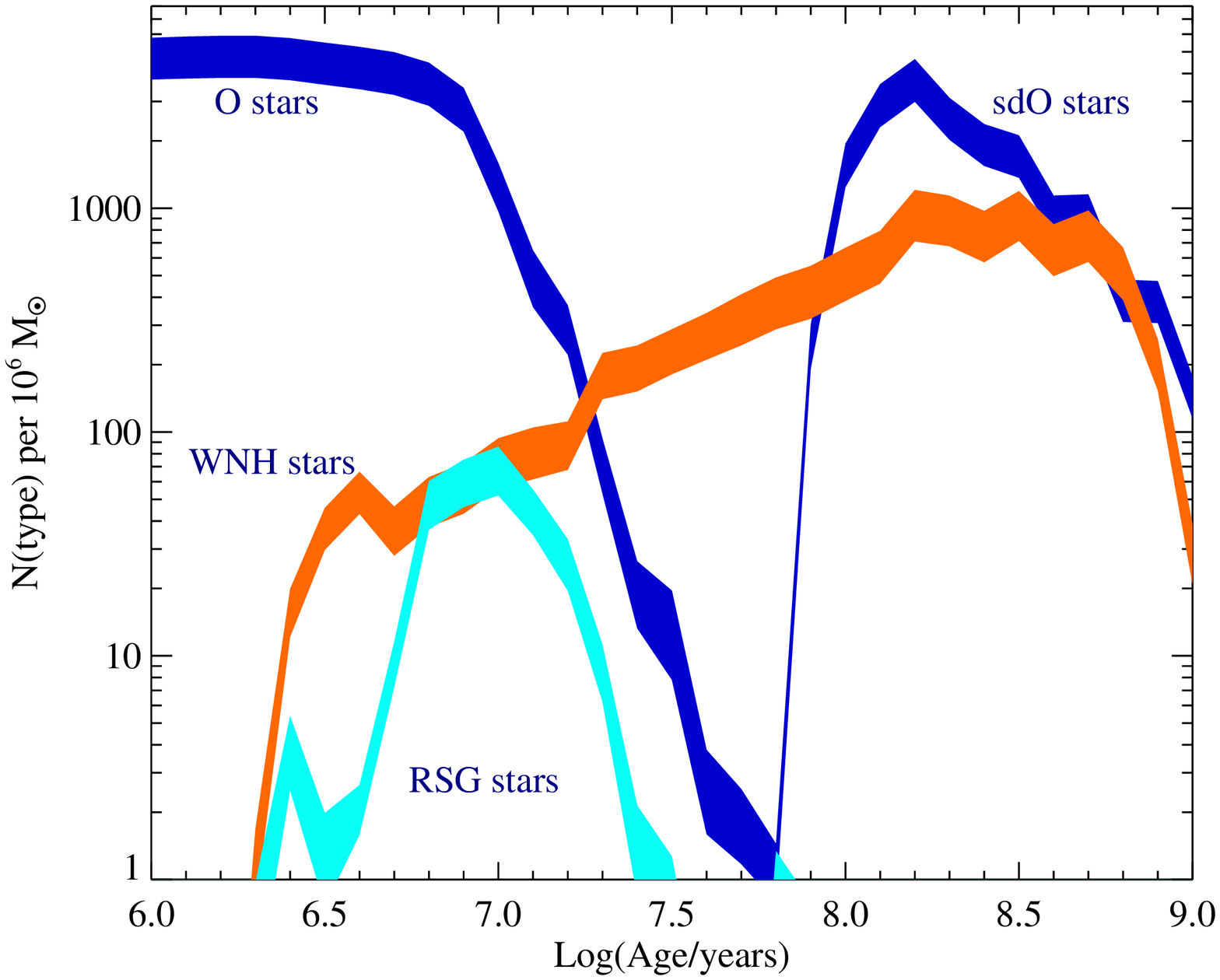}
	\includegraphics[width=\columnwidth]{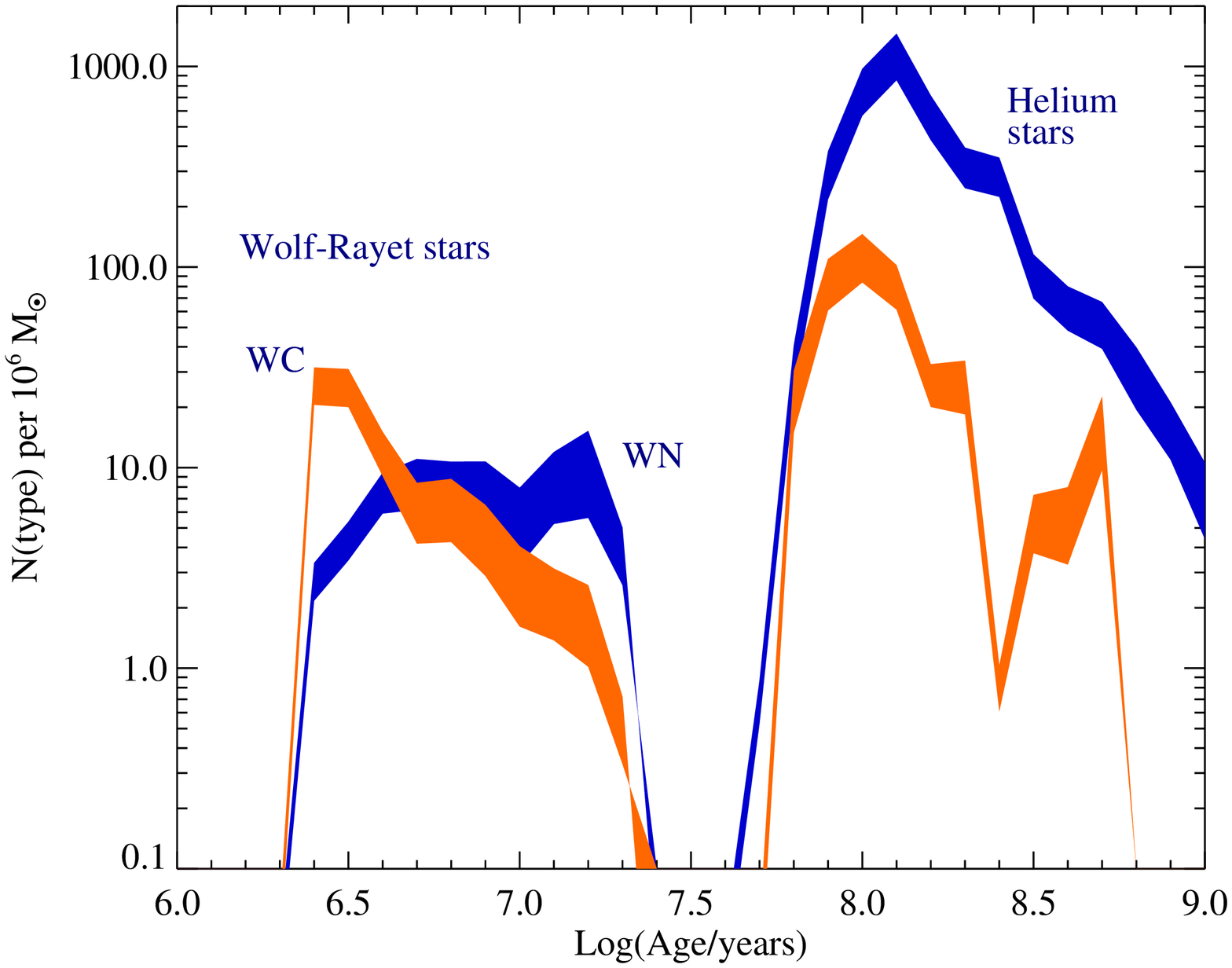}
	\includegraphics[width=\columnwidth]{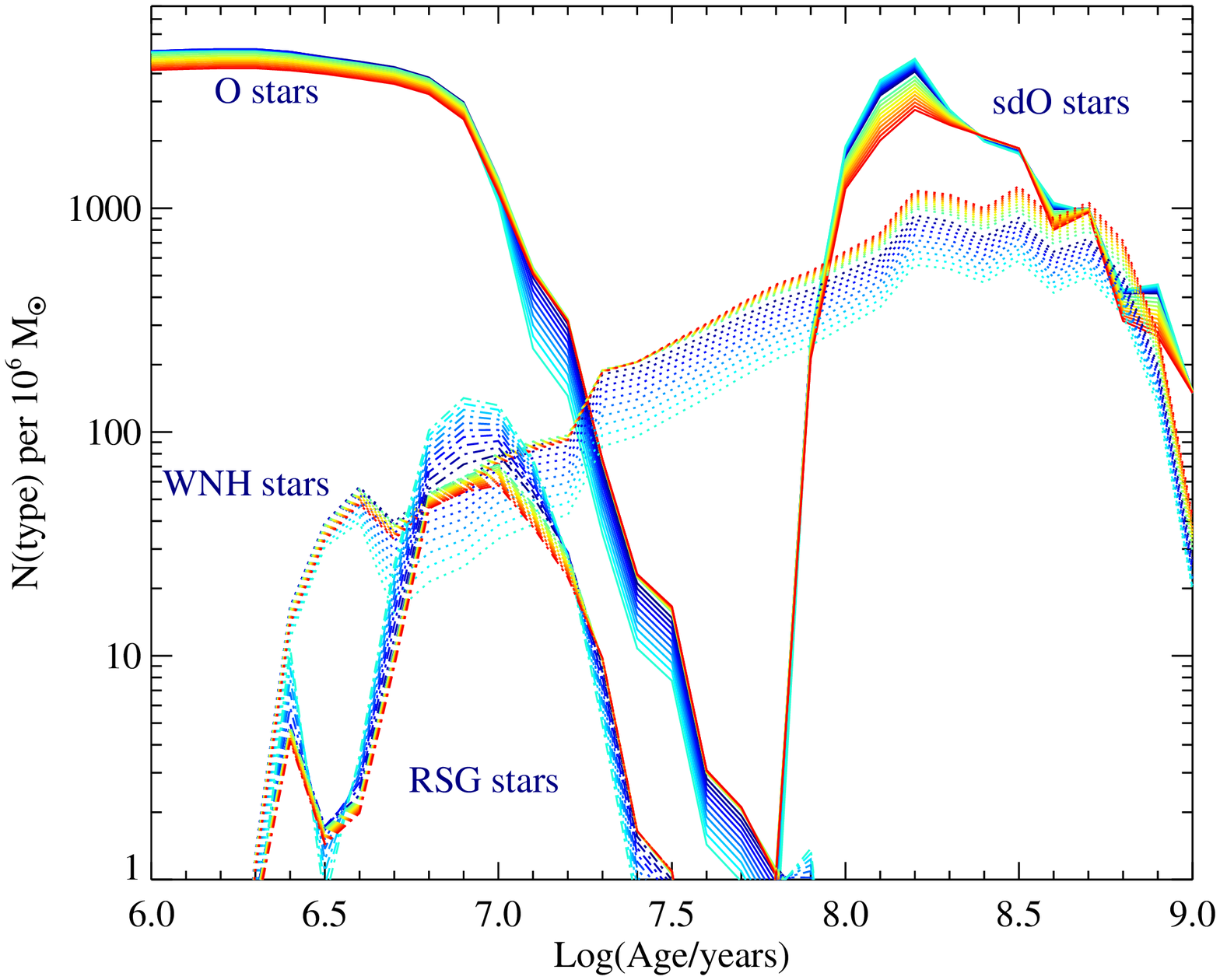}
	\includegraphics[width=\columnwidth]{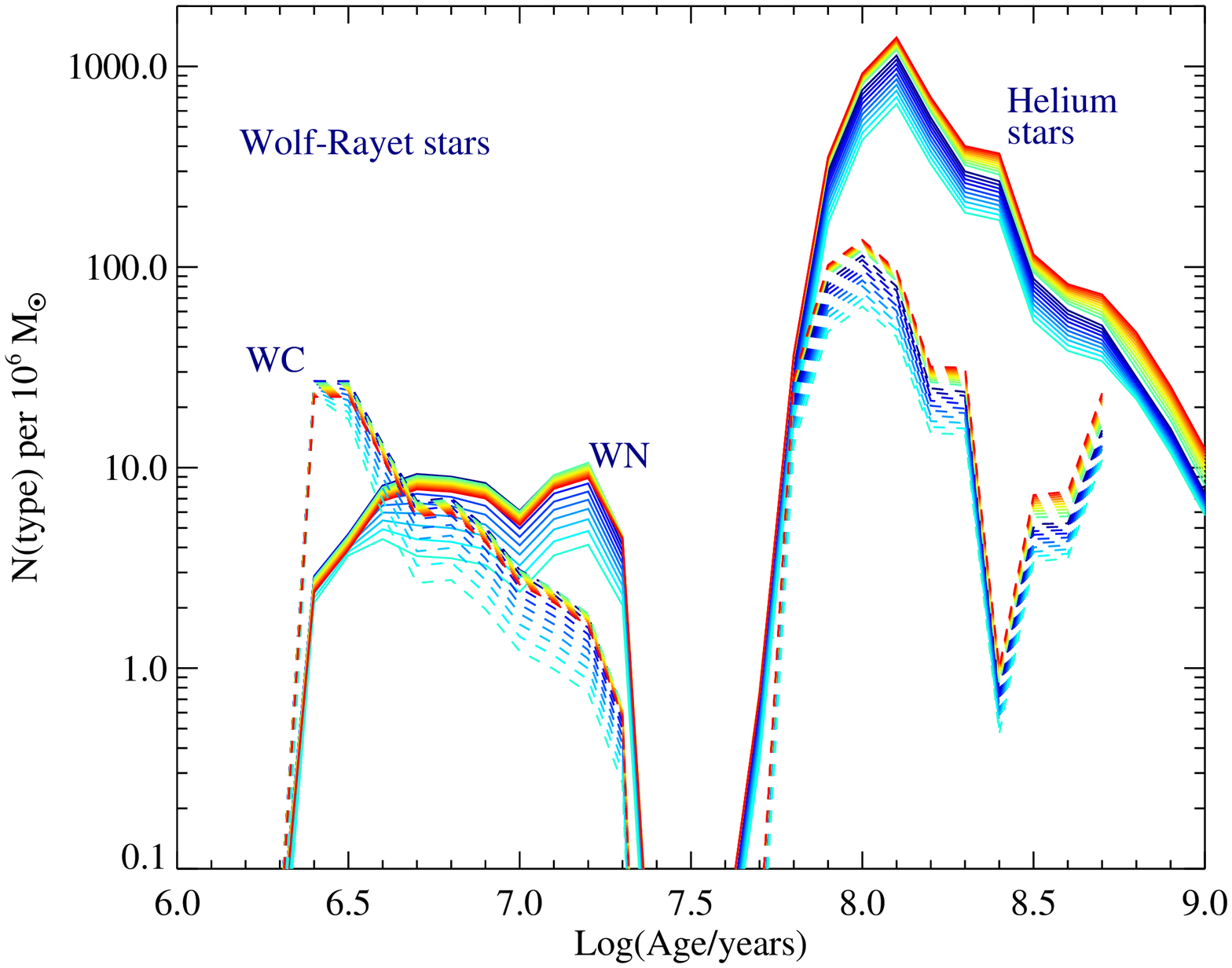}
    \caption{The distribution of stars between stellar types as a function of age, for SSP models at $Z=0.002$. Top: mean $\pm\sigma$ uncertainty range in number counts permitted by current uncertainties on binary parameters; Bottom: varying binary fraction, colour coded as in Fig.\,\ref{fig:fbase}.}
    \label{fig:nums}
\end{figure*}

Fig.~\ref{fig:nums} shows the effect of varying binary population parameters on the numbers of different stellar types as a function of stellar population age for a simple stellar population of $10^6$\,M$_\odot$ at Z=0.002. Number counts are presented for the dominant hot star types responsible for ionizing radiation in star-forming populations: type O main sequence stars, partially-stripped WNH stars and stripped-envelope Wolf-Rayet (WR) stars in the WC and WN sub-classes. All stellar models in BPASS are classified by effective temperature and surface composition \citep[see][]{2017PASA...34...58E}. Behaviour at other metallicities is qualitatively similar, although evolution tends to be a little faster as metallicity increases (due to stronger winds) and the number of luminous WR stars produced is lower at the highest metallicities. 

The top panels of Fig.~\ref{fig:nums} illustrate the $\pm1$\,$\sigma$ range of number counts which are permitted by current observational constraints on binary parameters (i.e. the variant (v) models of section \ref{sec:method}), while the lower panels show explicitly the dependence on binary fraction while keeping all other parameters fixed (the models of section \ref{sec:base}). In each case, the total population by stellar type is shown as a function of age. It should be noted that stripped and partially-stripped populations are almost entirely comprised of luminous, massive stars (log(L/L$_\odot$)>4.9) at early ages and slower evolving, often post-interaction, lower mass stars at late times, resulting in the characteristic double-humped time distribution. 

As the comparison of this figure with Fig.~\ref{fig:heii3} makes clear, the origin of time (and binary fraction) dependence in  He\,II line strength can be attributed almost entirely to the luminous (log(L/L$_\odot$)>4.9) population of classic Wolf-Rayet stars. Indeed, the time evolution of the He\,II stellar line equivalent width hints that WN stars likely dominate over WC stars (which peak earlier). These are present in the population in a well-defined temporal window between log(age/years)$\sim$6.3 and 7.3, and are strongly dependent on the binary fraction of massive stars at low metallicity, as is also seen in the equivalent width of the He\,II\,1640\AA\ emission line. At higher metallicities, where winds contribute significantly, the dependence on binary fraction is less marked. While other stellar populations may contribute weakly to this line at other times, the behaviour of the He\,II\,1640\AA\ line in these models is clearly dominated by this handful of very massive stars which have lost their envelope.

The hydrogen ionizing photon production efficiency at low metallicities may also receive a significant contribution from this population. As Fig~\ref{fig:xi} demonstrated, at Z=0.002 the early-time decline in ionizing flux is much gentler than that seen at Z=0.010 and Z=0.020, before abruptly dropping at log(age/years)$\sim$7.3. This would be consistent with Fig~\ref{fig:xi} primarily tracking the O star population (which dominates in both total luminosity and number counts) but being boosted at low metallicities by the WR population, before these drop sharply away.

In the case of low high-mass binary fractions, the decline in the number of  giant stars whose envelopes are stripped through Roche Lobe overflow is compensated for by an increase in the number of red supergiants (inferred types K and M with log(L/L$_\odot$)$>4.9$, labelled RSG on Fig.\,\ref{fig:nums}). These stars retain their envelopes through their expansion onto the giant branch and cool significantly, resulting in the redder colours seen in the optical in Figs.\,\ref{fig:specbase} and \ref{fig:newcol}. In common with \citet{2018ApJ...867..125D}, we find that the WR-to-RSG ratio is diagnostic of high mass binary fraction in resolved populations, and note that the relative fraction of type II core-collapse supernovae to stripped-envelope type Ib/c supernovae also tracks this ratio.

 The role of the lower luminosity hot population which arises primarily from binary interactions is less clear cut. Comparison of the timescales in Fig.~\ref{fig:nums} with Fig.~\ref{fig:stdevs} suggests that the binary initial period distribution affects these populations most strongly.  The ionizing photon production rate at early times is always dominated by the (relatively binary-insensitive) short-lived massive stars, but at late times it is the low luminosity population that dominates. In particular, the large population of partially stripped WNH stars are sensitive to binary fraction and show a marked increase from low numbers at early times, to a peak in population numbers at log(age/years)$\sim8.5$. While fully-stripped low luminosity helium stars also grow in number until their population is comparable to that of WNH stars, they are more peaked in time, falling away sharply after log(age/years)$\sim8.1$. By contrast the uncertainty in ionizing photon rate from the populations shown in Fig.~\ref{fig:stdevs} varies more slowly, suggesting that it is the partially- rather than fully-stripped stars which are dominating this uncertainty, and thus dominating the uncertainty in the ionizing photon rate from a continuously star forming population.

\section{Conclusions}\label{sec:conc}

In this paper we have explored the impact of empirical uncertainties on the parameter distributions in binary fraction as a function of initial primary mass, initial orbital period and initial mass ratio. We have sampled the range of values permitted by current empirical constraints on these distributions and also explored the effect of modifying the binary fraction as a function of primary mass into regimes outside of the current constraints. We have focused on unresolved stellar populations, since the majority of circumstances in which the products of SPS models are used do not permit the identification of individual stars, but rather considered the integrated light of stellar populations or galaxies. Our principle conclusions can be summarised as follows:
\begin{enumerate}
\item The current constraints on binary distribution parameters are relatively tight on the binary fraction as a function of mass and period, but allow more variation in mass ratio. Fitting a functional form to the empirical constraints as a function of mass or period prevents unphysical distributions and reduces the impact of uncertainties.
\item The inferred uncertainties on the ionizing photon flux from a young stellar population are $<0.1$\,dex, while those on the UV spectral slope and 1500\AA\ continuum luminosity are $<$0.03 and $<$0.03\,dex respectively. Low metallicity models show a slightly increased sensitivity to the binary parameters.
\item Uncertainty on the ultraviolet-optical continuum luminosity is typically 10-15\,per cent, with the uncertainties acting primarily on the continuum normalisation rather than slope or emission/absorption lines.
\item Photometric colours are typically uncertain at the $<0.05$\,mag level, except in a narrow range of ages.
\item The He\,II emission line at 1640\AA\ shows increased sensitivity to the binary fraction relative to the continuum, both due to stellar emission and nebular emission from ionized gas. However this emission is highly time dependent and small uncertainties in population age or metallicity will compromise attempts to interpret measurements.
\item The 1500-4000\AA\ colour relative to the 4000-7000\AA\ colour of the stellar continuum is also mildly sensitive to extreme binary fraction distributions. Again, age and metallicity uncertainties will complicate interpretation, as will uncertainties on dust and nebular gas reprocessing of the stellar continuum.
\item Most of the variation seen with binary parameters arises from changes in the number of stripped-envelope stars (particularly luminous Wolf-Rayet stars) in the population, with contributions also arising from partially-stripped WNH stars.
\end{enumerate}

While the absence of any clear and unambiguous binary parameter indicators in the integrated light of stellar populations is disappointing, the very small uncertainties permitted by current observational constraints on the nearby stellar population are reassuring: interpretation of these populations are unlikely to be compromised by binary parameter uncertainties, and is only mildly sensitive to massive star binary fractions ranging between 40 and 100 per cent. It is notable however that in every parameter, the lowest metallicity models show the largest sensitivity to the input binary parameter distributions. Since it is in low metallicity environments that binary fractions are also expected to vary, and for which the stellar atmospheres of hot stars are also most uncertain, these results that caution should be used in their interpretation, and are a reminder that efforts should be intensified to fully characterise very low metallicity populations wherever possible.

\section*{Acknowledgements}

ERS and AAC recieved support from United Kingdom Science and Technology Facilities Council (STFC) grant number ST/P000495/1 and studentship 1763016. JJE and HFS acknowledge support from the University of Auckland and the Royal Society Te Ap\={a}rangi of New Zealand under the Marsden Fund.
BPASS would not be possible without the computational resources of the University of Auckland's NZ eScience Infrastructure (NeSI) Pan Cluster and the University of Warwick's Scientific Computing Research Technology Platform (SCRTP).




\input{bpass_uncertainties.bbl}

\label{lastpage}


%
\cleardoublepage
\pagenumbering{arabic}
\renewcommand*{\thepage}{A\arabic{page}}
\appendix

\section{The Interaction of Binary Parameters and IMF (online only)}

As discussed in detail by \citet{2018PASA...35...39H}, the initial mass function of a stellar population presents significant difficulty, both conceptually and in terms of its observational determination. Increasing evidence suggests that there may be no such thing as a universal initial mass function, and those empirical functions determined from observations rely on stellar evolution and population synthesis models to infer the mass and age of individual stars and estimate the changes in the population which have occurred since the stars reached the zero-age main sequence. Such inference is typically made with single star population synthesis models, which may introduce systematic offsets, and suffer from additional uncertainties in mass and age determination from unresolved binaries, or high mass-ratio systems in which the secondary is not observed.

The application of an initial mass function also presents certain complications in binary population synthesis. Since the companion probability and mass ratio distribution is strongly dependent on primary mass, altering these parameters will result in higher or lower numbers of comparatively low mass companions for each star, altering the overall mass function. Thus each perturbation in binary parameters is also an effective perturbation on the IMF power law slope.

Kroupa and coworkers \citep[e.g.][]{1993MNRAS.262..545K,2001MNRAS.322..231K,2003ApJ...598.1076K,2013pss5.book..115K,2018arXiv180610605K} have extensively discussed the effects of stellar multiplicity on initial mass function estimates. They advocate the random pairing of stars drawn from a canonical single star IMF, with a period assigned statistically to each binary, in a so called Birth Binary population. However such an approach does not permit the mass ratio distribution to be  controlled or varied, and implicitly assumes that this distribution is determined only by the initial mass function (i.e. low mass companions are more common as a result of random pairings, simply because low mass stars are more common). By contrast, other binary population synthesis models including {\it BSE} \citep{2013ascl.soft03014H}, {\it binary\_c} \citep{2018arXiv180806883I,2018MNRAS.473.2984I} and {\it StarTrack} \citep{2008ApJS..174..223B,2019arXiv190207718B} treat the mass function as applying only to single and primary stars, with secondary stars lying outside of the initial mass function weighting and their mass determined instead by the mass ratio distribution.

\begin{figure}
	\includegraphics[width=\columnwidth]{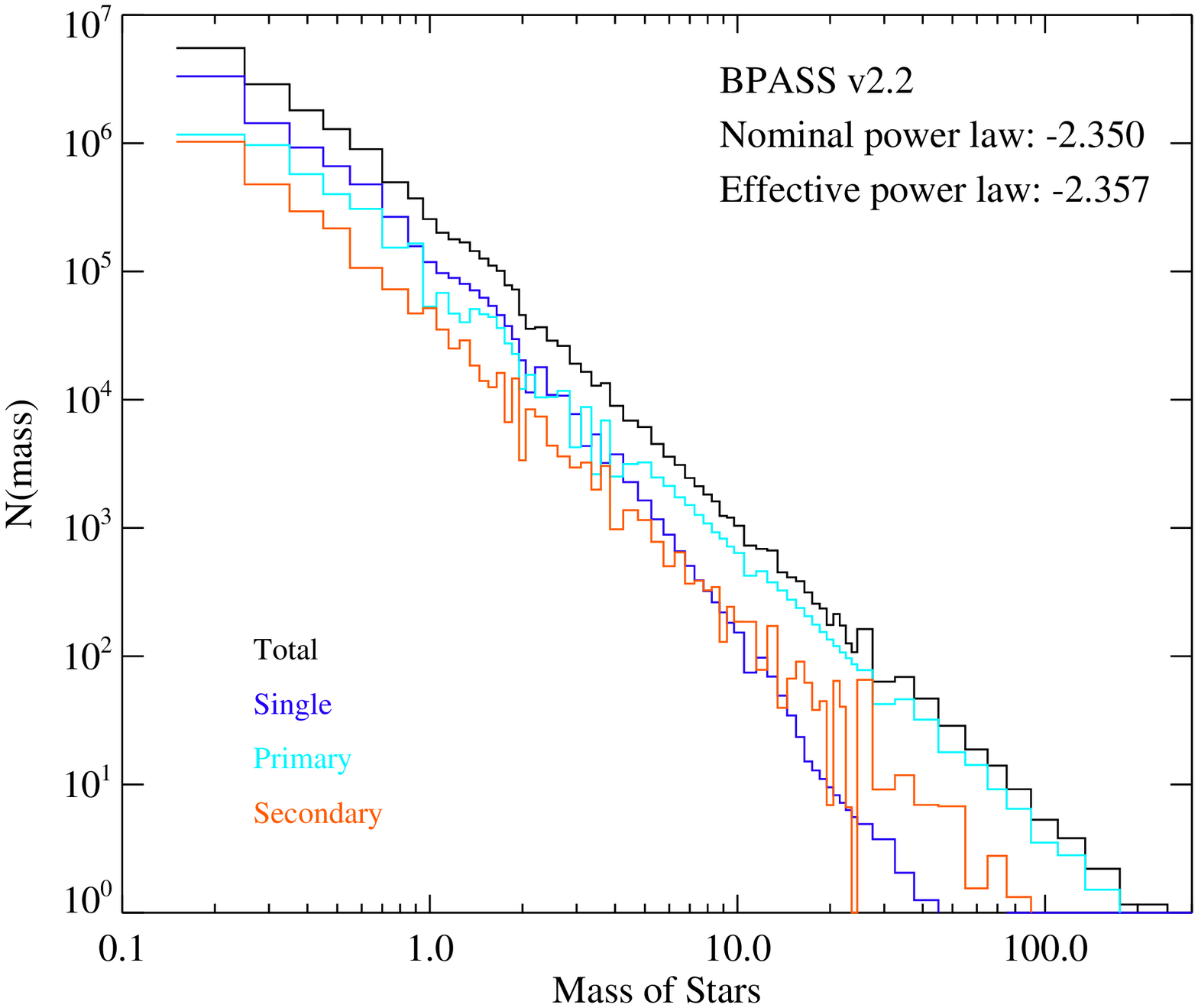}
    \caption{The role of single, primary and secondary stars in building the effective IMF using the canonical prescription for BPASS v2.2. The nominal IMF (applied to single stars and primaries) has a slope of -2.350 between 0.5 and 300\,M$_\odot$. Secondary stars perturb this, such that a power law fit between these mass limits yields an effective slope of -2.357.}
    \label{fig:imf_v2.2}
\end{figure}

\begin{figure}
	\includegraphics[width=\columnwidth]{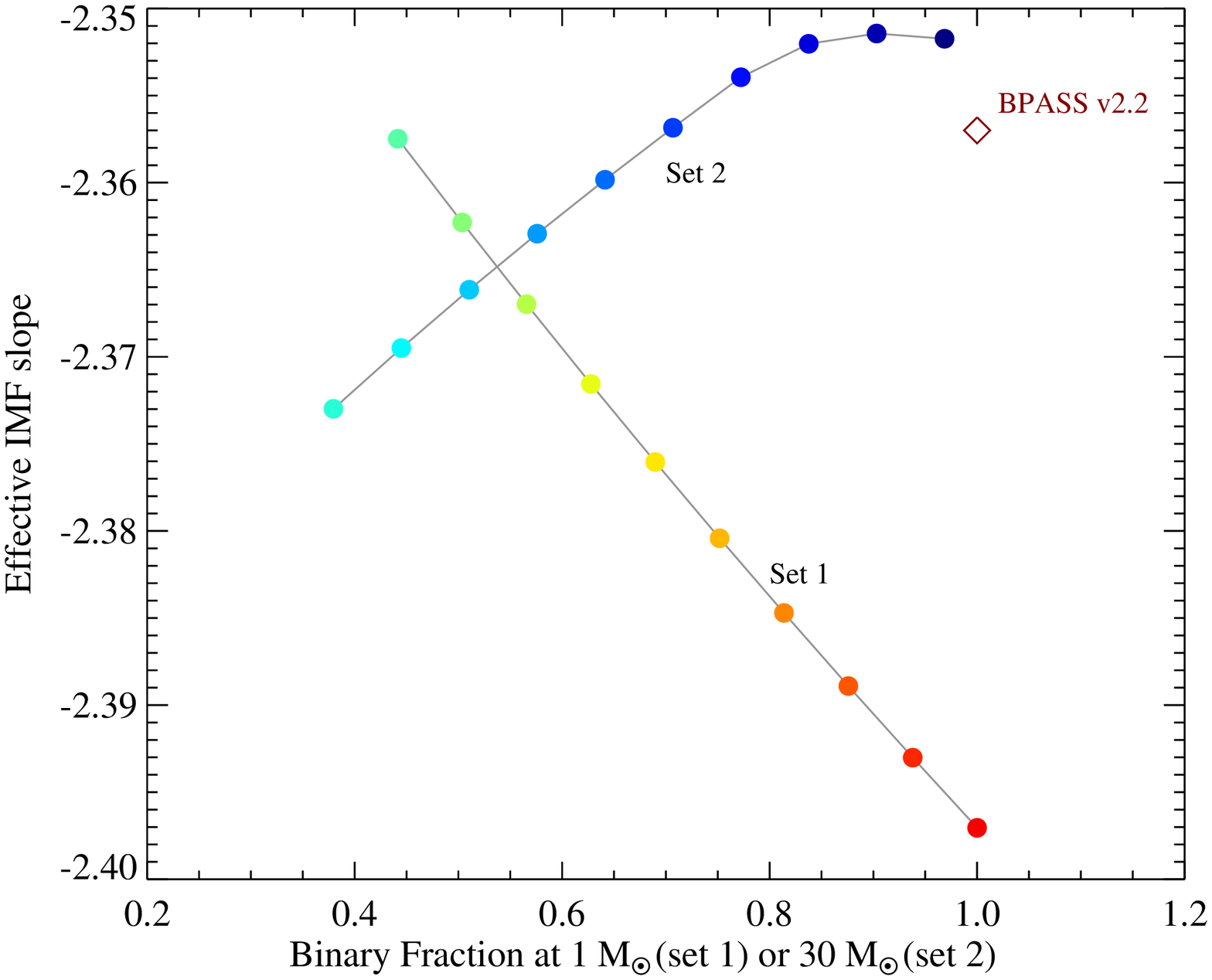}
    \caption{The effective IMF slopes (determined from the full stellar population between 0.5 and 300\,M$_\odot$) for the variant models described in section \ref{sec:base}. In each case, the nominal IMF has a slope of -2.350, and secondary stars act to reduce the IMF slope. The slope is shown against the binary fraction at 1\,M$_\odot$ for set 1 and the binary fraction at 30\,M$_\odot$ for set 2 and BPASS v2.2.}
    \label{fig:imf_base}
\end{figure}

The implementation of initial mass function in BPASS follows the latter approach. The IMF is presumed to apply to the sum of primary stars and single stars (which together dominate the mass), but not to the secondary stars. As discussed in section \ref{sec:nion}, during the population synthesis, primary stars are weighted by mass according to the selected IMF. The weighting for each primary mass bin is then subdivided between single and binary models, and the binary weighting is further subdivided between models with a range of periods and mass ratios. The mass of secondary companions are therefore not included in that initial IMF weighting, but are included in the final normalisation of the weighting scheme to a total of $10^6$\,M$_\odot$. The models are sampled statistically rather than stochastically, and as such may be inappropriate in situations where a low starburst mass results in a truncated or sparsely sampled upper mass limit to the initial mass function  \citepapp[although this only applies in stellar populations with total masses $\lesssim10^3$\,M$_\odot$, see][for discussion]{2012MNRAS.422..794E}.

Fig.~\ref{fig:imf_v2.2} illustrates this process for the canonical ``imf135\_300" version of BPASS v2.2. The number (i.e. final model weighting) of stars in each mass bin is presented separately for single stars, primaries and secondaries, as well as for the total population. The resultant lines are not entirely smooth - the result of the limited grid of mass ratios permitted for each primary star mass, which result in some secondaries being shifted to the nearest available model bin. A power law function of the form $N(M)\propto M^{-\alpha}$ is then fit to the total population between 0.5 and 300\,M$_\odot$ to produce an effective power law slope. For the canonical BPASS v2.2. model the nominal slope of -2.350 is altered to an effective power law slope of -2.357: an insignificant change compared to the large slope differences between IMFs being modelled in that data release (-2.7, -2.35, -2.0). 

The small change is largely the result of the dependence of binary fraction on primary mass. The high mass stars which have many companions are relatively few in number to begin with, while the abundant low mass stars have relatively few companions. Thus the perturbation to the IMF is small. The same cannot be said for all the variant prescriptions explored in this study. The extreme variants explored in section \ref{sec:base} include those where the binary fraction is near constant with mass, such that the fraction of low mass stars with companions approaches unity. Fig.~\ref{fig:imf_base} demonstrates the effect of these models on the effective initial mass function (again with a nominal slope of -2.35, and assuming a fixed mass ratio distribution). As the figure shows, the effect of either decreasing the high mass binary fraction, or increasing the low mass binary fraction is to steepen the effective IMF slope in the models, with the extreme cases shown in the figure reaching a slope of -2.40. 

\begin{figure}
	\includegraphics[width=\columnwidth]{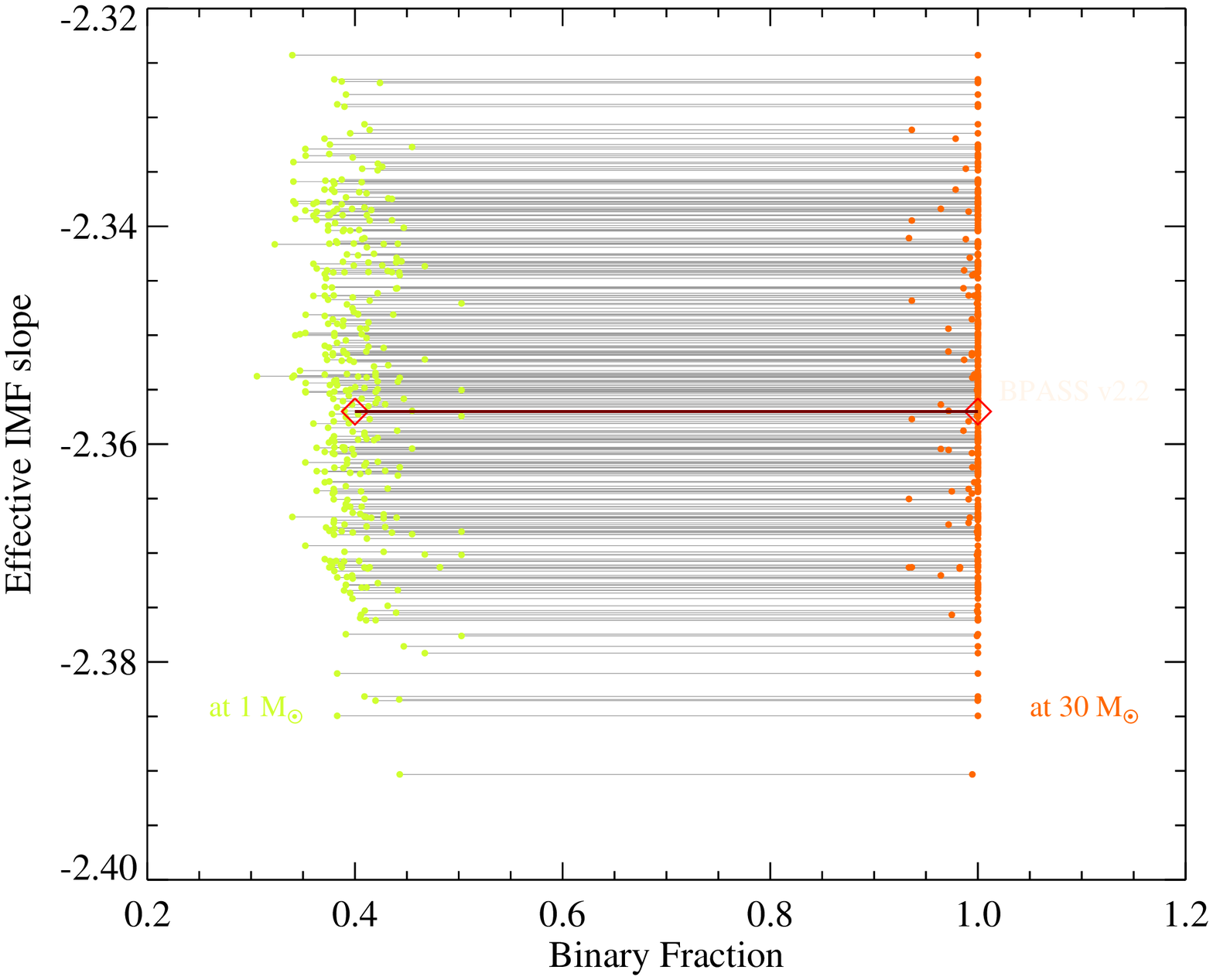}
    \caption{The effective IMF slopes (determined from the full stellar population between 0.5 and 300\,M$_\odot$) for the variant (v) models described in section \ref{sec:params}. In each case, the nominal IMF has a slope of -2.350, and secondary stars act to reduce the IMF. The slope is shown against the binary fraction at 1\,M$_\odot$ and at 30\,M$_\odot$, with the pairs of values at these two masses linked by a line. The BPASS v2.2 canonical distribution is shown by the thick line.}
    \label{fig:imf_mix}
\end{figure}

We have also calculated the effective IMF slope measured in the variant (v) models of section \ref{sec:params} and show these in Fig.~\ref{fig:imf_mix}. Note that in this case there is no clear trend in IMF slope with overall binary fraction for primary stars at either 1\,M$_\odot$ or 30\,M$_\odot$, since the binary fraction itself is varying over a relatively narrow range (as permitted by empirical constraints) and perturbations in the mass ratio distribution dominate the effective IMF slope determination.

\input{appendix.bbl}


\bsp	

\end{document}